\documentclass[usenatbib]{mnras}
\usepackage{epsfig}
\usepackage{amsmath,amssymb}
\usepackage{color}
\usepackage{bbding} 
\def\aj{AJ}                   
\def\apj{ApJ}                 
\def\apjl{ApJ}                
\def\apjs{ApJS}               
\def\aap{A\&A}                
\def\mnras{MNRAS}             
\newcommand{\Msolar}{\mbox{\,$\rm M_{\odot}$}}        
\newcommand{\Rsolar}{\mbox{\,$\rm R_{\odot}$}}        
\newcommand{\Lsolar}{\mbox{\,$\rm L_{\odot}$}}        

  \newcommand{\Teff}{\mbox{\,\em T$_{\rm eff}$}}         
  \newcommand{\sg}{\mbox{\,log $g$}}                     
 \newcommand{\teff}{\mbox{\,$T_{\rm eff}$}}      

  \newcommand{\kmsec}{\,\mbox{$\mbox{km}\,\mbox{s}^{-1}$}}    
  \newcommand{\kelvin}{\,\mbox{K}}                       
  \def\simge{\mathrel{\raise1.16pt\hbox{$>$}\kern-7.0pt
    \lower3.06pt\hbox{{$\scriptstyle \sim$}}}}           
  \def\simle{\mathrel{\raise1.16pt\hbox{$<$}\kern-7.0pt
    \lower3.06pt\hbox{{$\scriptstyle \sim$}}}}           

\title[BLAP theory]{Linear and non-linear models for large-amplitude radial pulsation in faint blue stars (BLAPs)}

\author[C. S. Jeffery]
       {C. S. Jeffery\thanks{E-mail: simon.jeffery@armagh.ac.uk} \\
Armagh Observatory and Planetarium, College Hill, Armagh BT61 9DG, Northern Ireland\\
}

\date{Accepted 2025 March 07, 
      Received 2025 March 07 ;
      in original form 2024 November 29}

\pagerange{\pageref{firstpage}--\pageref{lastpage}}
\pubyear{}

\begin{document}

\maketitle

\label{firstpage}

\begin{abstract}
The recent discovery of large-amplitude pulsations in faint blue stars (BLAPs) provides both challenges for stellar pulsation theory and opportunities to explore the late evolution of low-mass stars.  
This paper explores the radial-mode stability of stars across parameter space occupied by BLAPs. 
Models are constructed for homogeneous stellar envelopes and are agnostic of evolution. 
Linear non-adiabatic models demonstrate the major requirement for pulsations to be enrichment of iron and nickel in the driving zone to a few times the solar abundance. 
There is no constraint on mass. 
Non-linear models demonstrate that BLAP pulsations will be of large amplitude and will show strong shocks at minimum radius. 
A variety of light-curve shapes are found across the BLAP instability strip, accounting for the variety observed. 
Linearised period relations are derived from the non-linear models.
The phase of maximum luminosity relative to minimum radius is correlated with effective temperature (\Teff), preceding for cool stars and following for hot stars, and split if close to minimum radius.  
In both linear and non-linear cases, most models pulsate in the fundamental mode (F).
First-overtone (1H) pulsations are excited on the low luminosity blue side of the instability region and become more prevalent at higher mass. 
The period ratio $P_{\rm 1H}/P_{\rm F}=0.81$ contrasts with the classical Cepheid value (0.70 - 0.75).
The transition from F to 1H mode pulsations follows a period-mass relation; the F-mode pulsators adjacent to the transition show a reverse shock.  
At high \Teff\ some non-linear models show unstable overtone modes up to 5H and multi-mode behaviour. The linear and non-linear analyses concur on the red-edge of the instability region, but the non-linear blue edge is  hotter.  
\end{abstract}

\begin{keywords}
hydrodynamics, 
shock waves,
stars: chemically peculiar, 
stars: low-mass, 
stars: oscillations, 
stars: subdwarfs, 
\end{keywords}

\section{Introduction}              
\label{intro}

This paper presents a theoretical study of the physics of pulsating faint blue stars, 
in particular those which have been designated blue  large-amplitude pulsators (BLAPs). 
These stars are interesting because their masses 
appear to be less than that required for core helium ignition, 
and the amplitude of their light variations can substantially exceed 10\%.   
The phenomena associated with such pulsations, including shocks and chemical diffusion, 
provide potentially superb tests for stellar astrophysics. 

\begin{figure}
        \centering
                \includegraphics[width=0.47\textwidth]{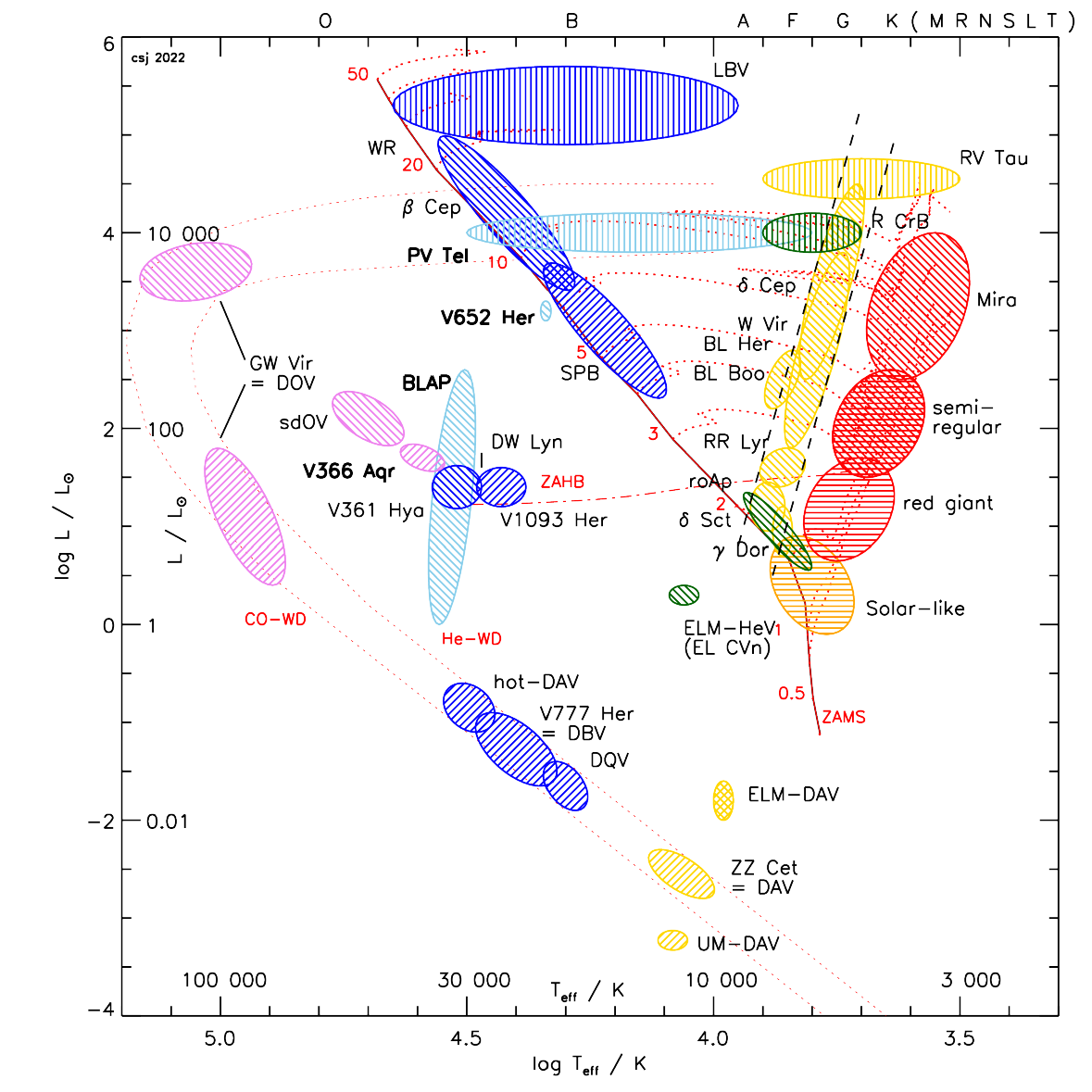}
\caption{The location of blue large-amplitude pulsating variables in the 
luminosity-effective temperature (or Hertzsprung-Russell) diagram in context with  the  approximate locations of other major classes of pulsating variable. 
Classes are coloured roughly by spectral type. 
The zero-age main sequence and horizontal branch, the Cepheid instability strip are marked. Evolution tracks are labelled by mass (\Msolar).
Shadings represent opacity-driven
p-modes ($\backslash\backslash\backslash$), g-modes (///) and strange modes ($|||$) and
acoustically-driven modes ($\equiv$).  
Approximate spectral types are indicated on the top axis.  
Based on figures by J. Christensen-Dalsgaard and subsequently by \citet{jeffery08.coast}.  }
        \label{f:puls_hrd}
\end{figure}

\begin{table*}
    \begin{center}
    \caption{Opacity mixtures used for linear and non-linear calculations. 
    Elemental abundances are given as mass fractions $\beta_i$ relative to the total metal fraction $Z$, rounded to 4 decimal places. 
    Cl, Ar, K, Ca, Ti, Cr, and Mn are included in the opacities but omitted from this table;  
    Their abundances scale with Si as defined in the original OPAL95 mixture \citep{grevesse93}.
    Type 2 tables are generated for $X = 0, 0.03, 0.10, 0.35$ and 0.70, for a specified value $Z$, and for multiple additional fractions of C and O (not used). 
    Type 1 tables are generated for the values of $X = 0, 0.1, 0.2, 0.35, 0.5, 0.7, 0.8, 0.9, 0.95$ and ($1-Z$), and for $Z=0.0, 0.0001, 0.0003, 0.001, 0.002, 0.004, 0.01, 0.02, 0.03, 0.06, 0.08$ and  $ 0.10$.
    }
    \label{t:mixes}
    \begin{tabular}{lrrrrrrrr}
    \hline
Code & \multicolumn{3}{c}{linear} & \multicolumn{3}{c}{non-linear} & Observed \\
\hline
Opacity & & & & &  & \\
Table  & Z02$^1$ & Z0242$^2$ & Z047 &  BP & FE$^3$ & N1$^4$ & BLAP-009$^5$  \\  
OPAL95 &  Type 2 & Type 2 & Type 2 & Type 1 & Type 1 &  Type 1 &  --- \\[1mm]
$X$  & --- & --- & --- & --- &  --- & --- &   0.475  \\
$Y$  & --- & --- & --- & --- &  --- &  --- &   0.478  \\
$Z$  & 0.02 & 0.0242 & 0.047 & --- &  --- &  --- &   0.047  \\[1mm]
$\beta_{\rm C}/{Z}$  & 0.1733 & 0.1028 & 0.2307 &  0.2307 & 0.1309 & 0.0256  & 0.2309 \\
$\beta_{\rm N}/{Z}$  & 0.0531 & 0.0315 & 0.2817 &  0.2817 & 0.1598 & 0.6429  & 0.2820 \\
$\beta_{\rm O}/{Z}$  & 0.4823 & 0.2860 & 0.1895 &  0.1895 & 0.1075 &  0.0499 & 0.1898 \\
$\beta_{\rm Ne}/{Z}$ & 0.0987 & 0.0585 & 0.0974 &  0.0974 & 0.0552 &  0.1170 & 0.0967 \\
$\beta_{\rm Na}/{Z}$ & 0.0020 & 0.0012 & 0.0019 &  0.0019 & 0.0011 &  0.0008 & --- \\
$\beta_{\rm Mg}/{Z}$ & 0.0376 & 0.0223 & 0.0407 &  0.0407 & 0.0231 & 0.0828  & 0.0402 \\
$\beta_{\rm Al}/{Z}$ & 0.0032 & 0.0019 & 0.0032 &  0.0032 & 0.0018 &  0.0017 & 0.0032 \\
$\beta_{\rm Si}/{Z}$ & 0.0405 & 0.0240 & 0.0382 &  0.0382 & 0.0217 &  0.0387 & 0.0381 \\
$\beta_{\rm P}/{Z}$  & 0.0004 & 0.0002 & 0.0004 &  0.0004 & 0.0003 &  0.0035 & 0.0042 \\
$\beta_{\rm S}/{Z}$  & 0.0211 & 0.0125 & 0.0091 &  0.0091 & 0.0052 &  0.0147 & 0.0091 \\
$\beta_{\rm Fe}/{Z}$ & 0.0718 & 0.4257 & 0.0913 &  0.0913 & 0.4616 & 0.0144 & 0.0917 \\
$\beta_{\rm Ni}/{Z}$ & 0.0045 & 0.0264 & 0.0046 &  0.0046 & 0.0255 & 0.0008 & ---  \\
\hline
Model & & & & &  & \\
Label & Z02 & Z0242 & Z047 &  BP & FE & N1 & BLAP-009  \\ 
$X$ &  0.3,0.475,0.7  & 0.3,0.475,0.7  & 0.475  &  0.475 &  0.475  &  0.475  & 0.475  \\
$Z^6$ &   0.02  & 0.0242 & 0.047 &  0.047 & 0.047 & 0.047 & 0.047 \\
Fe/Fe$_{\odot}$ &  1  &  7  &  3  &  3  & 15 & 0.47 & 3 \\
Ni/Ni$_{\odot}$ &  1  &  7  & 2.4 & 2.4 & 13 & 0.42 & --- \\
    \hline
    \end{tabular}\\
    1: \citet{grevesse93}, 2:  Z02$\times7$Fe,Ni,  3: BP$\times5$Fe,Ni,   4: \citet{jeffery22a},  5: \citet{bradshaw24} \\
    6: Values of $Z$ used in equation of state where $\mu$ is dominated by H, He and light elements.  
    \end{center}
\end{table*}

\subsection{Observational Background}

The first blue  large-amplitude pulsating variable stars  (BLAPs) were discovered as part of the OGLE survey by  \citet{pietrukowicz17} and identified as stars of low mass ($<0.4\Msolar$) and early B spectral type. 
A second group was discovered  by \citet{kupfer19}, having surface gravities somewhat higher and pulsation periods somewhat shorter than those of the first.  
Subsequently, over 100 BLAPs have now been identified in large-scale surveys 
\citep{mcwhirter22,ramsay22,borowicz23,borowicz24,pietrukowicz24}, 
as well as individual discoveries \citep{lin23,pigulski22,chang24} (Fig.\,\ref{f:puls_hrd}).   

Defining features include periods in the range 10 -- 80 min, I-band amplitudes in the range 0.1 -- 0.4 mag, 
and non-sinusoidal and variable light curves  \citep{pietrukowicz24}.
Effective temperatures (\teff) lie in the range 24 -- 34 kK, surface gravities ($g$) are between 4.0 and 5.5 dex (cm\,s$^{-2}$), 
and surface helium-to-hydrogen ratios are between --3 and --0.5 dex by number (ibid). 
There is some evidence of a period-amplitude correlation, and clear evidence of a period-gravity correlation.
The latter implies most BLAPs are likely fundamental-mode (F) radial pulsators and the remainder are first-overtone (1H) radial pulsators  (ibid.). 
Measurements of the rate of period change ($\dot{P} \equiv {\rm d}P/{\rm d} t$) show both positive and negative values (ibid.). 

Phase-resolved spectroscopy of OGLE-BLAP-009 provides detailed insight into the properties of a typical BLAP with a period of 32 min \citep{bradshaw24}. 
Notable features include the presence of a sharp dip at maximum light, and a steep acceleration in the radial-velocity around minimum radius followed by near constant deceleration over the remaining 80\% of the pulsation cycle. 
An abundance analysis of the mean spectrum indicates that the surface is slightly helium enriched (+0.4 dex), strongly nitrogen enriched  (+1.4 dex), 
and moderately metal-enriched ($\approx+0.6$ dex)  relative to solar. 
Combining multiple direct radius measurements, \citet{bradshaw24} find radius and mass constraints in strong agreement with a low-mass helium-core pre-white dwarf of $\approx0.30\Msolar$. 

\subsection{Theoretical Background}

Since their discovery, large-amplitude pulsations in faint-blue stars have been associated with the so-called `Z-bump' instability strip  \citep{pietrukowicz17}.
That is, pulsations are driven by the opacity or $\kappa$ mechanism via a peak in the opacity of iron-group elements at temperatures 
around $2\times10^5$ \kelvin\ in stars which have surface temperatures approximately between 20\,000 and 30\,000 \kelvin. 
This mechanism is also responsible, {\it inter alia}, for $\beta$ Cepheid pulsations in main-sequence B stars \citep{moskalik92}, V361\,Hya pulsations in subdwarf B stars  \citep{charpinet97a}
and V652\,Her pulsations in some extreme helium stars  \citep{saio93}. 
These classes include both radial and non-radial acoustic (p-) mode pulsations, and most classes include cases where the amplitudes exceed 0.1 mag.  
They are also associated with non-radial gravity (g-) modes which dominate  at lower effective temperatures, including the slowly-pulsating B stars \citep{dziembowski93} and the V1093\,Her sdB pulsators \citep{charpinet97b,fontaine03} 

The iron-peak opacity mechanism operates most efficiently at high luminosity-to-mass ratios ({\it e.g.} the  $\beta$ Cepheids) or where the opacity gradients $\partial \kappa/\partial T$ 
either side of the peak are exaggerated either by an increase in the abundance of one or more iron-group elements or by a reduction in the hydrogen abundance 
\citep{jeffery16a}. 

In their discovery paper, \citet{pietrukowicz17} examined two models, a 1\Msolar\ helium main-sequence star with an inflated envelope and a stripped red-giant (RG) core with a mass $\approx 0.3\Msolar$, favouring the latter. 
Whilst temperatures, gravities and periods for both models broadly matched those of observed BLAPs, neither model showed unstable radial modes, essentially because the radiative levitation of iron was not considered.  
\citet{romero18} also found models equivalent to the low-mass stripped RG cores would account for BLAPs, observing that these are the hot precursors of extremely low-mass white dwarf stars. 
\citet{byrne18b} included radiative levitation in their calculation of two BLAP models, one being a 0.46\Msolar\ pre-helium main-sequence star contracting from the tip of the giant branch, and the other being a 0.31\Msolar\ stripped RG or pre-white dwarf (pre-WD). 
Both models are unstable to pulsations and demonstrate clearly that BLAPs are excited by Z-bump driving in a low-mass star; timescale considerations favoured the pre-WD model.  
In their discovery of high-gravity BLAPs, \citet{kupfer19} showed that the stripped RG or pre-white dwarf (pre-WD) models with masses in the range $0.26 - 0.30\Msolar$ give temperatures, gravities and periods consistent with observations. 
They did include radiative levitation in their models and found that their non-adiabatic stability  agreed with those of \citet{byrne18b}.
Subsequently, \citet{byrne20} computed the evolution of pre-WD models over a range of masses; they included the radiative levitation of iron-group elements and calculated the pulsation stability of each model. 
The results show how iron accumulates in the Z-bump layer and pulsations are excited precisely in the region of $\teff-g$ space where both low- and high-gravity BLAPs occur. 
 
The stripped-giant or pre-WD model implies that the star should have a blue-straggler companion. 
Only one binary BLAP has been found so far \citep{pigulski22}.
\citet{byrne21} used binary-star population synthesis to examine the orbital-period distribution of BLAP-hosting binaries and showed that in the majority of cases, the BLAP is the primary star and only a small fraction would be in short-period systems. 

Another implication of the pre-WD model is that BLAPs should be contracting, implying a negative $\dot{P}$. 
Since several have a positive $\dot{P}$, the pre-WD model might not be  unique. 
Other proposals for the evolution of BLAPs include 
a core-helium burning star in the middle or late phase of helium burning \citep{wu18}, 
a helium-shell burning subdwarf which is expanding after the completion of core-helium burning \citep{lin23,xiong22}, 
the surviving companion of Type Ia supernovae \citep{meng20},
the product of the merger of a helium-core white dwarf with a low-mass main-sequence star \citep{zhang23}, 
and the merger of low-mass white dwarfs \citep{kolaczek24}
Of these, only \citet{zhang23} and \citet{kolaczek24}  established that their models for BLAPs are unstable to pulsation through the combined action of hydrogen depletion and radiative levitation of iron and nickel.   

Since the excitation of radial-mode pulsations is a consequence of the opacity structure of the stellar envelope, it follows that there may be multiple ways for a star of a given mass and initial composition to reach a configuration in which such a pulsation is excited. 
Thus studies that do not depend explicitly on any one evolution model are valuable; \citet{jadlovsky24} have computed a large grid of linear pulsation models over a wide mass, luminosity and temperature range (0.3 -- 1.1 \Msolar), but for a fixed chemical composition ($X=0.7, Z=0.05$); they identify instability regions, calculate period ratios for fundamental and first overtone modes, and provide theoretical period relations. 
Again, the general overview provided by \citet{jeffery16a} is instructive. 

\subsection{Objectives}
\label{s:objectives}

The global properties of BLAPs, including periods, temperatures, gravities, radii, both as individual and ensemble provide powerful constraints on likely families of evolution models. 
To succeed, the latter must a) pass through the observed BLAP parameter space, b) excite pulsations in the observed instability regions, and
c) expand or contract at a rate commensurate with observed values of $\dot{P}$. 
At present, more than one evolution path leads to envelope properties commensurate with the pulsation properties of BLAPs (the structure condition). 
Appropriate diffusion calculations that include radiative levitation can demonstrate which of these will have sufficient opacity in the driving zone to drive pulsations (the opacity condition). 
Neither approach maps the overall parameter space where BLAPs might be found, nor quantifies the relationship between pulsation period, amplitude and internal structure, including opacity, nor accounts for the diverse shapes of BLAP light curves. 

Several questions arise relating to the pulsations themselves, and these affect how they may be used as diagnostics of evolution. 
For example:
Observed stability boundaries are increasingly well established; how do these boundaries constrain the physics?  
Is the fundamental radial mode always excited and if not, when do first or higher overtone pulsations dominate?  
Why are BLAP amplitudes so large, particularly in comparison with the V361\,Hya (sdB) pulsators? 
In view of their large amplitude, are linear diagnostics reliable? 
Do sharp dips at light maximum, and high accelerations at minimum radius indicate the presence of shocks?
How can light and radial-velocity curves be used as diagnostics?  

The quality of recent data and variety of light and radial-velocity curves demands explanation with non-linear models. 
This paper therefore aims to (i) map the BLAP instability strip as a function of envelope composition, (ii) compare results for linear and non-linear pulsation calculations, (iii) explore the relationships between BLAP amplitude, mass (or luminosity) and envelope composition (or opacity), and (iv) interpret the diversity of BLAP light curve shapes. 
To do so, it will be agnostic of evolution and will only consider the pulsation properties of homogeneous stellar envelopes. 

\begin{figure*}
\begin{center}
\vspace{-20mm}
\includegraphics[width=0.32\textwidth]{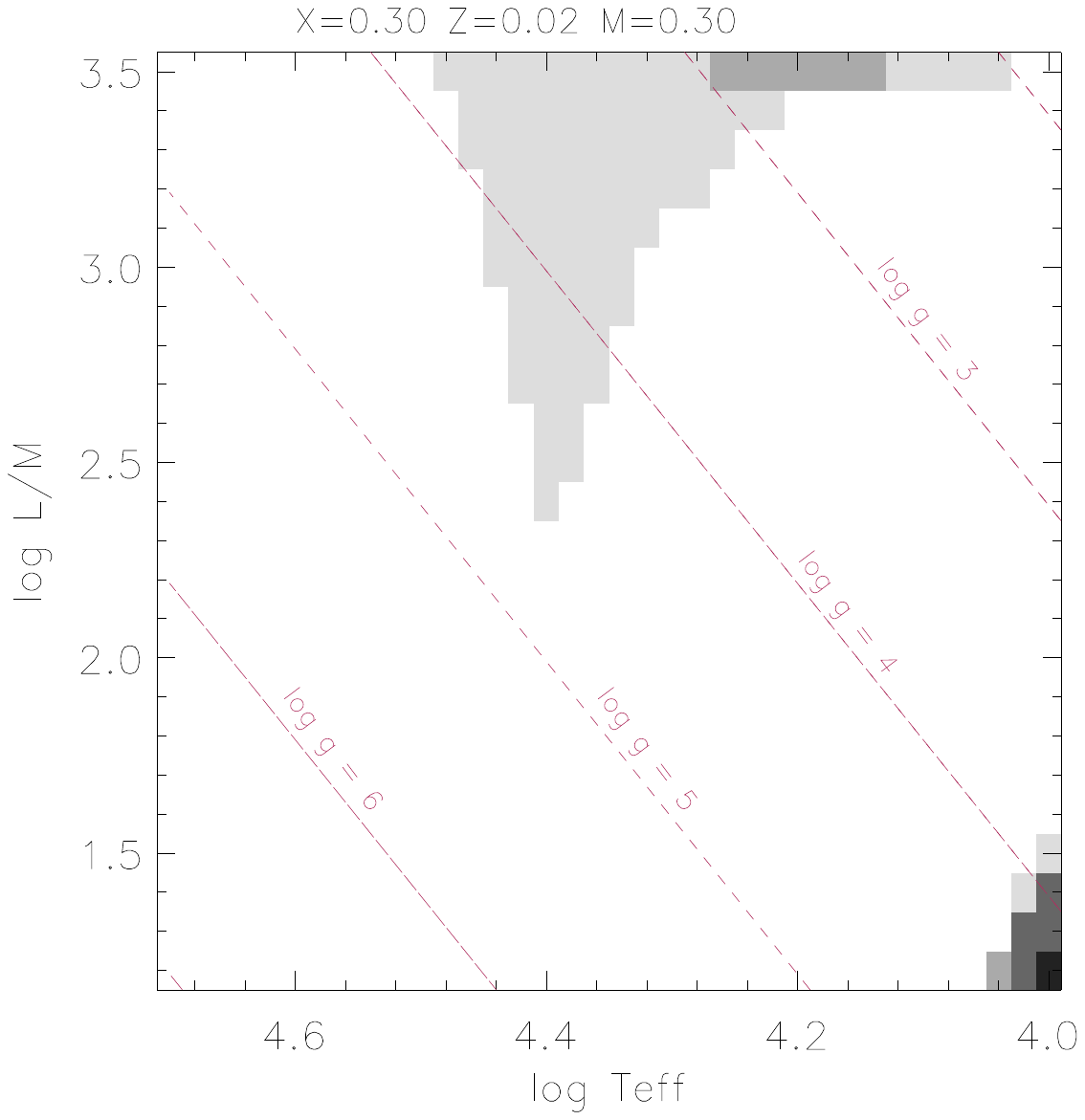}
\includegraphics[width=0.32\textwidth]{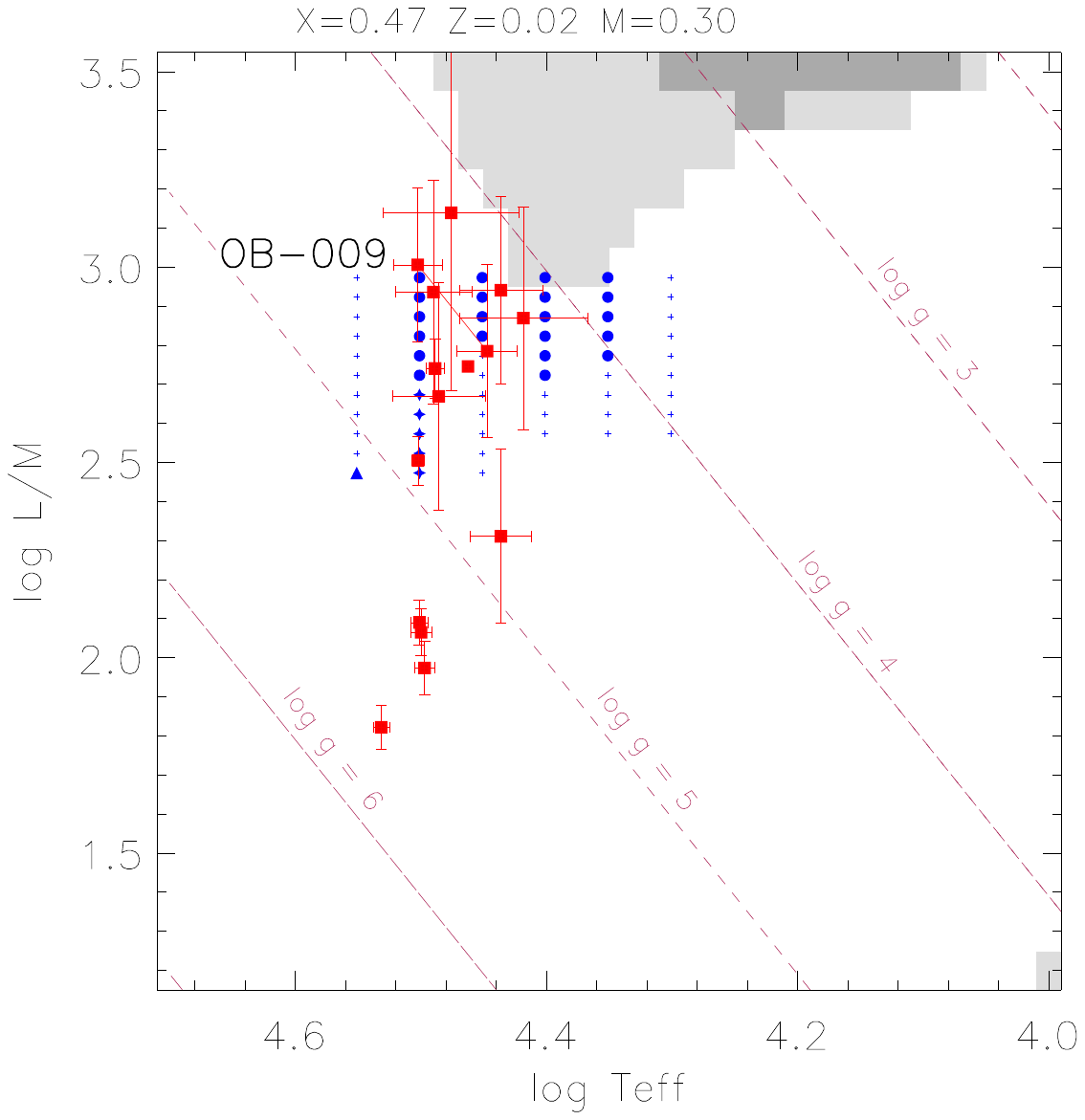}
\includegraphics[width=0.32\textwidth]{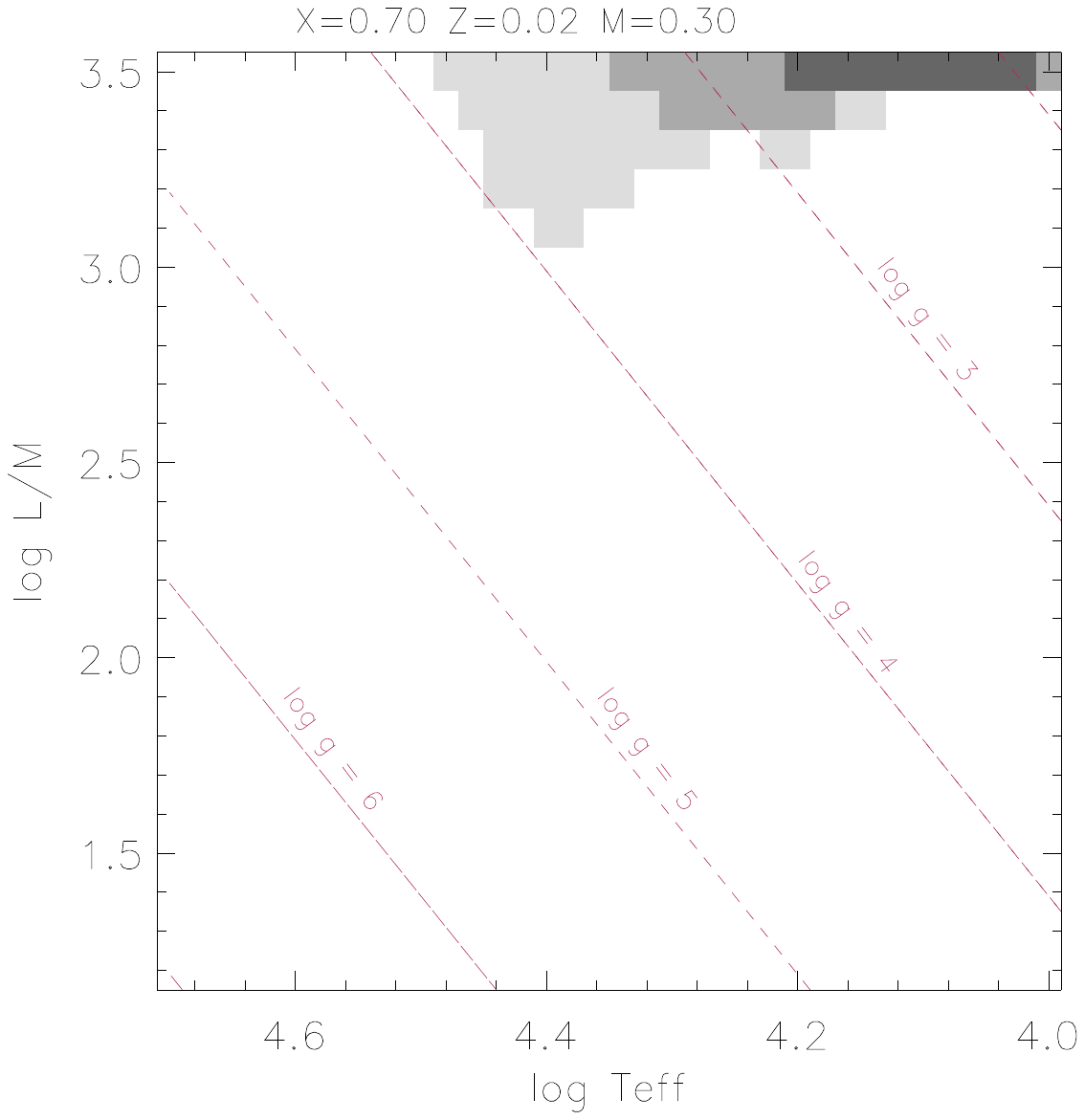}\\
\caption[Unstable modes as f(X) at Z=0.02]
{Unstable  modes in linear non-adiabatic  and non-linear pulsation models of stars with homogeneous
envelopes for selected compositions and masses, as labelled. 
This figure compares models with mass $M=0.30$\Msolar, metallicity $Z=0.02$ (approximately solar) and hydrogen abundances (by mass fraction) $X=0.30, 0.475$ and $0.70$. 
The second value corresponds to that measured for OGLE-BLAP-009 by \citet{bradshaw24}. 
The {\it number} of unstable radial modes in the {\it linear} non-adiabatic model is represented by grey scale contours, with the lightest shade marking the instability boundary
(one unstable mode), and the darkest shade representing five or more more unstable modes. 
 Blue symbols indicate {\it non-linear} models (\S\,\ref{s:nl_models}) showing large-amplitude radial pulsations; $\bullet$: F-mode, $\blacklozenge$: 1H, $\blacktriangle$: 2H, $\blacktriangledown$: 3H, \FourStar: $k>3$, $+$: low-amplitude and stable models. 
Red squares with error bars (selected panels) represent the observed positions of BLAPs as tabulated by \citet{zhang23}. 
Two positions are shown for OGLE-BLAP-009 (labelled OB-009), connected by a solid red line, indicating $T_{\rm eff}$ given by \citet{pietrukowicz17} and \citet{bradshaw24}. Surface gravities are the same in both cases.
 Broken (maroon) diagonal lines represent contours  of constant surface gravity at  $\log g = 6, 5, \ldots, 2$.   
}
\label{f:nmZ02}
\end{center}
\end{figure*}

\begin{figure*}
\begin{center}
\vspace{-20mm}
\includegraphics[width=0.32\textwidth]{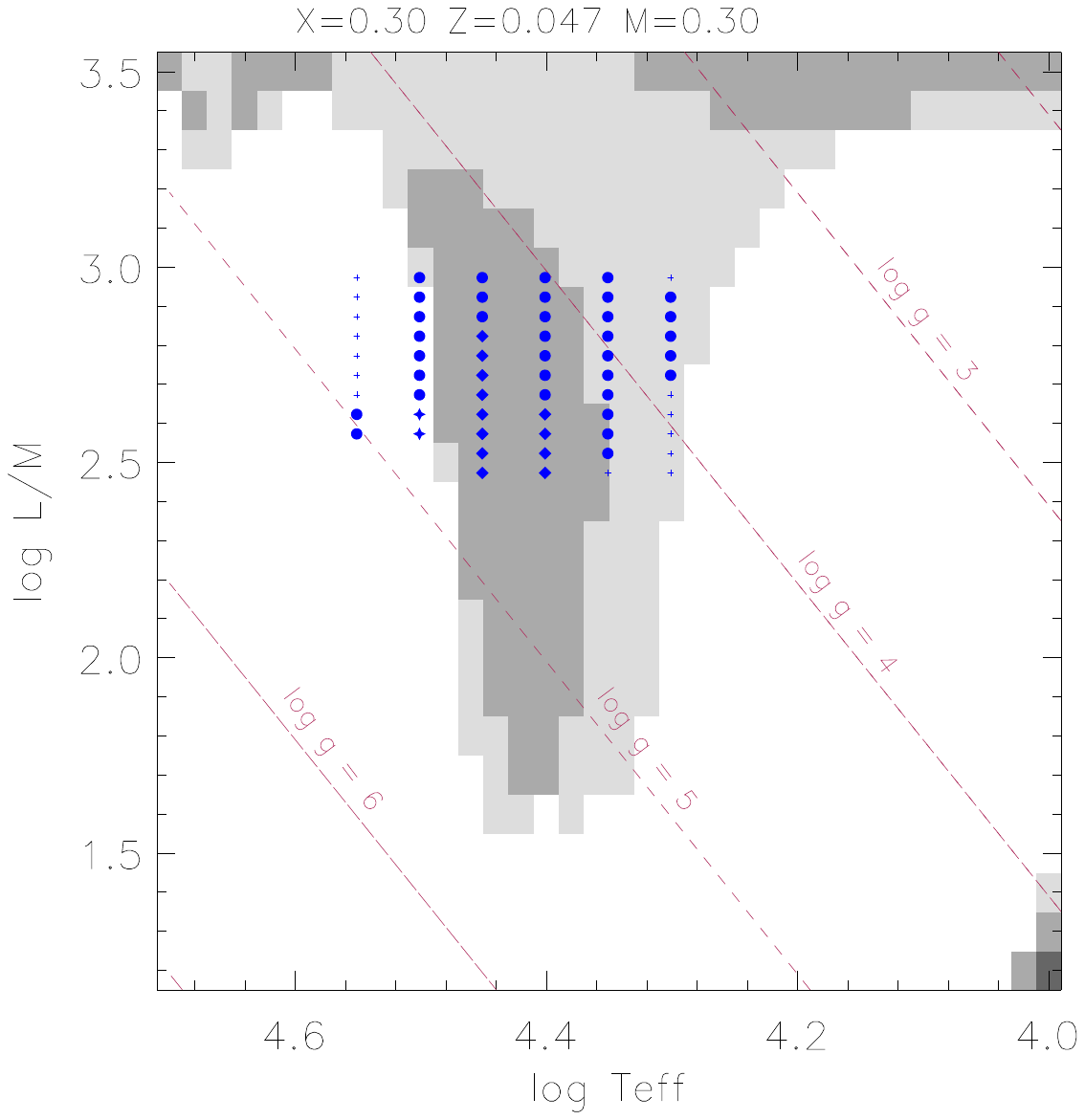}
\includegraphics[width=0.32\textwidth]{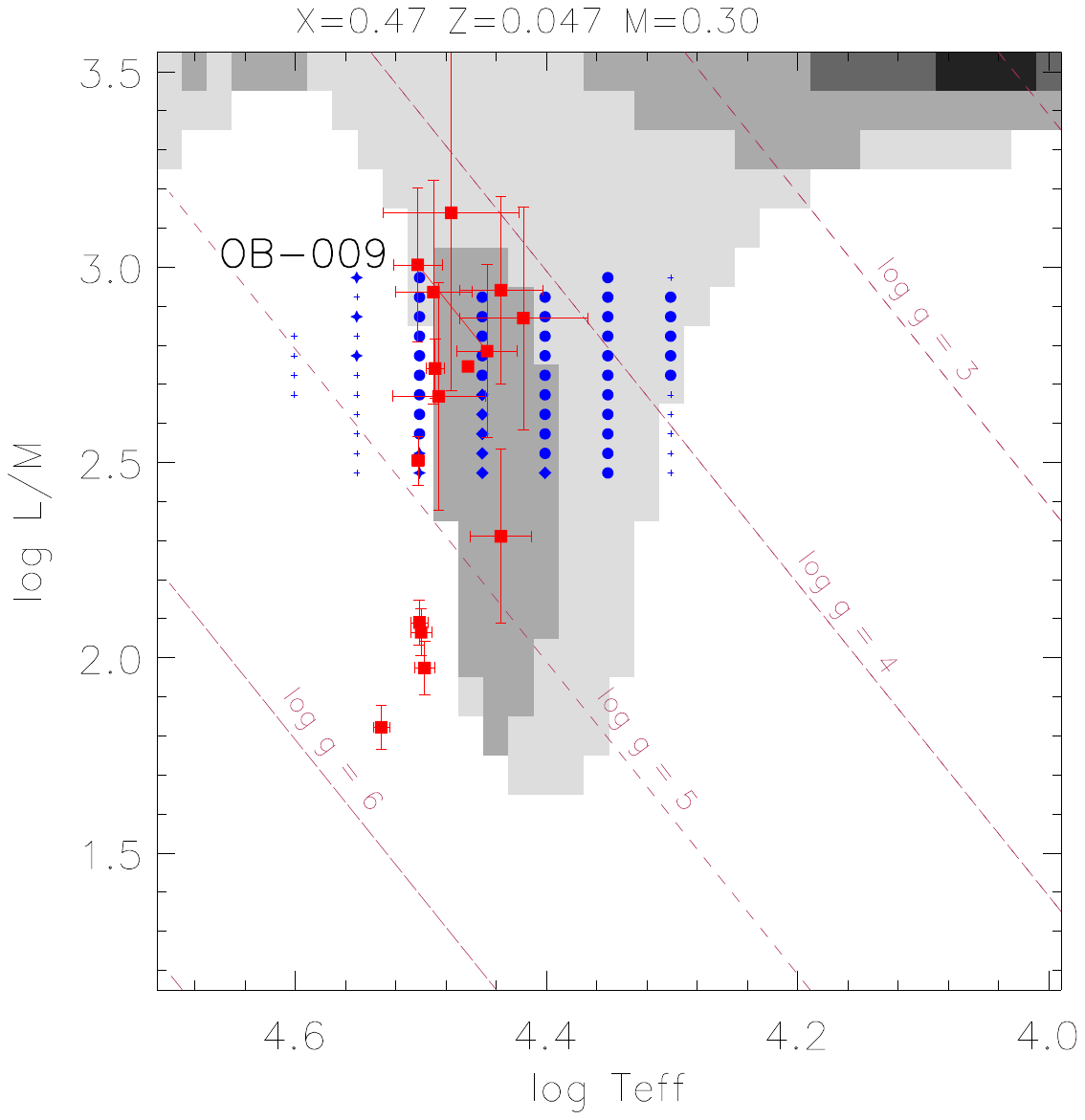}
\includegraphics[width=0.32\textwidth]{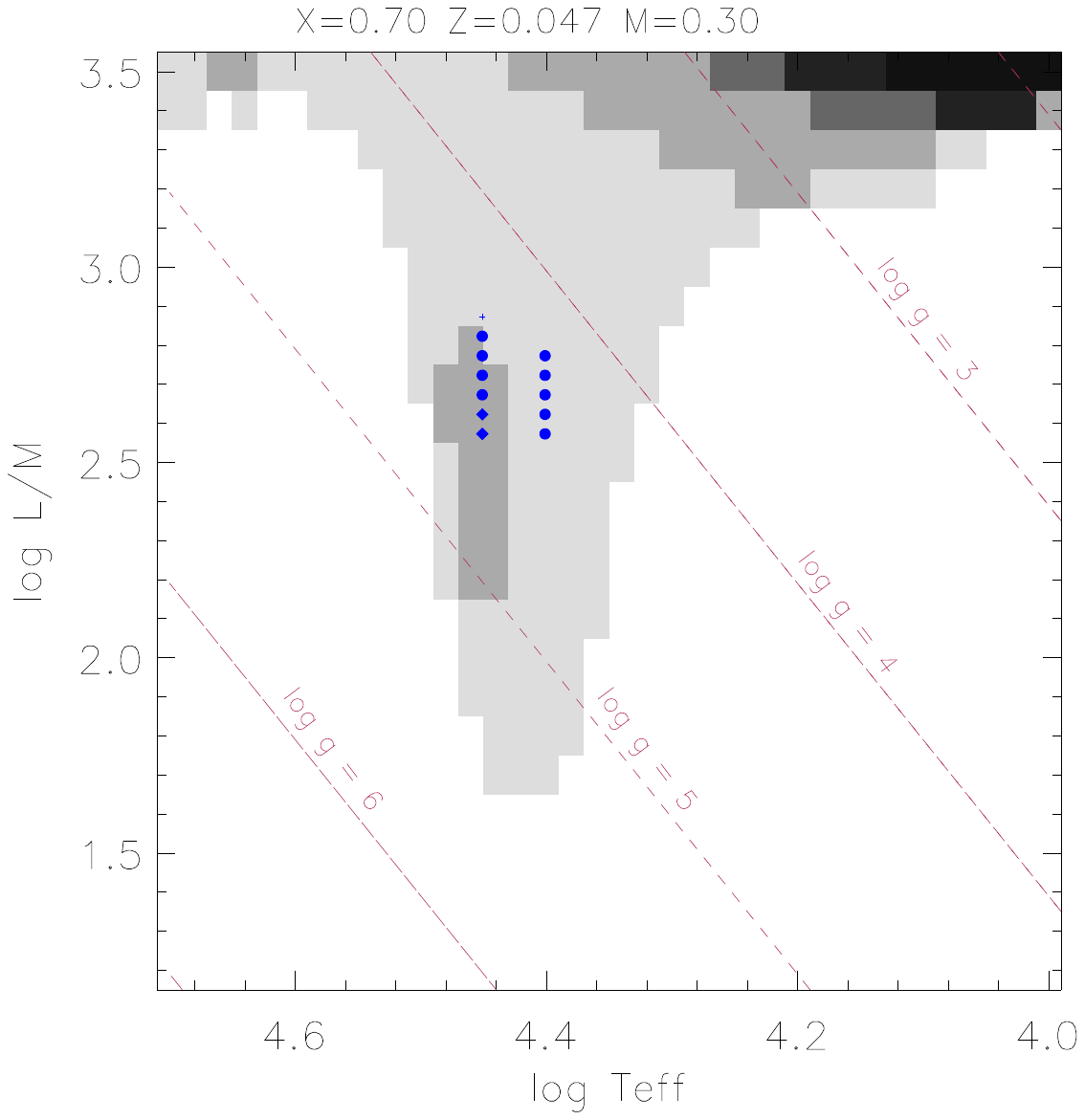}\\
\caption[Unstable modes as f(X) at Z=0.047]
{As Fig.\,\ref{f:nmZ02} for models with mass $M=0.30$\Msolar, metallicity $Z=0.047$ and hydrogen abundances (by mass fraction) $X=0.30, 0.475$ and $0.70$. 
The distribution of metals has been selected to match that measured for OGLE-BLAP-009 by \citet{bradshaw24}.    
}
\label{f:nmZ047}
\end{center}
\end{figure*}

\begin{figure*}
\begin{center}
\vspace{-20mm}
\includegraphics[width=0.32\textwidth]{nmodes/nmodes_x475z02m0.30_00_opal.pdf}
\includegraphics[width=0.32\textwidth]{nmodes/nmodes_x475z047BPm0.30_00_opal.pdf}
\includegraphics[width=0.32\textwidth]{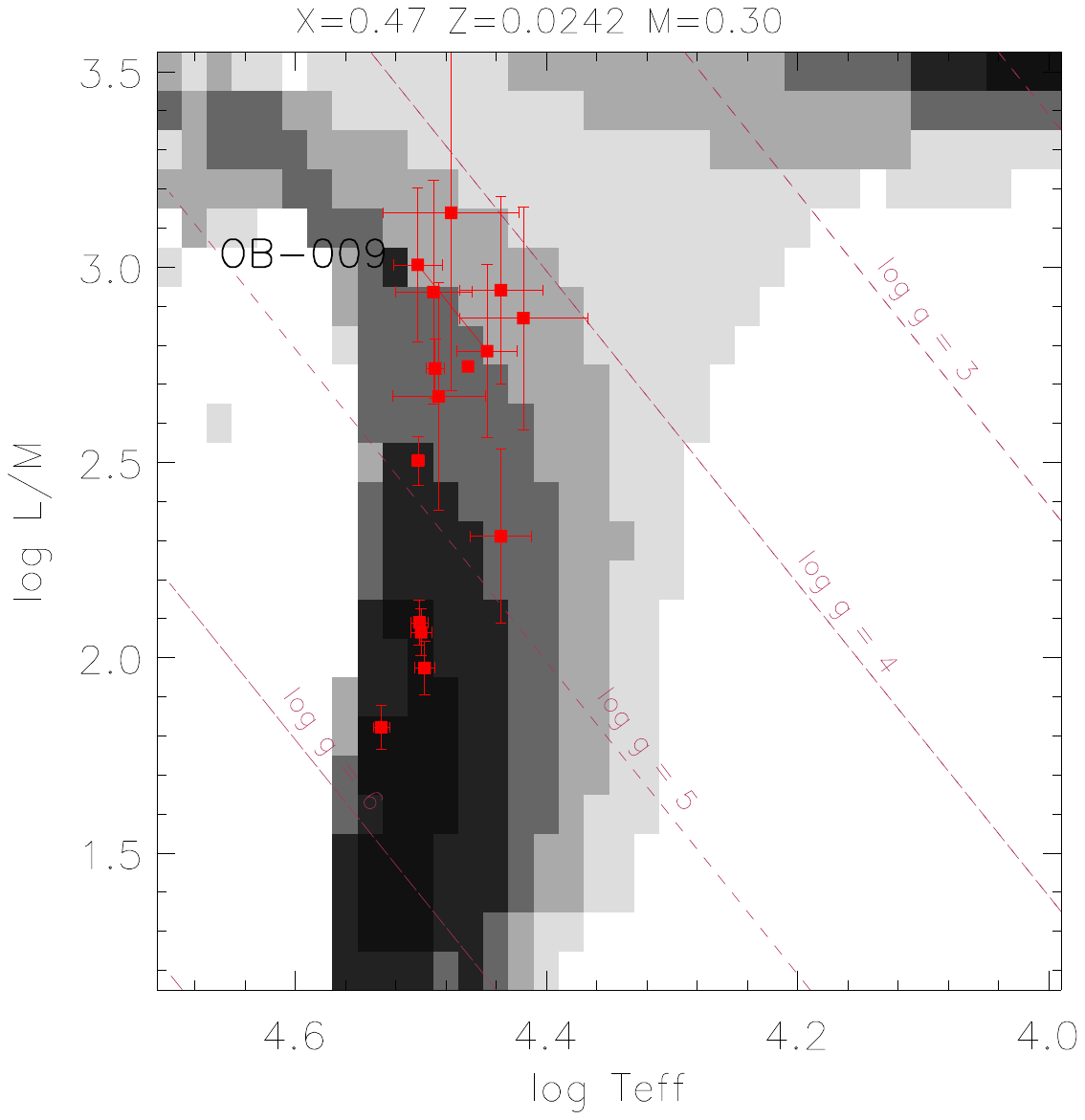}\\
\caption[Unstable modes as f(Z)]
{As Fig.\,\ref{f:nmZ02} for models with mass $0.30$\Msolar, hydrogen abundance (by mass fraction) $X=0.475$, and metallicity.  $Z=0.02, 0.047$ and 0.0242. 
The fraction of iron and nickel relative to other metals is solar for $Z=0.2$, $4\times$\,solar for $Z=0.047$, and { $7 \times$\,}solar for $Z=0.0242$. 
}
\label{f:nmZ}
\end{center}
\end{figure*}

\begin{figure*}
\begin{center}
\vspace{-20mm}
\includegraphics[width=0.32\textwidth]{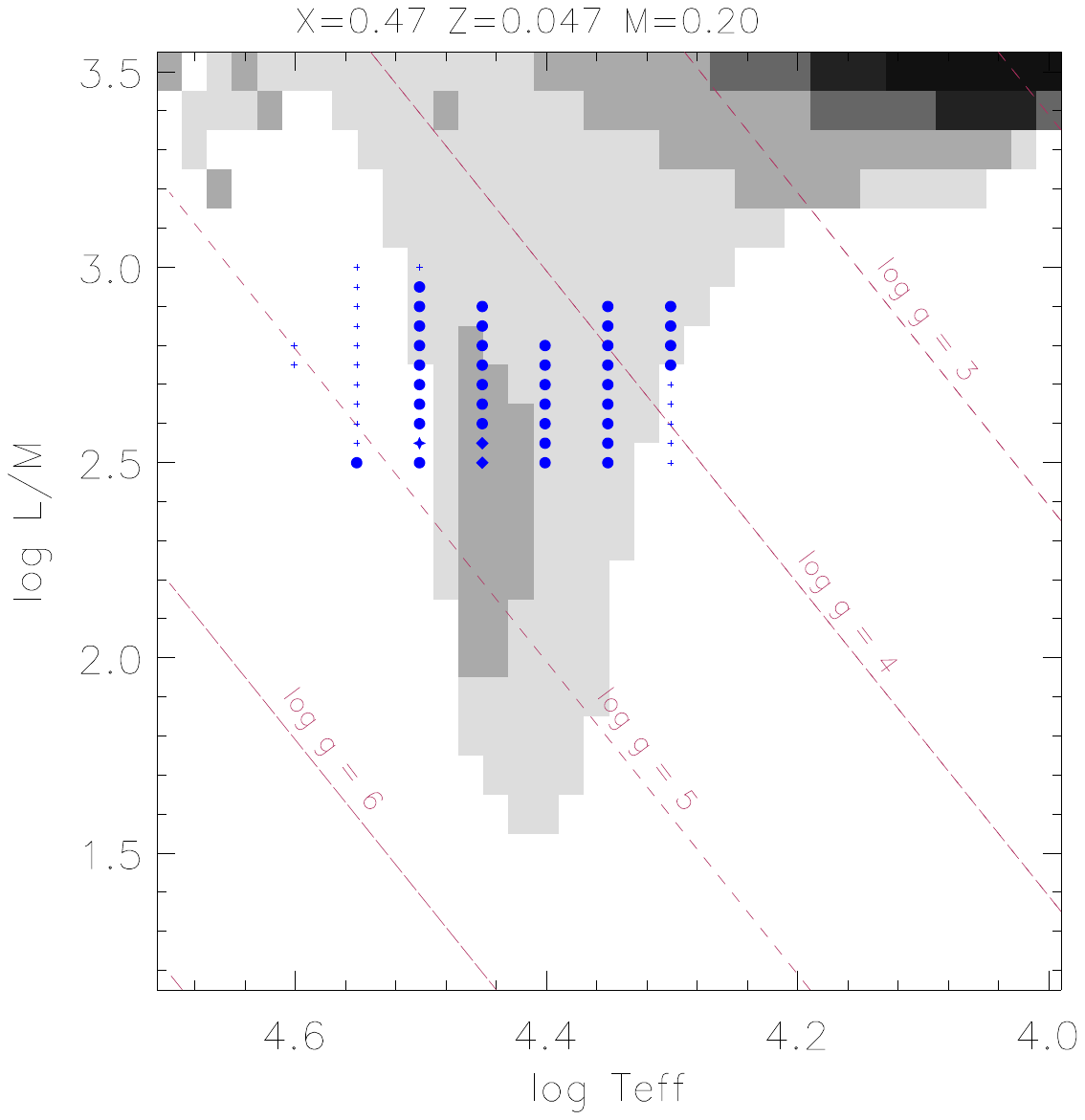}
\includegraphics[width=0.32\textwidth]{nmodes/nmodes_x475z047BPm0.30_00_opal.pdf}
\includegraphics[width=0.32\textwidth]{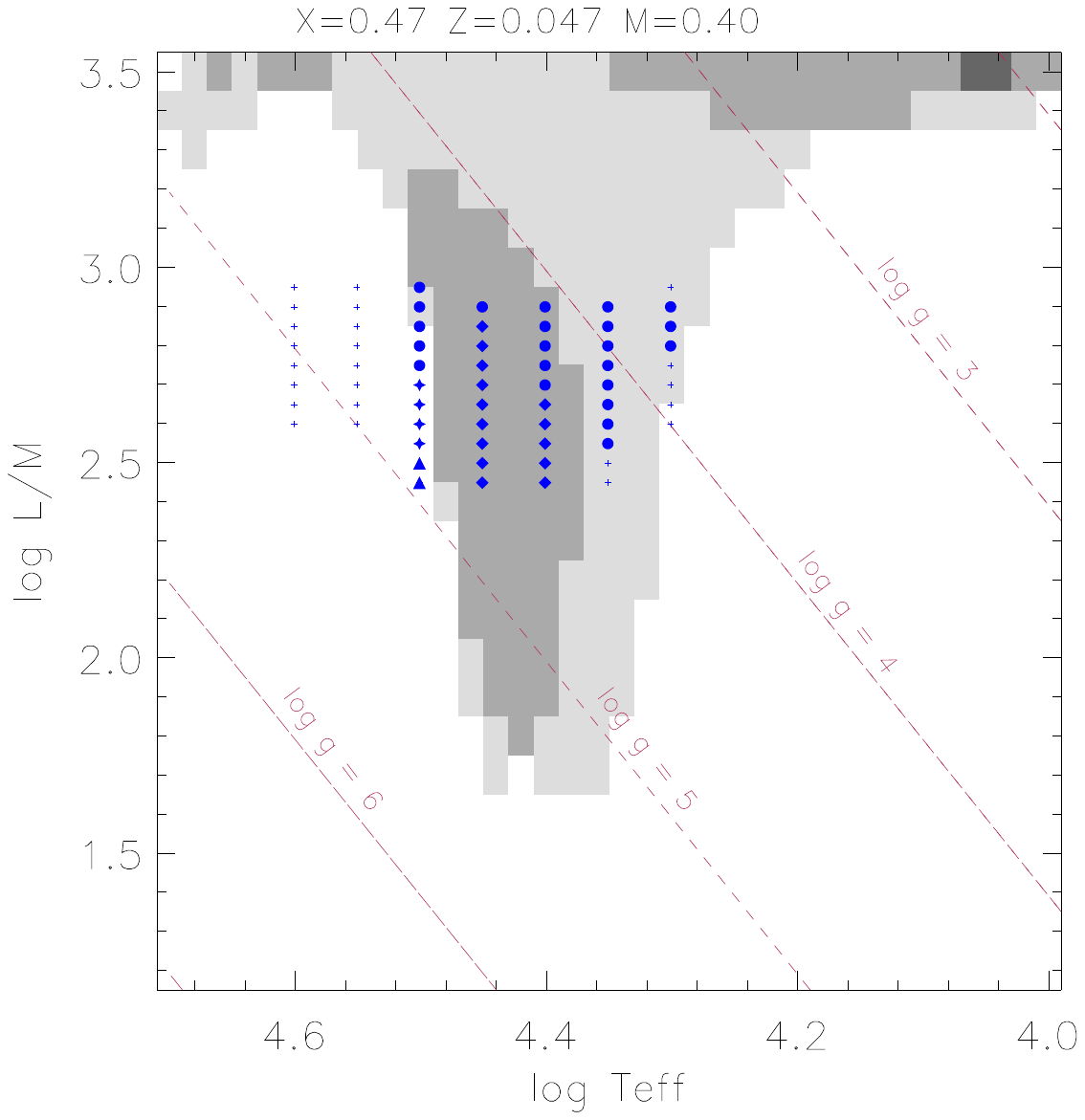}\\
\vspace{-20mm}
\includegraphics[width=0.32\textwidth]{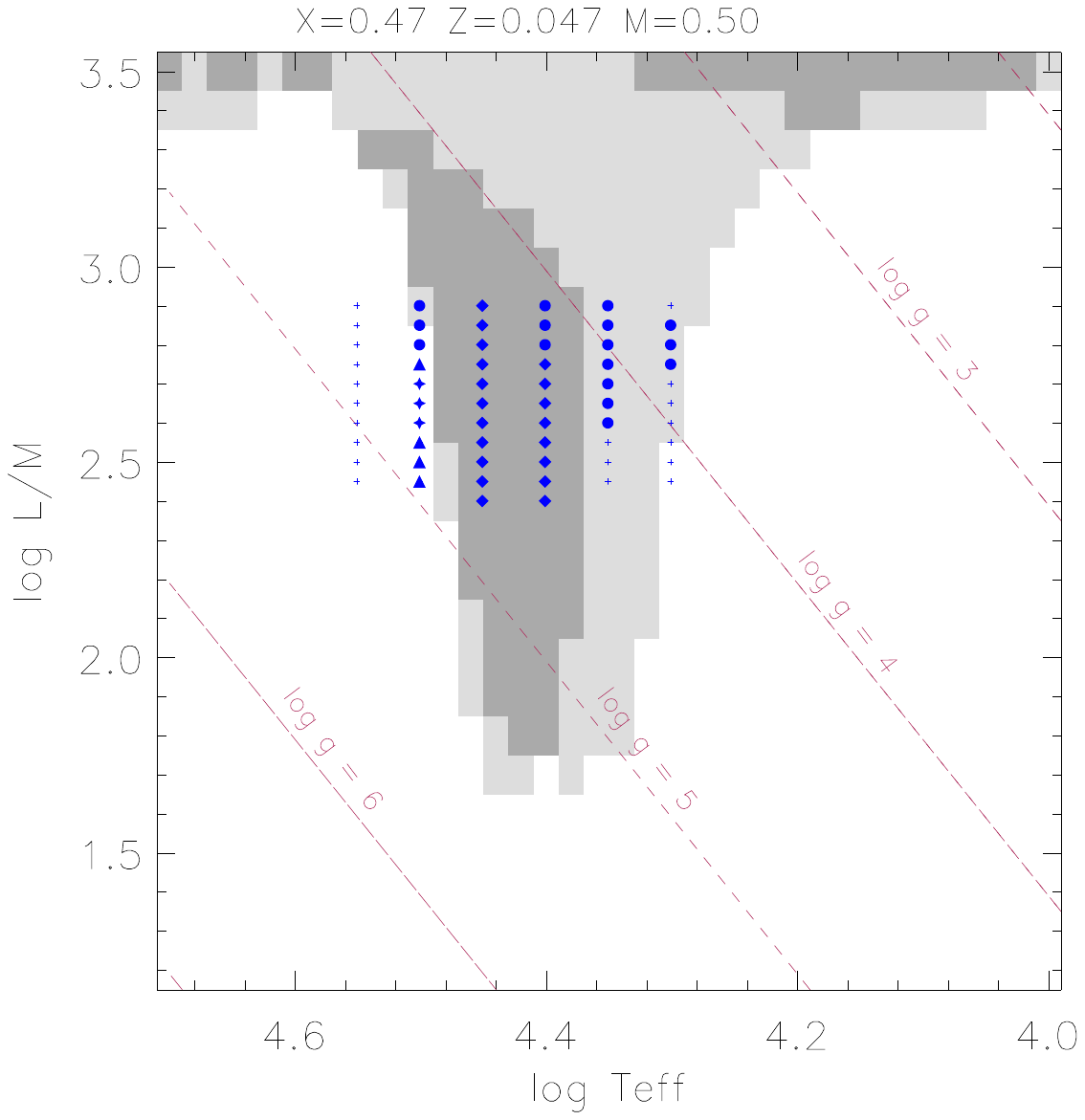}
\includegraphics[width=0.32\textwidth]{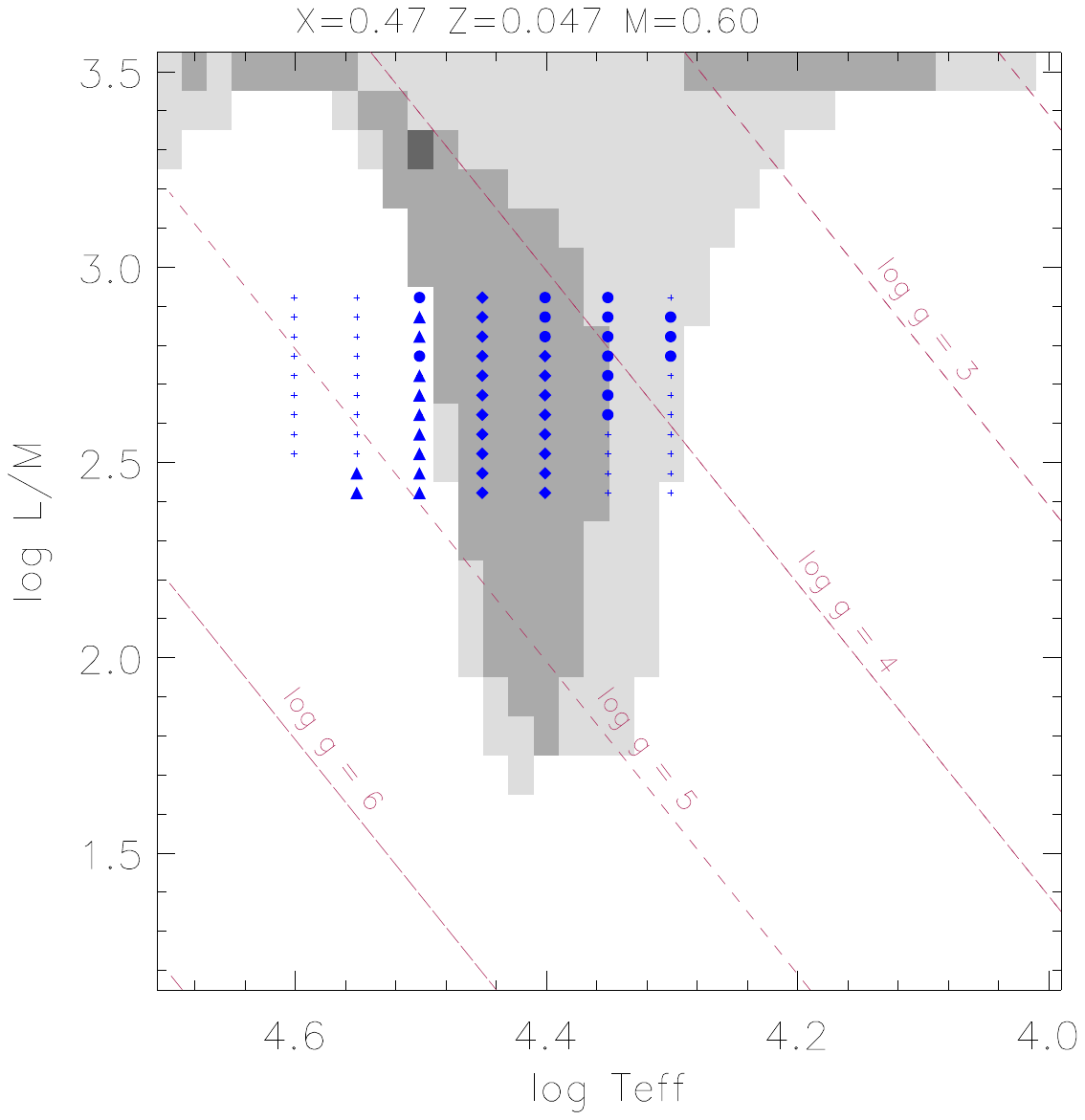}
\includegraphics[width=0.32\textwidth]{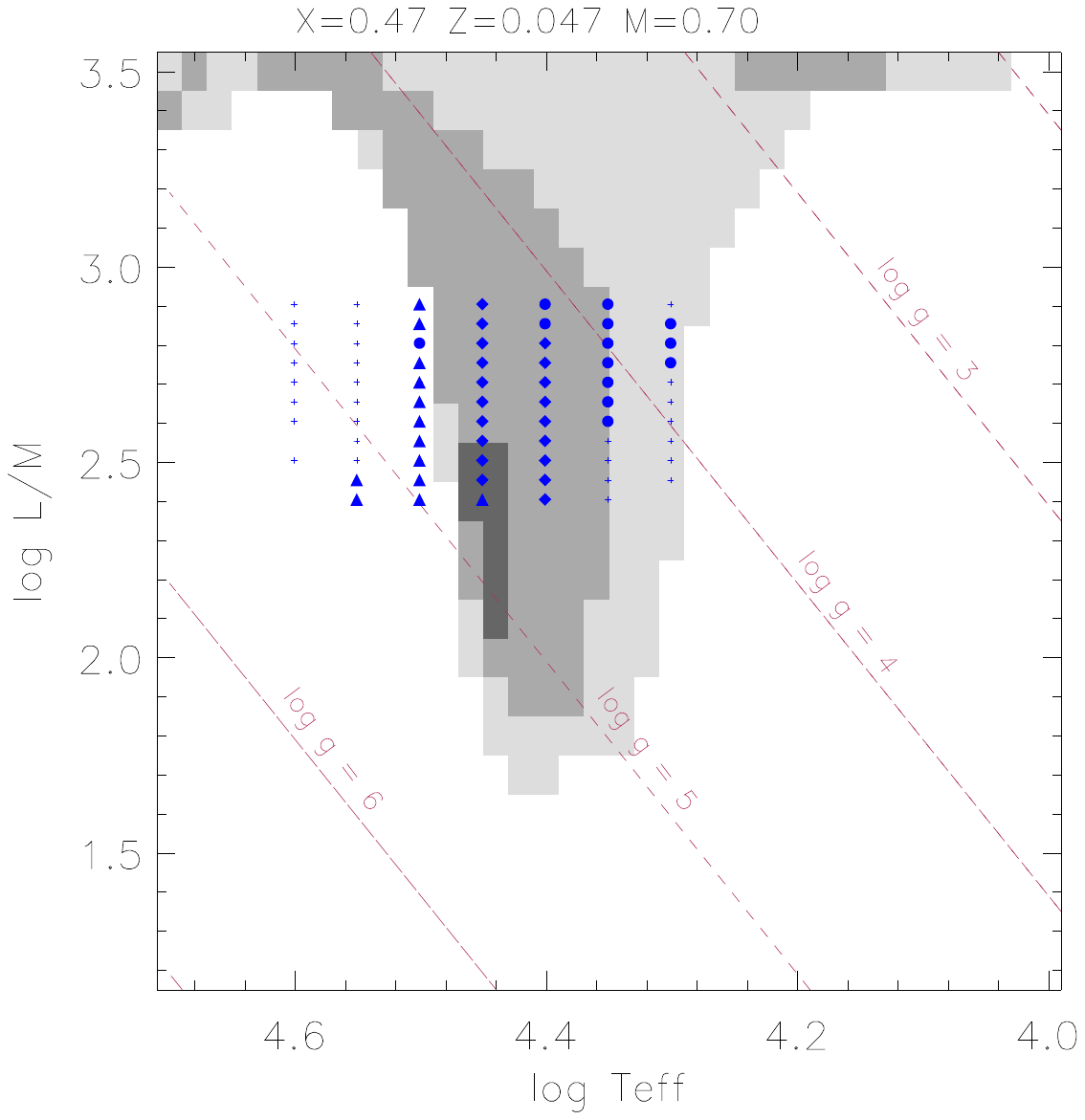}\\
\caption[Unstable modes as f(M)]
{As Fig.\,\ref{f:nmZ02} for models with mass $M=0.20, 0.30, \ldots 0.70$\Msolar, metallicity $Z=0.047$ and hydrogen abundance (by mass fraction) $X=0.475$.
This figure demonstrates how instability as a function of $T_{\rm eff}$ and $L/M$ depends only weakly on $M$.     
}
\label{f:nmM}
\end{center}
\end{figure*}

\section{Linear non-adiabatic models}
\label{s:linear}
\subsection{Method}

A linear non-adiabatic analysis of stability against pulsation was  carried out using  methods described by \citet{saio83b} and \citet{jeffery06c,jeffery06b}; 
it closely follows the large-scale study of pulsation as a function of chemical composition by  \citep{jeffery16a}. 

A grid of stellar envelopes covering a range of chemical mixtures was calculated for { stars with total mass} $M/\Msolar = 0.20 (0.05) 0.40 (0.10) 0.70$ (lower / interval / upper), 
effective temperatures $\log T_{\rm eff}/{\rm K} = 4.00 (0.02) 4.70 $,  and luminosity-to-mass ratios  $\log (L/\Lsolar)/(M/\Msolar) = 1.0 (0.1) 3.5$. 
{ As before \citep{jeffery06c,jeffery16a}, the envelope models integrate the time-independent spherically-symmetric stellar-structure equations inward from the surface where $L$ and $\Teff$ are given, an Eddington grey atmosphere yields the pressure where temperature $=\Teff$, and radius $R$ is defined by $L=4\pi\sigma R^2\Teff^4$.   
The outer boundary is set at the Rosseland mean optical depth $\tau = 10^{-3}$. 
The integration uses the logarithm of pressure as the independent variable, and is halted when the fractional mass $m\leq M/10$ or the fractional radius $r\leq R/100).$ 
}
For convection, the codes assume a standard mixing-length theory with the ratio mixing-length to pressure scale height $l/H_p = 1.5$; the models considered here are too hot for convection to be significant. 
Opacities are based on OPAL95 \citep{iglesias96} ``Type 2'' tables \footnote{Opacity tables were generated from the OPAL website {\tt https://opalopacity.llnl.gov} during 2023. The site is not available at the time of writing (2024 November).}.
Type 2 tables comprise some 60 subtables for a single metal mixture $Z$, but for sufficient H:He:C:O combinations that stellar evolution calculations can be executed through to  hydrogen and helium exhaustion.

In considering  models for the pulsation of BLAPs, the choice of the envelope composition can be influenced by the following factors:
\begin{itemize}
\item a few measurements of surface helium-to-hydrogen ratio.
\item a measurement of the surface distribution of heavier elements on OGLE-BLAP-009 \citep{bradshaw24}.
\item a prediction of the driving-zone iron+nickel enrichment due to radiative levitation  \citep{byrne20}. 
\end{itemize}

For benchmark comparisons, a metallicity of $Z=0.02$  was adopted, as in \citet{jeffery16a} (label Z02). 
Since \citet{byrne20} demonstrated that enhanced metals are necessary to drive BLAP pulsations, a Type 2 table was constructed with a nominal metallicity $Z=0.0242$ (label Z0242).  
This mixture was augmented in iron and nickel only, relative to the Z=0.02 mixture.  
The metal component of the Type 2 mixtures is shown in Table\,\ref{t:mixes}. 

\citet{bradshaw24} report an even more iron-rich surface composition for OGLE-BLAP-009.
A Type 2 table was constructed with $Z$ component matched to the abundances of observed species in OGLE-BLAP-009, and with unobserved species scaled to match the mean observed metallicity relative to solar of +0.61 dex (by number) (Table\,\ref{t:mixes}, label Z047).

{ Up to 3 hydrogen abundances $X$ were used for each metal mixture (see Table\,\ref{t:mixes}). These give a nominal grid of $[X]\times[Z]\times[M]\times[L/M]\times[\Teff]$ or $3\times3\times8\times31\times36 = 80\,352 $ models. 
In practice, some $X$ and/or $M$ values were omitted for some metallicities. 
Including unreported experimental grids, a total of 56\,916  model envelopes were computed for this study.    }
 
For each model envelope,  up to 17 eigenfrequencies ($n=0 - 16$) were located and stored, including the real and imaginary components
$\omega_{\rm r}$ and $\omega_{\rm i}$, the period $\Pi$ and the number of nodes in the eigensolution. 
Modes with $\omega_{\rm i} < 0 $ were deemed unstable, {\it i.e.} pulsations could be excited. 
{ \citet{jeffery06c} found that 10 eigenfrequencies ($n\leq 9$) were sufficient  to identify the instability boundaries for hot subdwarfs. 
\citet{jeffery16a} considered $n\leq 16$ over a much larger region of the $L/M - \Teff$ plane. 
For $L/M - \Teff$ appropriate to BLAPs, unstable modes were only found up to $n=2$ except for mixture Z0242, where unstable modes up to $n=6$ were found. } 

The results are presented  as contour plots representing the number of unstable modes as a function of 
$(T_{\rm eff}, L/M)$ for each composition (Figs.~\ref{f:nmZ02}, \ref{f:nmZ047}, \ref{f:nmM}, and \ref{f:nmZ}).
These contour plots provide an overall instability boundary since each includes  pulsations in both low- and high-order modes.  
The  instability boundaries for modes with $n = 0, 1$, or more nodes, {\it i.e.} the instability boundaries for the fundamental radial (F), and for the first (1H) and higher overtone pulsations are shown, also as a function of $(T_{\rm eff}, L/M)$ in  Appendix A, Figs.\, \ref{f:p_X} -- \ref{f:p_mass}. 

\subsection{Radial-Mode Instability}

Figure\,\ref{f:nmZ02} demonstrates the effect of the hydrogen abundance on the extent of the Z-bump instability strip for 0.30\Msolar\ stars with solar metallicity (mixture Z02). 
As hydrogen increases towards a solar value, the instability region shrinks to become almost negligible. 
Only the F-mode is excited, and the blue edge is too red to account for observed BLAPs.
Figure\,\ref{f:nmZ047} is the same for an envelope composition matched to the observed surface composition  of OGLE-BLAP-009 \citep[mixture Z047:][]{bradshaw24}.
In this case the instability region is enlarged by the increased contribution to opacity by iron and nickel; more than one mode may be excited and nearly all of the observed BLAPs can be accounted for.  
Figure\,\ref{f:nmZ} demonstrates the contribution of iron and nickel abundances, assuming a mass of 0.30\Msolar\ and the OGLE-BLAP-009 hydrogen abundance $X=0.475$.  
On the left is the standard solar composition $Z=0.02$, in the middle is the observed metal mixture $Z047$ and on the right is a standard solar mixture of metals with the exception that iron and nickel are enhanced to the values predicted by \citet{byrne20} (mixture Z0242).
The latter are $\approx7\times$ solar and make the zone extremely unstable.
Figure\,\ref{f:nmM} demonstrates the effect of mass; with mixture Z047 and $X=0.475$, BLAPs lie in the predicted instability zone at all masses in the range 0.2 -- 0.7 \Msolar.  

{ \citet{jeffery16a} have already shown that the instability zone in the region of $L/M - \Teff$ space that applies to hot subdwarfs and early B stars is relatively insensitive to envelope hydrogen abundance ($0.3 < X < 0.7$) and stellar mass ($0.20 \leq M/\Msolar \leq 0.70$). 
BLAPS share this space and insensitivity to $X$ and $M$, but are } extremely sensitive 
to the envelope iron and nickel abundance, with a value $\approx3\times$ solar being sufficient to excite pulsations. 

\begin{figure}
\centering
\includegraphics[width=0.98\linewidth]{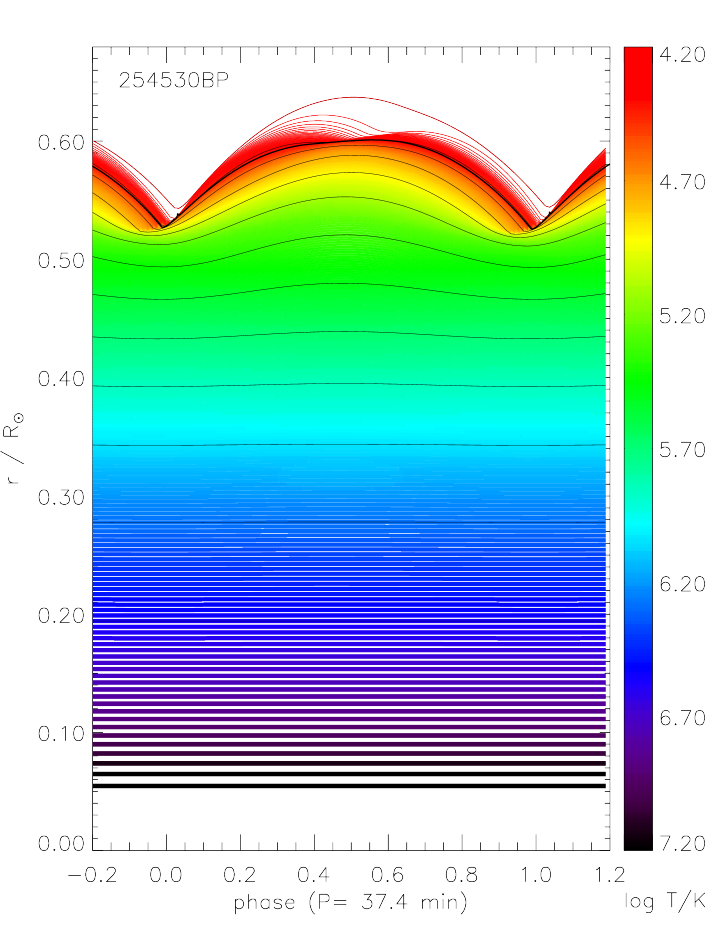}
\caption{The position $r$ of mass zones in a model interior throughout a pulsation cycle. This figure shows model 254530BP, having $M = 0.30\Msolar$, $\log L/\Lsolar = 2.25$,  $\log \Teff/{\rm K} = 4.45$ and mixture BP (see text). 
The mass zones are colour-coded for temperature. 
The base of the photosphere, defined by Rosseland mean optical depth $\tau_{\rm Ross}=2/3$, is shown by a bold dark line. The position of every 20th zone is marked by a thin black line. Fig.\,\ref{f:surf} shows parts of this model in more detail. 
}
\label{f:guts}
\end{figure}

\section{Non-linear Models: Methods}
\label{s:nl_models}

\subsection{Method}
\label{s:nl_methods}
Non-linear hydrodynamic models from BLAPs were computed with the  radial pulsation code {\sc puls\_nl} \citep{bridger83}.
This code was based on the prescription by \cite{christy67} and substantially modified by \cite{montanes02} to incorporate more modern opacities.  
A further description was given  by \citet{jeffery22a}. 
In summary, a system of differential equations is written in a finite-difference form and 
solved as an initial value problem. 
As initial value, a finely-zoned envelope in hydrostatic 
equilibrium is constructed, given the stellar mass $M$, luminosity $L$, effective temperature \Teff, and composition $X, Y, Z$ as free parameters. 
This model envelope is rezoned onto a Lagrangian grid with $n$ mass zones. 
A small velocity perturbation { $u(r) = 10^4 r^4 {\rm cm\,s}^{-1}$ is applied to all models; $r$ is the fractional radius of each mass zone.  
This perturbation is } allowed to grow until a stable pulsation is obtained. { Convergence was estimated by comparing successive light maxima by eye. 
Examination of a few models so identified showed peak values to differ by less than 0.5\% (and often $< 0.1$\%). Exceptions will be discussed in \S\,\ref{s:weird}.}
The solution is a time-dependent description of pressure, temperature, velocity and displacement as a function of mass or radius within the envelope (Fig.\,\ref{f:guts}).

\begin{figure*}
\centering
\includegraphics[width=0.48\linewidth]{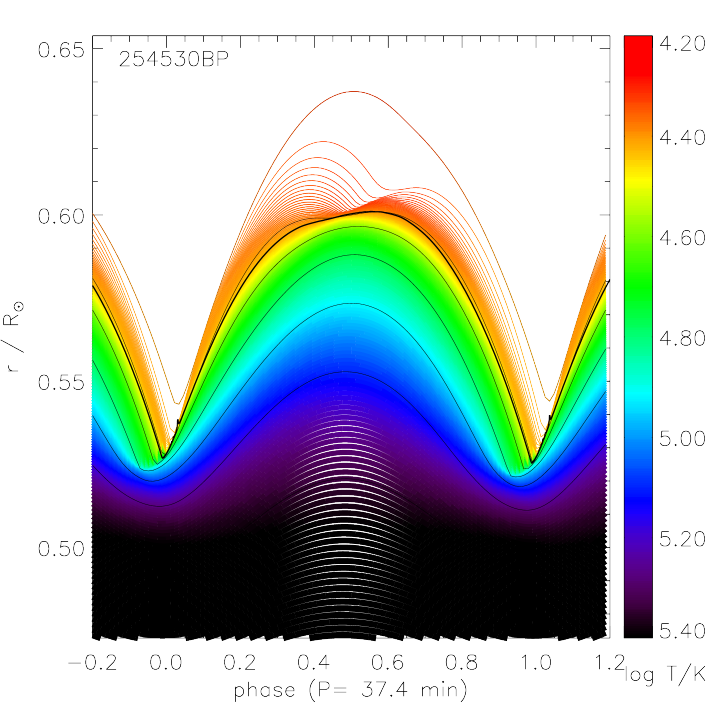}
\includegraphics[width=0.48\linewidth]{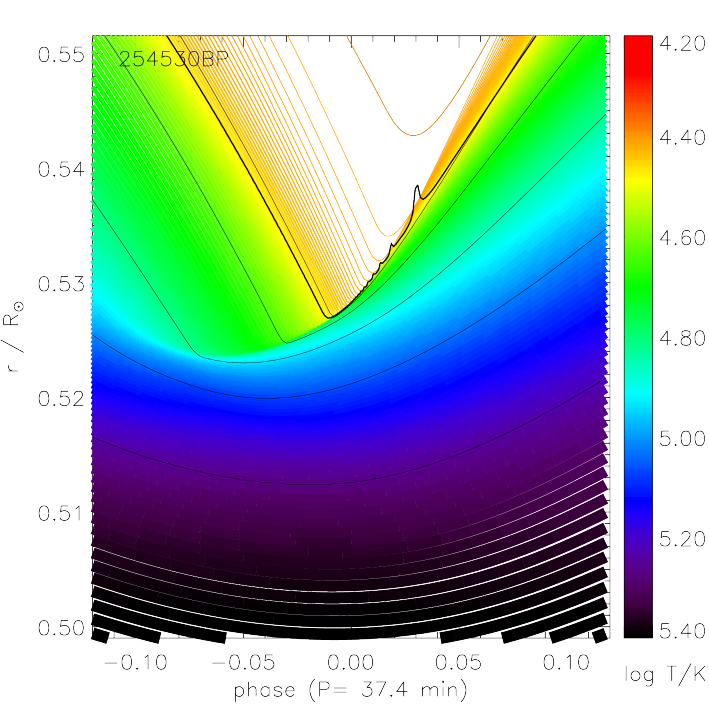}
\\
\includegraphics[width=0.48\linewidth]{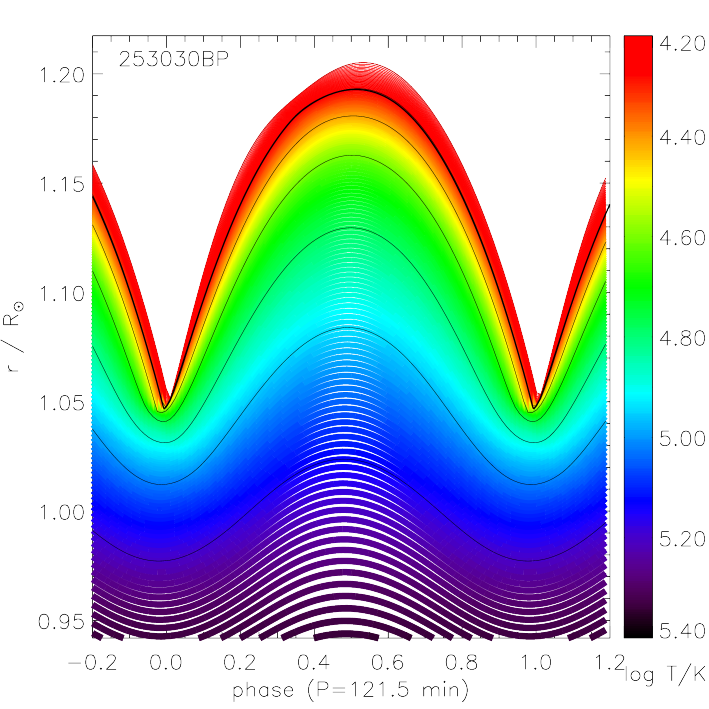}
\includegraphics[width=0.48\linewidth]{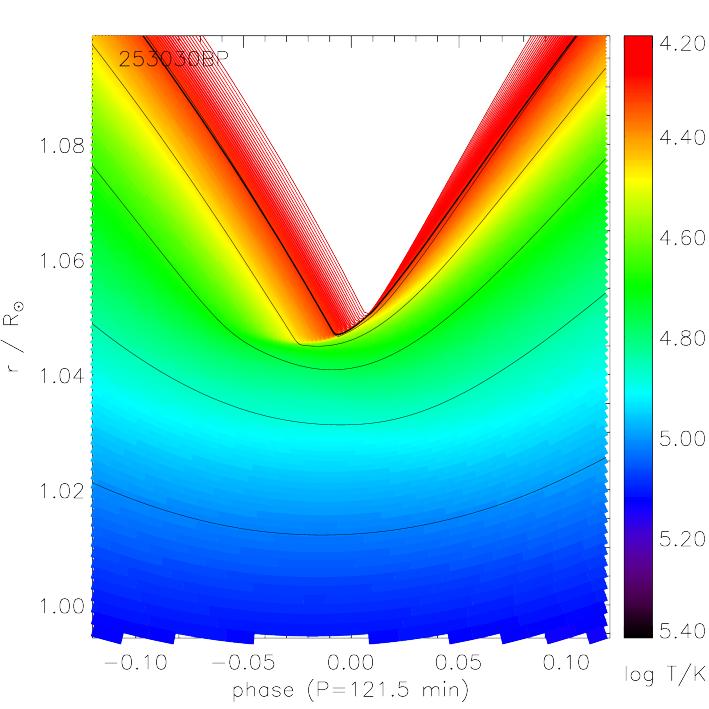}
\caption{Top left: As Fig.\,\ref{f:guts} for outer layers including the opacity driving zone. Top right: layers close to the surface during passage through minimum radius.
Bottom: As above for model 253030BP with $\log \Teff/{\rm K}=4.30$, demonstrating the absence of a reverse shock and prompter compression of the outer layers.  
}
\label{f:surf}
\end{figure*}

An artificial viscosity dissipation  \citep{stellingwerf75} damps any shocks produced in outer layers during a supersonic contraction and rebound at minimum radius (Fig.\,\ref{f:surf}).
This prevents sharp discontinuities in the physical variables between adjacent zones which can lead to numerical singularities causing the calculation to fail. 
Following \citet{jeffery22a}, default values of the viscosity dissipation constants $C_Q=2.0$ and $\alpha_v = 0.1 c_s$ were adopted initially, as these yielded convergence for most models of V652\,Her. 
However, since the model grid covers a larger range of parameter space, $C_Q$ and  $\alpha_v$ had to be adjusted over significant ranges to obtain converged pulsating models in a number of cases. 
Typically,  $C_Q$ and $\alpha_v$ were increased for high luminosity models and reduced for low-luminosity models, with the constraints  $1\leq C_Q \leq 4$ and $0 \leq \alpha_v \leq 0.5$.  

In general, models were computed for 100 runs of 24\,000 time steps { each, or 2\,400\,000 time steps in total. } 
The local time step is adaptive; one run would typically cover 3 -- 5 cycles of an F-mode oscillation. 
To check for or ensure convergence, this would occasionally be extended to 200 ({ 81 models: 4\,800\,000 time steps} ) or 800 runs (2 models: 19 million time steps). 

\subsection{Opacity}
\label{s:nl_opac}

In contrast to the linear code, {\sc puls\_nl} uses Type 1 OPAL95 (or OP) opacity tables \citep{iglesias96}. 
Each table provides  for a fixed distribution of elements within the ``metal'' component $Z$. 
Metal and hydrogen abundances in the ranges  $0<Z<0.1$ and $0<X<1-Z$ are provided by interpolation between 120 sub-tables (see Table\,\ref{t:mixes} caption).  
Since $Z$ represents a fixed distribution, care has to be taken when considering changes to only one or two elements, such as iron and nickel, without altering others. 
Table\,\ref{t:mixes} shows the Type 1 mixtures adopted. 

\subsection{Chemical composition}
\label{s:nl_comp}

The choice of initial hydrostatic model makes assumptions about the internal structure and previous evolution of the star. 
For the current calculations, the envelope is {\it assumed} to be chemically homogeneous and free of  energy sources. 
If homogeneous, one choice for the chemical composition is provided by surface abundance analyses such as that for OGLE-BLAP-009 \citep{bradshaw24} (Mixture BP: Table\,\ref{t:mixes}). 
Mixtures BP and Z047 are equivalent, but are used to distinguish the use of Type 1 and Type 2 opacity tables. 

A second choice is guided by theoretical diffusion calculations \citep{byrne18b}, which indicate a larger iron and nickel content. 
Mixture FE (Table\,\ref{t:mixes}) is equivalent to BP except that the iron and nickel fractions are increased by a factor 5, and other metals are reduced in proportion. 
All {\sc nl\_puls} calculations attempted with this mixture were { numerically } unstable and failed.  

Given that hydrogen is known to inhibit pulsations \citep{jeffery16a}, two sets of models were calculated with hydrogen increased and reduced relative to the reference mixture (BP), in which the hydrogen mass fraction $X=0.475$. 
Referring to Table\,\ref{t:mixes}, these mixtures are identical to mixture BP in terms of $Z$, but with $X=0.90$ (mixture XX) and $X=0.353$ (mixture HE), respectively. 

A set of control models was constructed using chemical mixture N1 as adopted  by \citet{jeffery22a} in their study of V652\,Her. 
Mixture N1 is nitrogen-rich and carbon and oxygen poor and represents CNO-processed material. 
Within N1, the relative contribution of iron and nickel to $Z$ is approximately 1/5 solar but, after scaling up to $Z=0.047$, the absolute contribution is closer to 1/2 solar.  
Models were calculated for the same hydrogen and helium mass fractions as for the reference (BP) mixture (Table\,\ref{t:mixes}). 


\begin{figure}
\centering
\includegraphics[trim=0cm 2cm 9cm 12cm, width=\columnwidth]{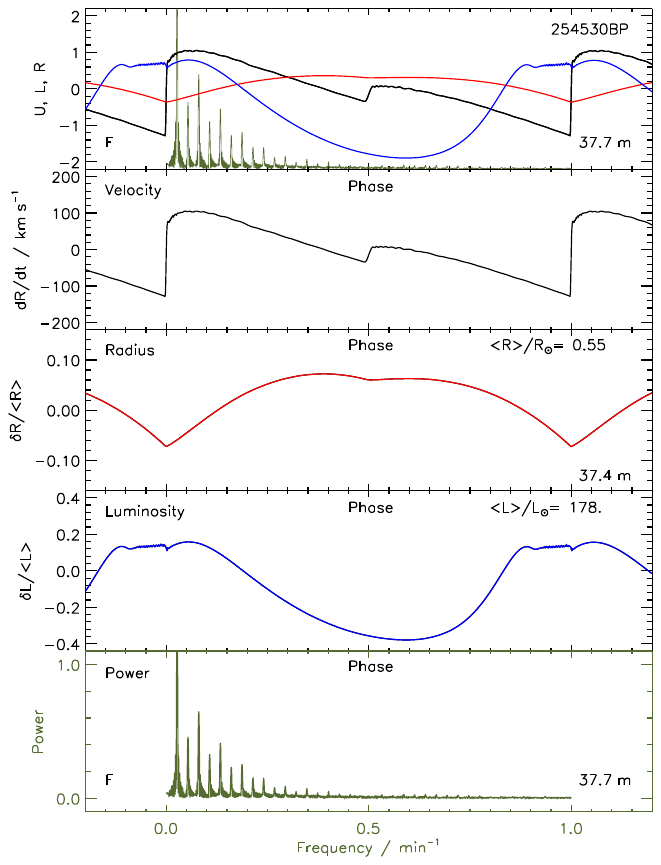}
\caption{Surface properties for a pulsation cycle of a model with $M/\Msolar=0.30$, $\log L/\Lsolar=2.25$, $\log \Teff/{\rm K} = 4.45$, and mix BP.  
From bottom upwards: the luminosity power spectrum (olive green) as a function of frequency (min$^{-1}$), and relative luminosity ($L$, blue), relative radius ($R$, red) ), and radial velocity ($U \equiv {\rm d}R/{\rm d} t$, black), as a function of pulsation phase. 
The top panel shows all four variables superimposed with the model label shown top right.
Phases 0 and 1 are defined to be at consecutive minima in the radius.
The equilibrium luminosity and radius are shown in the appropriate panels,
The period obtained from successive values of minimum  radius is shown lower right in the radius panel.
The period given by the highest peak in the luminosity power spectrum is shown lower right in the power spectrum and composite panels.
The pulsation mode is indicated lower left in the  power spectrum and composite panels.
In order to show all variables on a common scale in the top panel, the relative luminosity and radius curves are both scaled by a factor of 5, the velocity curve is scaled by 1/100, and the power spectrum $p(\nu)$ is shown in arbitrary units offset to the lower limit of the panel.
Thus, a fractional change $\delta L/\langle L\rangle = 0.2 \equiv 20\%$ corresponds to a displacement of $+1$ on this scale.
Likewise, a fractional change $\delta R/\langle R\rangle = 0.1 \equiv 10\%$ also corresponds to a displacement of $+0.5$. 
Equivalent  panels are used at reduced sizes in Fig.\,\ref{f:mods30} and Figs.\,\ref{f:mods20}--\ref{f:mods30N1} to illustrate the variation of pulsation behaviour with $L$, $\Teff$,  $M$, $X$ and $X_{\rm Fe,Ni}/Z$. 
}
\label{f:254530BP}
\end{figure}

\begin{figure*}
\centering
\vspace*{-5mm}
\includegraphics[width=0.97\textheight,angle=90]{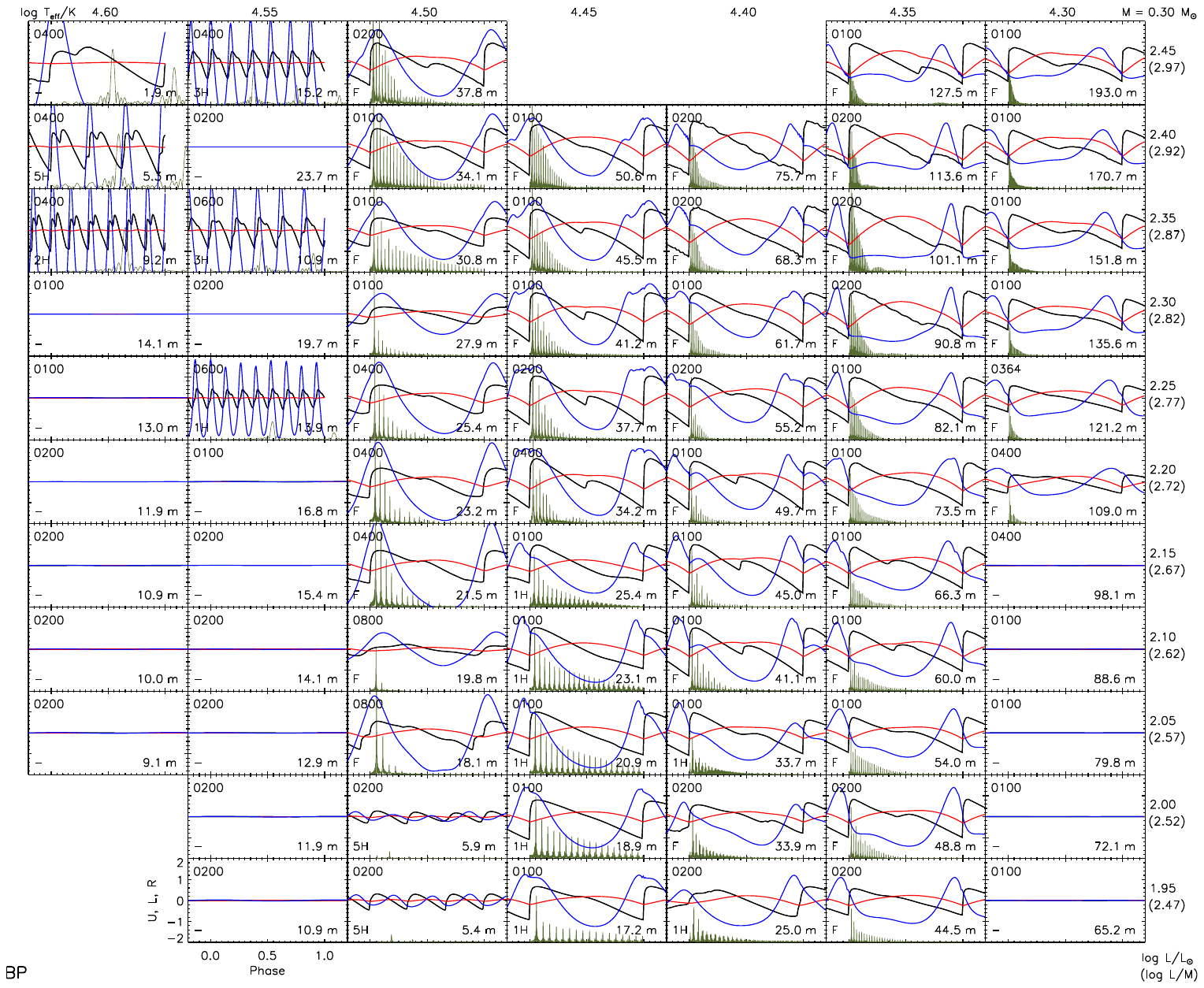}
\vspace*{-20mm}
\caption{Luminosity ($5 \delta L/\langle{L}\rangle$, blue), radius ($5 \delta R/\langle{R}\rangle$: red), radial velocity ($U \equiv {\rm d}R/{\rm d} t / \kmsec$: black) and the luminosity power spectrum (olive green) as a function of phase for models with $M/\Msolar=0.30$ and mix BP. 
{ $\delta L, \delta R$ and $U$ have been resampled to 200 bins cycle$^{-1}$. }
The input parameters $M$, \Teff,  $L$ (and $L/M$) associated with each model are shown along the top and right hand axes respectively. 
Phase 0 is defined to be at radius minimum. 
The period in minutes obtained from the power spectrum is shown at bottom right of each panel. 
The associated pulsation mode is shown bottom right. 
If the pulsation is weak or absent (mode `--'), the predicted F-mode period is shown. 
The number shown top left of each panel represents the number of runs completed for each model, each run representing 24\,000 time steps. 
Similar panels for models with  other masses and mixtures are given in an Appendix.  
The axes and scales for all panels are identical; for clarity only axes in the lower-left panel are labelled. The mass and chemical mixture labels are shown upper right and lower left, respectively. 
}
\label{f:mods30}
\end{figure*}

\begin{figure*}
\centering
\includegraphics[width=55mm]{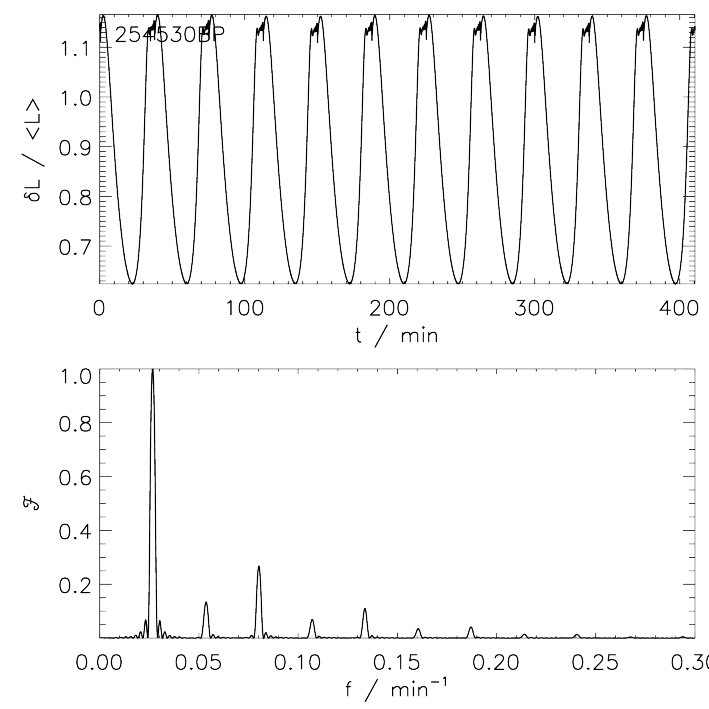}
\includegraphics[width=55mm]{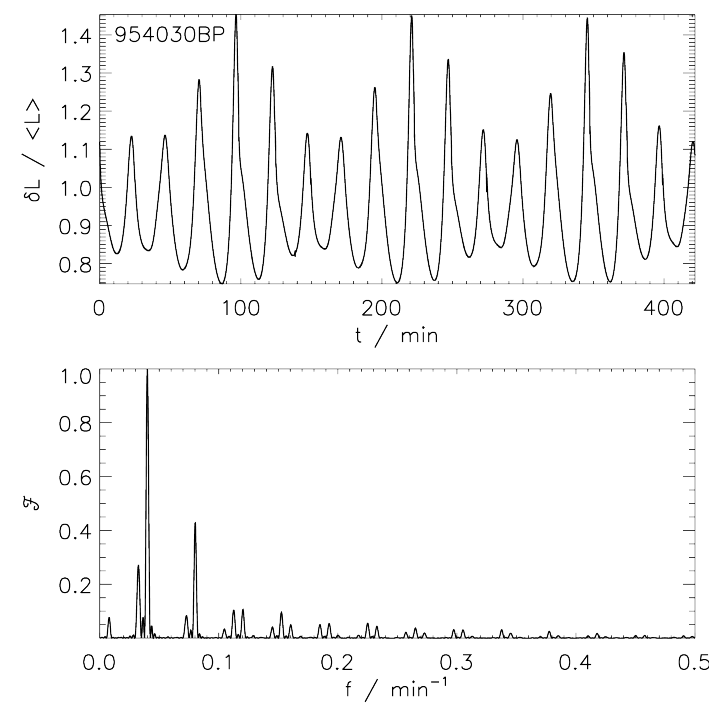}
\includegraphics[width=55mm]{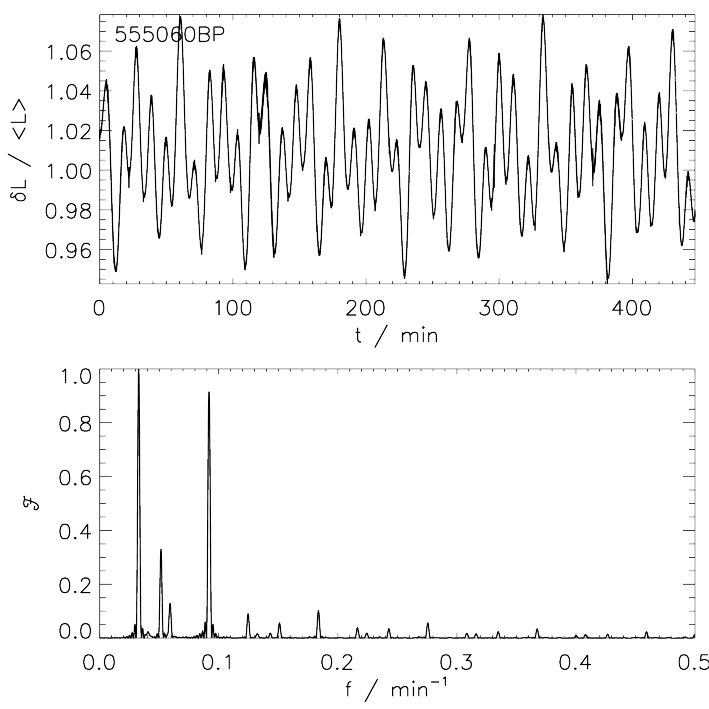}
\caption{Lightcurves and Fourier transforms ($\mathcal{F}$) of three BLAP models showing showing multi-mode pulsations, normalized to $\langle L\rangle$ and $\mathcal{F}_{\rm max}$, respectively. 
The reference model 254530BP shows the F mode and harmonics. 
Model 954030BP shows both F and 1H modes excited with periods 31 min and 25 m, respectively ($f=0.0325, 0.0400\,{\rm min^{-1}}$). Model 555060BP shows large amplitude in both F  ($f=0.0329\,{\rm min^{-1}}$) and 4H modes ($f=0.0919\,{\rm min^{-1}}$), with the 2H mode ($f=0.0515\,{\rm min^{-1}}$) also excited. Combination frequencies are also evident. 
}
\label{f:multi}
\end{figure*}

\subsection{Example}
\label{s:nl_example}
Figure\,\ref{f:guts} illustrates the internal behaviour of a model having $M/\Msolar=0.30$,  $\log L/\Lsolar=2.25$, $\log \Teff/{\rm K} = 4.45$, and mix BP over a full pulsation cycle.
This is occasionally referred to below as the reference model 254530BP and was chosen for its proximity to the parameters of OGLE-BLAP-009 \citep{bradshaw24}. 
Fig.\,\ref{f:surf} shows expanded segments of the same model emphasizing the surface layers throughout the cycle and around minimum radius. 
Both figures are colour coded for temperature. 
The opacity maximum for iron-group elements occurs around $\log T/{\rm K} = 5.3$. 

Fig.\,\ref{f:254530BP} summarises the surface properties luminosity, radius and radial velocity over a full pulsation cycle.  
This model is broadly similar to earlier models for the pulsating extreme helium star V652\,Her in that it shows a very rapid outward acceleration followed by a smoother near ballistic inward acceleration \citep{jeffery22a}.
There are two significant differences. 
First, the luminosity shows a broad maximum characterized by two distinct peaks with a small dip in between. 
This double peak or broad peak with a dip seen in some BLAP models is associated with the passage of a shock wave at minimum radius.  
Second, the inward acceleration shows a reversal - a second outward acceleration at around phase = 0.5.
Inspection of Fig.\,\ref{f:surf} shows that during the expansion phase, the outer layers of the photosphere expand more rapidly than underlying layers. Being less well supported they start to contract earlier, effectively producing a second or reverse shock as the contracting and expanding parts of the atmosphere collide. 

\begin{table*}
\begin{center}
\caption{Summary of non-linear pulsation models -- full table given in \S\,\ref{s:app2} (online).
Columns and units as labelled. 
$P_{\rm nl}$ is the period measured from successive radius minima, $P_{\rm ft}$ is from the strongest peak in the Fourier transform of the light curve.
$\Delta L, \Delta R, \Delta u$ are the pulsation amplitudes in luminosity, radius and surface velocity, respectively. 
$\Delta \phi$ is the phase difference between radius minimum and light maximum. 
$k$ is the number of radial nodes of the dominant pulsation mode; hence $k=0,1$ and 2 correspond to F, 1H and 2H respectively. 
$k=8$ flags an overtone $k>5$ and { $k=-$} flags a low-amplitude pulsation for which $P_{\rm ft}$ was unphysical and is here set to 0.  }
\label{t:models}
\begin{tabular}{lrrrr rrrrr rrrrr}
\hline
 model  & $M$    & $\log L$ & $\log \Teff$ & $\log R$ & $\log g$  & $\log \bar{\rho}$ & $\log L/M$ & $P_{\rm nl}$ & $P_{\rm ft}$ &  $\Delta L$ &  $\Delta R$ &  $\Delta u$   & $\Delta \phi$ & $k$\\
$llttmmNN$& \Msolar & \Lsolar & K     & \Rsolar & cm\,s$^{-2}$  & $\bar{\rho_{\odot}}$ & \Lsolar/\Msolar & min   &  min   & $L_{\star}$ & $R_{\star}$ & \kmsec &      &     \\
\hline
954020BP&0.20&1.95&4.40&-0.31&4.35& 0.22&2.67& 40.4& 40.6&0.18&0.49&224& 0.14&0\\
954520BP&0.20&1.95&4.45&-0.41&4.55& 0.52&2.66& 27.5& 27.5&0.14&0.62&230&-0.06&0\\
103020BP&0.20&2.10&4.30&-0.03&3.80&-0.60&2.82&118.7&123.2&0.14&0.33&144& 0.19&0\\
103520BP&0.20&2.10&4.35&-0.13&4.00&-0.30&2.86& 82.6& 82.6&0.24&0.44&187& 0.14&0\\
104020BP&0.20&2.10&4.40&-0.23&4.20&-0.00&2.84& 54.9& 55.1&0.26&0.49&251& 0.10&0\\
\hline
\end{tabular}
\end{center}
\end{table*}

\subsection{Model grid}
\label{s:nl_grid}

The model grid parameters are designed to span the observed properties of BLAPs in general, and OGLE-BLAP-009 in particular. 
Since non-linear models for BLAPs may be more sensitive to mass than linear models, grids were computed for $M= 0.2, 0.25, 0.3, 0.35, 0.4, 0.5, 0.6$ and 0.7 \Msolar, 
and for effective temperatures $\log \teff / {\rm K} = 4.30, 4.35, 4.40, \ldots 4.60$.  
Luminosities were taken from a subset of $\log L / {\Lsolar} = 1.80, 2.00, \ldots 2.75$ chosen to cover the range 
$2.3 < \log (L/M) / {\Lsolar/\Msolar} < 2.9$. 
The extent and resolution of the grids are illustrated in Figs.\,\ref{f:nmZ02}--\ref{f:nmM}.
In all, over 700 non-linear models were computed up to run 100. 

For convenience, non-linear models are given labels in the form $ llttmmNN $ where 
$ll = {\rm frac}(\log L/\Lsolar)$, 
$tt = {\rm frac}(\log \Teff/{\rm K})$, 
$mm = {\rm frac}(M/\Msolar)$, 
$NN$ is a 2-letter or letter and integer combination identifying the chemical composition, the choice of opacity table and other properties of the  calculation.    
For example, the label 254530BP represents a model with 
$\log L/\Lsolar = 2.25$, $\log \Teff/{\rm K}=4.45$, $M = 0.0.30\Msolar$, and mixture BP (Table\,\ref{t:mixes}). 
For clarity: $\log L/\Lsolar =1.ll$ if $ll \geq 80$ and $\log L/\Lsolar=2.ll$ if $ll \leq 79$.

Figure\,\ref{f:mods30} represents the model grid for $M/\Msolar=0.30$ and mixture BP, with each model plotted in the same format as the top panel of Fig.\,\ref{f:254530BP}.  
Each panel shows the principle pulsation period for each model as determined (a) from two successive minima in the radius curve and (b) from Fourier analysis of the light curve obtained from the last 3 runs of each model (72\,000 time steps).  
Discrepant values occur when the pulsation is not well converged, when beating between modes occurs, or when the strongest peak in the power spectrum is an harmonic rather than the principle component of a mode.  

Whilst most models in the predicted instability zone show well-behaved regular pulsations, some do not.  
The following cases are identified:\\
1) Well-behaved converged pulsations: {\it e.g.} 254530BP: $\log \Teff/{\rm K}=4.45$, $\log L/\Lsolar = 2.25$.\\
2) As above, but on inspection the amplitudes are slowly increasing and indicate incomplete convergence. It would be unnecessarily expensive to ensure complete convergence in all cases. \\ 
3) Stable or heavily damped pulsations: {\it e.g.} 005530BP: $\log \Teff/{\rm K}=4.55$, $\log L/\Lsolar = 2.00$. The initial perturbation is either completely damped or of very low amplitude. \\
4) Overtone pulsations: {\it e.g.} 255530BP: $\log \Teff/{\rm K}=4.55$, $\log  L/\Lsolar = 2.25$. 
Nearly-regular large-amplitude oscillations with periods significantly shorter than the fundamental (F) radial mode  occur at the blue edge of the instability strip. 
These are likely first or higher overtone pulsations (1H, 2H, \ldots). 
Inspection of the above model indicates compression around $\log T/{\rm K} \approx 6$ in phase with surface expansion, whilst comparison with the predicted F-mode period indicates that these are 1H radial mode pulsations.\\
5) Multiperiodic pulsations: {\it e.g.} 954030BP: $\log \Teff/{\rm K}=4.40$, $\log L/\Lsolar = 1.95$ . Nearly-regular large-amplitude oscillations with periods close to the fundamental radial mode which do not repeat perfectly from cycle to cycle. A cursory glance suggests these have not converged, but inspection of a longer segment of the light curve and its Fourier transform shows that two or more modes are present (Fig.\,\ref{f:multi}).
These multi-mode pulsators can be identified from individual FTs in Fig.\,\ref{f:mods30} with sufficient magnification. \\
6) Unstable solutions:  {\it e.g} 454030BP: $\log \Teff/{\rm K}=4.40$, $\log  L/\Lsolar = 2.45$ (not shown). In some cases, no stable solution was found after an extensive search through $(C_Q,\alpha_v)$ space. 
In such cases, the model pulsation may be of such large amplitude that it is numerically unstable. 

Model grids for other masses are shown in Figs.\,\ref{f:mods20}--\ref{f:mods70}, and for other chemical mixtures in Figs.\,\ref{f:mods30XX}--\ref{f:mods30N1}. 
To provide a comparison of light and velocity curve shapes and the extent of non-linear instability, the grids are complete for $\log (L/M) / (\Msolar/\Lsolar) \approx 2.80$ and $\log \teff/{\rm K} = 4.40, 4.45$. 
Since they are expensive to compute and manage, the grids are not all complete.   
A summary of the models and their principle pulsation properties is given in Table\,\ref{t:models} (abbreviated) and Tables\,\ref{a:models}ff (online). 

{ The reference grid (mixture BP) contains a nominal $[M]\times[L/M]\times[\Teff]$ or $8\times11\times7 = 616 $ locations. 
Complete coverage was not attempted at every mass. 
Some subgrids were larger than this; 14 values of $L$ were used for $M=0.30\Msolar$.  
Models for mixtures XX, HE, and N1  were only computed for $M=0.30\Msolar$.  
Overall, 792 non-linear models were computed for this study. }

\subsection{Mode Identification}

It will be shown in \S\,\ref{s:fits} that the non-linear models follow a well-defined relation which can be used to predict the period of the fundamental mode $P_{\rm F}$ for a given model. 
The model periods $P$ fall into discrete  groups with $P/P_{\rm F}\approx 1, 0.81, 0.66, \ldots$. 
Empirically, $P_k/P_{\rm F} \approx 0.81^k$, which allows the principal order $k$ to be estimated from the Fourier transform. 
F, 1H, 2H and probably up to 5H modes are identified in \ref{f:mods30} and \ref{f:mods20}--\ref{f:mods30N1} with confidence; higher order modes may be less certain. 
Where more than one mode is present, the periods measured from successive radius minima and from the FT may refer to different modes.

\section{Non-linear models: Results}
\label{s:results}

\subsection{Systematics at 0.30 M$_{\odot}$}
\label{s:system30}

In considering first the $0.30\Msolar$ models the following trends are evident (Fig.\,\ref{f:mods30}).\\
1) At  $\log \Teff/{\rm K}=4.35$, a simple  light maximum {\it precedes} minimum radius by up to 0.15 cycles. 
The light curve is relatively flat between steeply peaked maxima. \\
2) At $\log \Teff/{\rm K}=4.50$, a simple  light maximum {\it follows} minimum radius by up to 0.05 cycles. 
Whilst the maxima are sharply peaked, the light curve between maxima is more sinusoidal \\
3) In between these temperatures, the light maximum has a complex structure, with lesser or greater peaks appearing either side of minimum radius.
In some cases, light maximum may appear double ({\it e.g.} $\log \Teff/{\rm K}=4.40$, $\log L/\Lsolar = 2.15$).
Consider, for example, the models for $\log \Teff/{\rm K}=4.45$. 
At low luminosity, light maximum precedes minimum radius by 0.1 cycles, but there is a slow drop to minimum light thereafter. 
With increasing luminosity, that slow drop is replaced by a second peak in the light curve 0.05 cycles after minimum radius, whilst the original progressively weakens. \\
4) In several cases ({\it e.g.} model 404530BP) light maxima show a short sharp dip that coincides with radius minimum. Similar features have been found in models for V652\,Her \citep{fadeyev96,jeffery22a} and is probably directly associated with  shock passage. \\ 
5) In {\it most} cases, the predicted pulsations are of large amplitude and a shock is evident from the rapid acceleration seen in radial velocity. 
\citet{jeffery22a} already showed that shocks may not occur if the amplitude is sufficiently low ({\it e.g.} $\log \Teff/{\rm K}=4.50$, $\log L/\Lsolar = 2.10, 2.30$). The first example was checked very carefully for convergence by allowing the model to run 8 times longer than standard. \\
6) Additional shocks appear in a number of circumstances: {\it e.g.} $\log \Teff/{\rm K}=4.45$, $\log L/\Lsolar = 2.20$, $\log \Teff/{\rm K}=4.50$, $\log L/\Lsolar = 2.05$). 
The most obvious is a reverse shock which occurs approximately half-way through the cycle (cf. Fig.\,\ref{f:surf}) which would be visible as a kink in the radial-velocity curve. 
These appear localized to a particular band of models running from low \Teff\ and $L$ to higher \Teff\ and $L$ or, equivalently, models of similar period $P\approx 40$\,min  (see Fig.\,\ref{f:mods30}).
Reverse shocks will be discussed further in \S\S\,\ref{s:mass} and \ref{s:shocks}. 
Other instances of shocks will be discussed in \S\,\ref{s:weird}.  \\
7) It is not immediately obvious from the light curves that at least two and probably more pulsation modes are represented by the models, being the fundamental (F), first (1H) and higher ($k$H) overtone radial modes. 
The multiplicity becomes apparent in an analysis of the periods (see below); overtone modes generally appear in lower $L/M$ and higher $\Teff$ models, {\it i.e.} on the blue side of the instability region. 
In Fig.\,\ref{f:mods30}, models  with $\log \Teff = 4.40$ and $\log L/\Lsolar \leq 1.95$ and  with $\log \Teff = 4.45$ and $\log L/\Lsolar \leq 2.15$ are 1H pulsators. \\
8) The last two points are correlated. The reverse shock occurs in the most compact models pulsating in the F-mode. Models with lower $L/M$ ratios which {\it do not} show the reverse shock are pulsating in the 1H-mode.  \\
9) It was anticipated that amplitude would reflect the luminosity-to-mass ($L/M$) ratio. 
This property is not demonstrated strongly by the $0.30\Msolar$ grid, but is evident for grids which extend to lower $L/M$ ratios (Figs.~\ref{f:mods35},\ref{f:mods40}).  
At sufficiently small $L/M$ ratios, depending on \Teff, the amplitudes  become small to vanishing. \\ 
10) Generally it was  difficult to converge models with  $\log (L/M) / (\Msolar/\Lsolar) \simge 2.9$. None were attempted for $\log (L/M) / (\Msolar/\Lsolar) > 3.0$. It would probably be necessary to reduce the iron+nickel contribution to the opacity to converge models at higher $L/M$ ratios. \\
11) For $\log \Teff/{\rm K}>4.50$, the models display a variety of behaviours and some are difficult to explain. High overtone and multi-mode pulsations may be excited. These models are discussed further in \S\,\ref{s:weird}. 

\begin{figure}
\centering
\includegraphics[width=0.98\linewidth,trim={0 2.2cm 0 -0.1cm}]{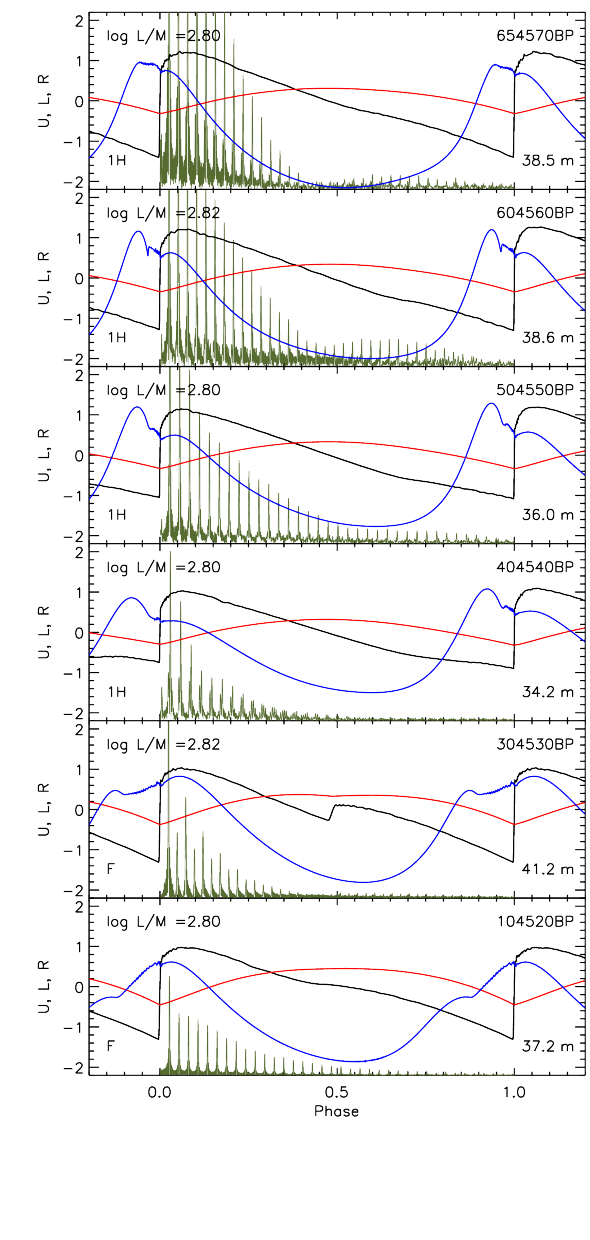}
\caption{
As Fig.\,\ref{f:254530BP} (top panel) for models with  $\log \Teff=4.45$,  $\log L/M\approx2.8$, $P\approx 35 - 40$\,min and $M = 0.2 - 0.7 \Msolar$.  
 }
\label{f:mcomp}
\end{figure}

\subsection{Effect of mass}
\label{s:mass}

As in Fig.\,\ref{f:nmM}, Figs.\,\ref{f:mods20}--\ref{f:mods70} show similar ranges in \Teff\ and $L/M$ ratio but for different masses. 
As anticipated from the linear analysis, the properties of pulsation models for a given \Teff\ and $L/M$ ratio vary slowly with mass. 
The general properties described above are obtained for models from 0.2 to 0.7 \Msolar\ (Fig.\,\ref{f:mcomp}). 
One significant difference is that the locus of modes with a strong reverse shock shifts towards lower \Teff\ or longer period with increasing mass. 
Providing $\log L/M>2.6$ the reverse shock appears at $P\approx 25, 35, 41, $ and 50\,min  at $M=0.20, 0.30, 0.35$ and 0.40 \Msolar\ respectively , extending to $P\approx 90$ min at $M=0.70\Msolar$ (Figs.\,\ref{f:mods30}, \ref{f:mods20} -- \ref{f:mods70}).
This correlation also reflects the period at which the models switch from F to 1H mode pulsations.  
It also explains why only one of the models in Fig.\,\ref{f:mcomp} (304530BP) shows the reverse shock. 
Significantly, the FT for the next model (404540BP) shows that both 1H and F modes are excited, explaining why consecutive light curve maxima have different values  (Fig.\,\ref{f:mcomp}). 
The same figure provides evidence that the F mode persists weakly in higher-mass 1H pulsators. 
Observing a velocity bump around maximum radius would offer a powerful diagnostic of mass, as would the detection of a double-mode (F+1H) BLAP. 

\subsection{Effect of chemical composition}

In contrast to mass, the effect of chemical mixture is  profound (Figs.\ref{f:mods30XX}--\ref{f:mods30N1}).

Grids XX and HE demonstrate the effect of increasing or reducing the hydrogen content, respectively.
Since hydrogen is effectively a damping agent for Z-bump driven pulsations, these grids should show weaker or stronger pulsation respectively. 
This is realised for the low $L/M$ hydrogen-rich (XX) models ($\log \Teff/{\rm K}=4.40, 4.45$, $\log L/\Lsolar = 1.95, 2.00$), where the amplitudes are smaller and shocks are not seen in all cases. 
The effect is less obvious for the hydrogen-poor (HE) models, but the increase in $Y$ from 0.48 to 0.60 is probably not very significant. 

Grid N1 demonstrates the effect of reducing the iron and nickel content without changing the hydrogen and helium content. 
As in the linear case, the size of the instability strip is markedly reduced, although it is interesting that small-amplitude high-overtone oscillations appear to be excited around $\log \Teff/{\rm K}=4.50$.

\begin{figure}
\centering
\includegraphics[width=0.98\linewidth]{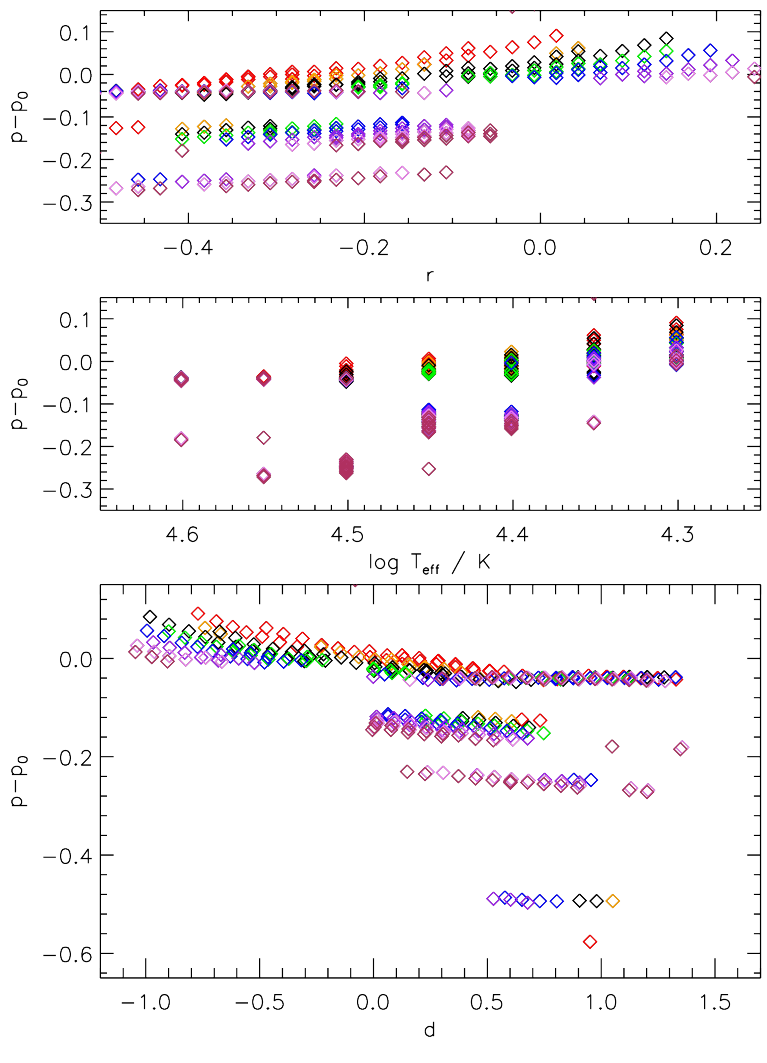}
\caption{The normalized model pulsation period { $p-p_0 \equiv \log P/P_0$} as determined from the power spectrum of the luminosity curves and as functions of radius $r=\log R/ \Rsolar$, effective temperature $\log \Teff$ and mean density $d = \log \bar{\rho}/\bar{\rho_{\odot}}$. The normalization has been carried out such that { $p_0 = 1.725-d/2$ }.  
All quantities are shown as logarithms (base 10).
All models are for mixture BP. 
Models with different masses are plotted in red (0.20\Msolar), orange (0.25 \Msolar), black (0.30 \Msolar), green (0.35 \Msolar),  blue (0.40 \Msolar), purple (0.50 \Msolar), violet (0.60 \Msolar) and maroon (0.70 \Msolar). 
{\it Bona fide} F and 1H mode pulsators are easily identified in the bottom panel as lying on one or other of the two lower sequences. 
}
\label{f:p_trends}
\end{figure}

\begin{figure}
\centering
\includegraphics[width=0.98\linewidth]{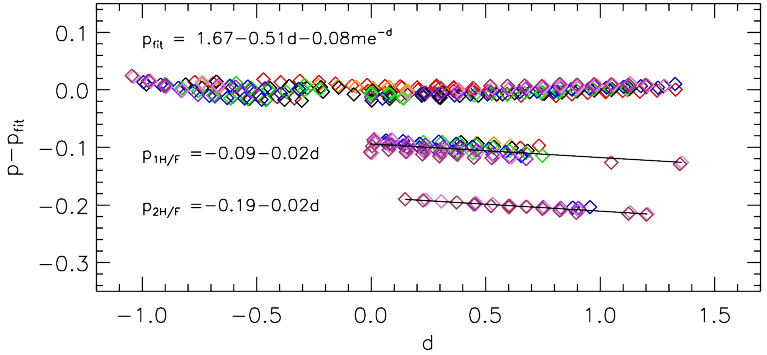}
\includegraphics[width=0.98\linewidth]{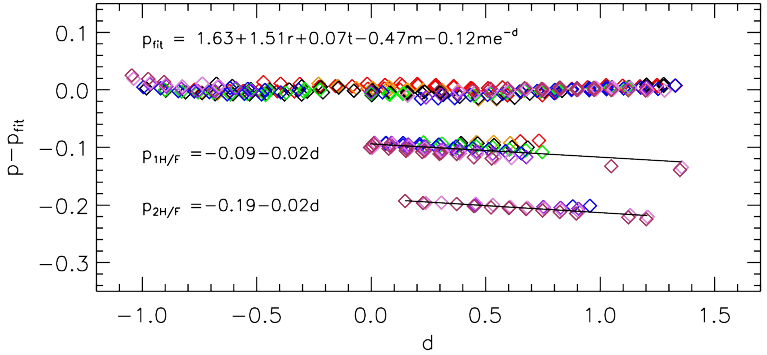}\\
\caption{
Top: Residuals from  for the best-fit period mean-density relation obtained from non-linear radial pulsation models for BLAPs. 
All quantities are shown as logarithms (base 10). 
All models are for mixture BP. 
The symbols are colour-coded for mass as in Fig.\,\ref{f:p_trends}. 
The form of the polynomial fit for F-mode pulsations is shown above the data. 
The polynomial fits for 1H and 2H-mode pulsations are shown as a ratio with the F-modes and are indicated by a solid black line. 
Bottom: As above for  the best-fit period - radius - mass - temperature relation.
 }
\label{f:p_fits}
\end{figure}

\begin{figure*}
\centering
\includegraphics[width=0.98\linewidth]{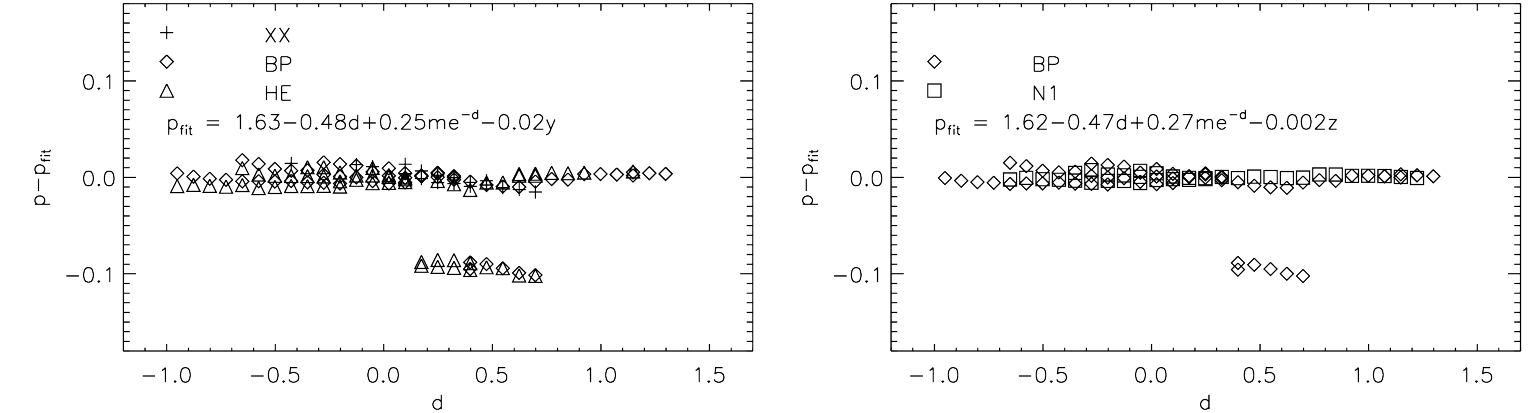}
\caption{As Fig.\,\ref{f:p_fits} with residuals from the polynomial fits in $y$ and $z$.  
Models for different composition are shown with symbols as labelled.}
\label{f:p_xyzfits}
\end{figure*}

\subsection{Period relations}
\label{s:fits}

Observationally, there is a long tradition of deriving relationships between the periods of radially pulsating stars and properties such as radius, temperature, luminosity and mass \citep{leavitt12,payne30,benedict07}.
The rationale is embedded in pulsation theory through, for example, the period  mean-density relation \citep{ritter79,eddington18}.

The same applies to BLAPs, where a 
 tight period-gravity relation $\sg = -1.14(5) \log P + 6.30(7)$ has been identified \citep{pietrukowicz24}.   
Such relations exist because pulsation period $P$, mean density $\bar{\rho}$, mass $M$, radius $R$ and surface gravity $g$ are closely correlated; by identity, $g \propto \bar{\rho} R$, $\bar{\rho} \equiv M/R^3$ and, from the pulsation-period  mean-density relation, 
$P \propto \sqrt(1/\bar{\rho}) \propto \sqrt(R/g) $. 

To quantify correlations between various quantities in the pulsation models, it is  convenient to work in logarithms (base 10 assumed). The following notation is adopted for this subsection only:\\
Period: $ p = \log P / {\rm min} $\\
Mass: $ m = \log M / \Msolar $\\
Radius: $ r = \log R / \Rsolar $\\
Luminosity: $ l = \log L / \Lsolar $\\
Effective Temperature: $ t = \log T_{\rm eff} / {\rm T}_{\rm eff \odot} $\\
Mean density: $ d = \log \bar{\rho} / \bar{\rho}_{\odot} $\\
Helium / Hydrogen ratio (by mass): $ y = \log Y / X $\\
Iron and Nickel Abundance (by mass fraction): $ z = \log X_{\rm Fe+Ni} $\\

Fig.\,\ref{f:p_trends} shows correlations between $p$, $r$, \Teff\ and $d$. 
To better separate trends, $p$ is normalized { to $p-p_0$} assuming a simple period mean-density relation { $p_0 = q-d/2$, where $q\equiv\log Q$ where $Q$ is the constant of proportion in the $P-\bar{\rho}$ relation. 
Its value has been obtained using a representative model 304030BP ($p=1.788$, $d=-0.127$ $\rightarrow q=1.725$) to establish $p-p_0$ throughout.} 
Since $d \propto -3r$ ($\bar{\rho} \propto R^{-3}$), the top and bottom panels show equivalent trends. 
Fig.\,\ref{f:p_trends} also shows groups of models with residual { $p-p_0 \approx -0.1$}, $-0.18$, $-0.42$  and $-0.5$ respectively (bottom panel).
These may be identified with models pulsating in the first (1H), second (2H) and higher overtones. 
Departure from a linear period-mean density relation is apparent at all masses, but is most marked at low mass (0.20\Msolar) and reduces with increasing mass.
At all masses, the departures increase exponentially as $e^{-d}$; {\it i.e.} as the stellar envelope becomes more extended, the density drops, amplitudes increase and both non-adiabatic and non-linear effects become increasingly important.  
The small scatter in $p-p_0$ for $\log \Teff > 4.5$ is due to the absence of strongly excited modes; nearly all pulsations excited are of small amplitude and the linear approximation is likely to be good.
A small zero-point offset is set by the arbitrary choice of model used to normalize $p_0$.   

Parametric fits to the model periods were constructed for various combinations of linear and quadratic terms in $r$, $t$, $m$, $d$, $y$ and $z$.   
Fits to $l$ and $g$ were not considered since these correlate with $r$, $t$ and $m$.  
Moreover, $r$ and $t$ can be obtained directly from the distance and spectral energy distribution or (for pulsating stars) from Baade's method and $g$ is more difficult to measure precisely.
Although successful for a restricted mass range (0.2 -- 0.4\Msolar), the mass dependency evident in Fig.\,\ref{f:p_trends} was not successfully captured over the full range 0.2 -- 0.7 \Msolar.  

Rather, after introducing an inverse exponential term in $d$, a correlation with $d$ and $m$ of the form
\begin{equation}
p_{\rm fit} = 1.67 -0.51 d -0.08 m e^{-d}, \chi^2 =  3.6(-5)
\label{eq:den_best}
\end{equation}
successfully yields a fit to 314 F-mode periods (Fig.\,\ref{f:p_fits}). 
$\chi^2$ is a merit figure derived from the fit. 
This elegantly and precisely expresses the models in terms of the standard period-mean-density relation $ p \approx a - 0.5 d $ and a non-linear mass-dependent correction.  
The periods of 85 1H and 30 2H modes can be obtained from: 
\begin{equation}
    \log P_{\rm 1H}/P_{\rm F} = -0.09 -0.02 d,  \chi^2 =  4.6(-5),
\label{eq:den_best_1h}
\end{equation}
\begin{equation}
    \log P_{\rm 2H}/P_{\rm F} = -0.19 -0.02 d,  \chi^2 =  4.3(-6).
\label{eq:den_best_2h}
\end{equation} 
{ Jackknife analyses were carried out for all regressions to obtain standard deviations $\sigma_{\alpha}$ for coefficients $\alpha$. 
The largest values of $\sigma_{\alpha}/\alpha$ in each of Eqns.\,\ref{eq:den_best}-\ref{eq:den_best_2h} are 0.3\% (coefficient of $m e^{-d}$), 1.2\% ($d$) and 0.9\% ($d$), respectively.
$\sigma_{\alpha}/\alpha<$0.3\% in all other cases.  }  

To investigate other dependencies, we solved for terms in $r$, $t$ and $m$ separately, but again included a term in $me^{-d}$ to fit the dependency on mass:
\begin{multline}
p_{\rm fit} =  1.63 +  1.51 r +  0.07 t -0.47 m -0.12 m e^{-d}, \\
\chi^2 =  2.1(-5).
\label{eq:rtm_best}
\end{multline}
Again, this fit is illustrated in Fig.\,\ref{f:p_fits} and still closely tracks the standard period mean-density relation  $ p \approx a + 1.5 r - 0.5 m $.
It is valid over the ranges: $-0.5 \leq r \leq 0.25$, 
$4.3 \leq \log \Teff \leq 4.6$ and  $ 0.20 \leq M/\Msolar \leq 0.70$. 
Effective temperature plays only a minor role.
The merit figure $\chi^2$ represents a small improvement over Eq.\,\ref{eq:den_best}. 
The periods of 1H and 2H modes can be obtained from: 
\begin{equation} 
\log P_{\rm 1H}/P_{\rm F} =  -0.09 - 0.02 d,  \chi^2 =  4.0(-5),
\label{eq:rtm_best_1h}
\end{equation}
\begin{equation}
\log P_{\rm 2H}/P_{\rm F} =  -0.19 - 0.02 d,  \chi^2 =  1.3(-5). 
\label{eq:rtm_best_2h}
\end{equation}
In more familiar terms $P_{\rm 1H}/P_{\rm F} = 0.81 (\bar{\rho}/\bar{\rho}_{\odot})^{-0.02}$ and 
$ P_{\rm 2H}/P_{\rm 1H} = 0.81$.
{ The largest values of $\sigma_{\alpha}/\alpha$ in each of Eqns.\,\ref{eq:rtm_best}-\ref{eq:rtm_best_2h} are 1.3\% ($r$), 2.2\% ($d$) and 2.0\% ($d$), respectively.}

It is noted that for classical Cepheids in the Milky Way and other galaxies, observed values of $P_{\rm 1H}/P_{\rm F} \approx 0.71 - 0.75$ and $P_{\rm 2H}/P_{\rm 1H} \approx 0.81$ \citep{pietrukowicz21}. 
Inspection of the {\it linear} models shows that, for $M=0.30\Msolar$, $\log \Teff/{\rm K} = 4.46$, chemical mix BP, and $\log L/M$ in the range $2.20 - 3.00$,  $P_{\rm 1H}/P_{\rm F}$ varies monotonically from $0.75 - 0.82$. 
Hence the mode identifications for the non-linear BLAP models are robust. 

Fits were obtained for models with $M=0.30\Msolar$ and the four chemical mixtures BP, XX, HE and N1. 
Extending the period-density relation, these were done piecewise to find the coefficients of $y$ and $z$ separately to be 0.02 and 0.002 (Fig.\,\ref{f:p_xyzfits}), { with 
$\sigma_{\alpha}/\alpha = $0.9\% ($y$) and 3.7\% ($z$), respectively.}
Such small values imply that composition has an almost negligible effect on the pulsation period.

\citet{pietrukowicz24} report a period-luminosity relation for BLAPs by analogy with the Cepheid $P-L$ relation. The latter arises because a) Cepheids  have roughly the same \Teff\ and hence $L\propto R^2$, b) Cepheids evolve at approximately constant luminosity from the main-sequence, where $L\propto M^3$ and c) pulsations occur on a dynamical timescale $P\propto R^{3/2}M^{-1/2}$. 
Eliminating $M$ and $R$ gives $P\propto L^{7/12}$ or $\log P \propto 0.58 \log L$.
\citet{pietrukowicz24} assert  $M$ and $\Teff$ to be similar for all BLAPs and use observations of OGLE-BLAP-009 \citep{bradshaw24} to derive $\log P \propto 0.35 \log L$. 
This can be compared with the models.
A fit to the model periods using $l$, $t$ and $m$ as free parameters was obtained. Again, we introduce a term in $me^{-d}$ to ensure a good fit across the mass range, and find
\begin{multline}
p_{\rm fit} =  1.63 +  0.75 l -2.95 t -0.47 m -0.12 m e^{-d}, \chi^2 =  2.1(-5)  
\label{eq:ltm_best}
\end{multline}
This is identical to Eq.\,\ref{eq:rtm_best} with $r=(l-4t)/2$. 
{ The fractional $\sigma_{\alpha}/\alpha$ are 0.04\%, 0.04\%, 0.03\%, 0.09\% and 0.25\%.  }
The large temperature exponent indicates there should be a significant contribution from $t$ to any $p-l$ relation derived from observation.
It is difficult to recover anything like the \citet{pietrukowicz24} $p-l$ relation without asserting an additional correlation between $t$ (or $r$) and $l$. 

\begin{figure}
\centering
\includegraphics[width=0.98\linewidth]{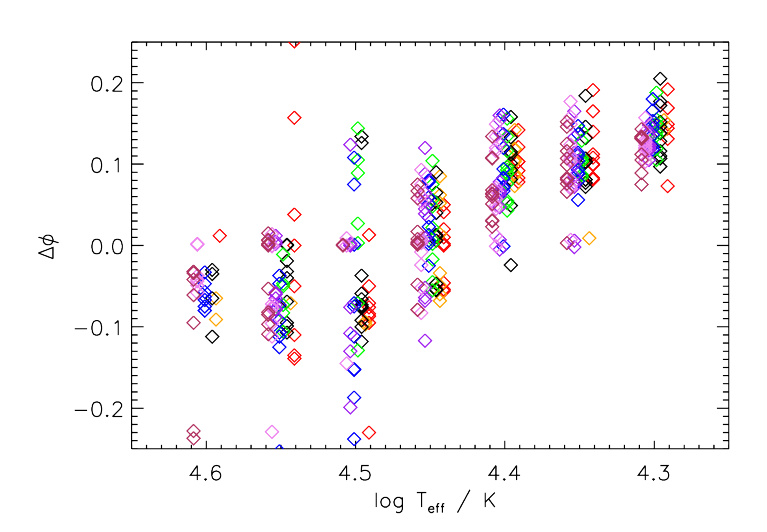}
\caption{The phase difference between maximum light and minimum radius as a function of effective temperature. 
Only one maximum is recorded for each light curve, even if the maximum has a compound or double-peaked structure. 
The models are colour-coded for mass as in Fig.\,\ref{f:p_fits} and displaced in \Teff\ for visibility. 
}
\label{f:dphi}
\end{figure}

\begin{figure*}
\centering
\includegraphics[width=0.48\linewidth]{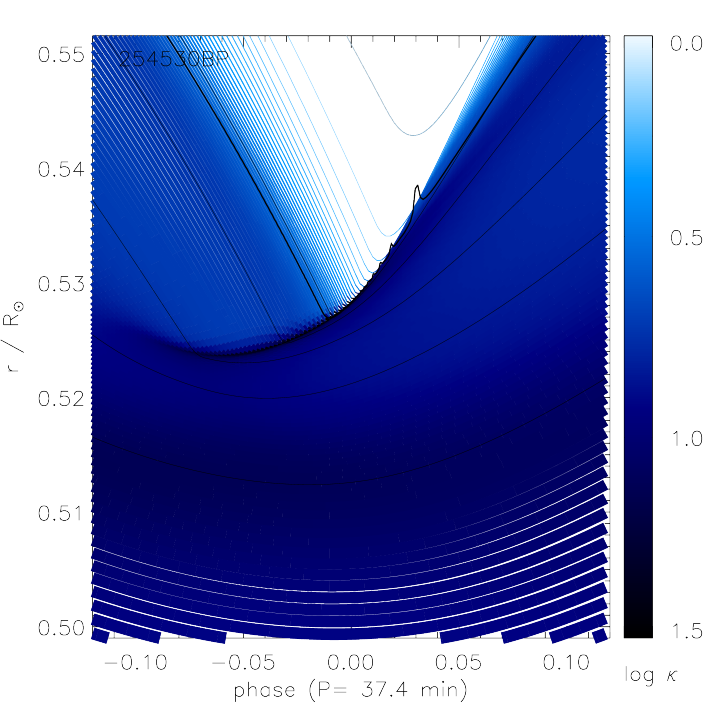}
\includegraphics[width=0.48\linewidth]{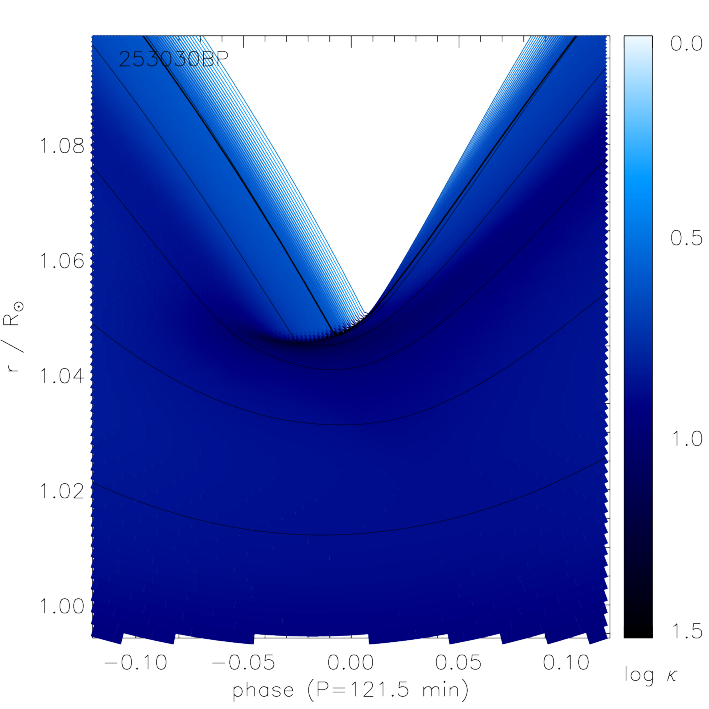}
\caption{As Fig.\,\ref{f:surf} for outer layers close to the surface during passage through minimum radius, but showing the radiative opacity ($\kappa$) for models 254530BP (left) and  253030BP (right).
The dark high $\kappa$ zone close to the surface visible in both models corresponds to the He{\sc ii/iii} ionization region. The lower dark zone visible in model 254530BP corresponds to the iron-nickel opacity bump at $\log T/{\rm K}\approx5.3$. The corresponding region lies below the plot window for model 253030BP. 
}
\label{f:surf_k}
\end{figure*}

\subsection{Light curve diversity}

Fig.\,\ref{f:mods30} demonstrates the striking diversity of light curves  referred to in \S\,\ref{s:system30}. 
Remarkably, these broadly reflect the diversity of BLAP lightcurves observed by \citet{pietrukowicz24}, although the comparison is imperfect since the models give luminosity, whereas observed BLAP lightcurves cover a restricted wave band. 
In the models, lightcurves with flat bases and narrow peaks correspond to longer-period pulsations (cooler stars); light maxima precede minimum radius.
Lightcurves with rounded bases correspond to shorter periods (hotter stars); light maxima follow minimum radius. 
Lightcurves with compound or double maxima tend to be found 
in between, with a peak either side or a sharp dip at the time of minimum radius.  
 
Fig.\,\ref{f:dphi} illustrates this dichotomy in phase shift between in more detail by plotting 
\begin{equation}
 \Delta \phi \equiv \phi(R_{\rm min}) - \phi(L_{\rm max})    
\end{equation}
as a function of \Teff\ for all of the mix BP models.
Only one light maximum is plotted per model, even if more than one {\it local} maximum is present. 

There may be several ways to interpret the behaviour of $\Delta \phi$.
For example, the definition of $\phi(R_{\rm min})$ depends on which mass zone is used to define radius minimum ($r_{\rm min}(m)$). 
In terms of phase, the latter propagates outward more indirectly in the hotter model, even though it is moving more quickly over a shorter distance (see Fig.\,\ref{f:surf}).  
Thus, for model 254530BP, the phase of $r_{\rm min}(m)$ relative to the photosphere varies from near zero at depth to $-0.07$ ($\log T \approx 4.9$) and then back through zero at the photosphere to $+0.03$ in the outer atmosphere.  
In contrast,  $r_{\rm min}(m)$ for model 253030BP varies by no more than $-0.02$ and $+0.01$ in phase. 

Another view is to look at the opacity (Fig.\,\ref{f:surf_k}). In terms of fractional radius, the principal iron-nickel driving zone is closer to the stellar surface in the hotter star. This will result in more non-linear coupling between the driving zone and the stellar surface, and hence explain the large phase variation in $r_{\rm min}(m)$.
More consequentially, comparing the distribution in temperature and opacity above the He{\sc ii/iii} ionization region, we see more material, and hence more total opacity, before $R_{\rm min}$ and less after $R_{\rm min}$ in the hotter model 254530BP than in the cooler model 253030BP. 
The phase at which the hottest material is visible from outside the star appears therefore to be a consequence of the transparency of the outer layers; this could also be a consequence of the smaller depth of the driving zone beneath the surface.   

\begin{figure*}
\centering
\includegraphics[width=0.32\linewidth]{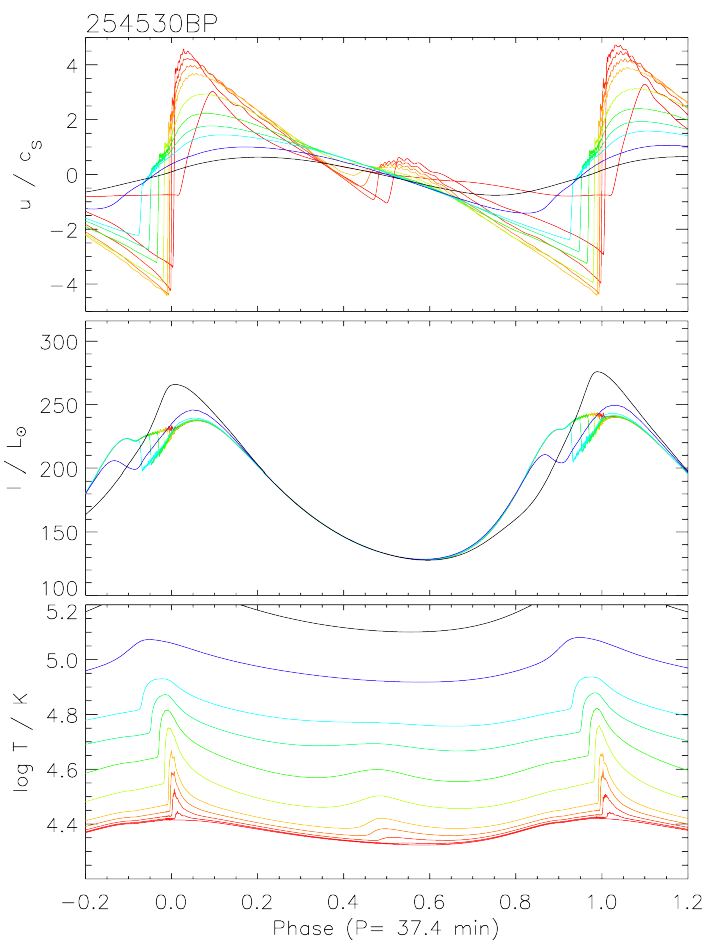}
\includegraphics[width=0.32\linewidth]{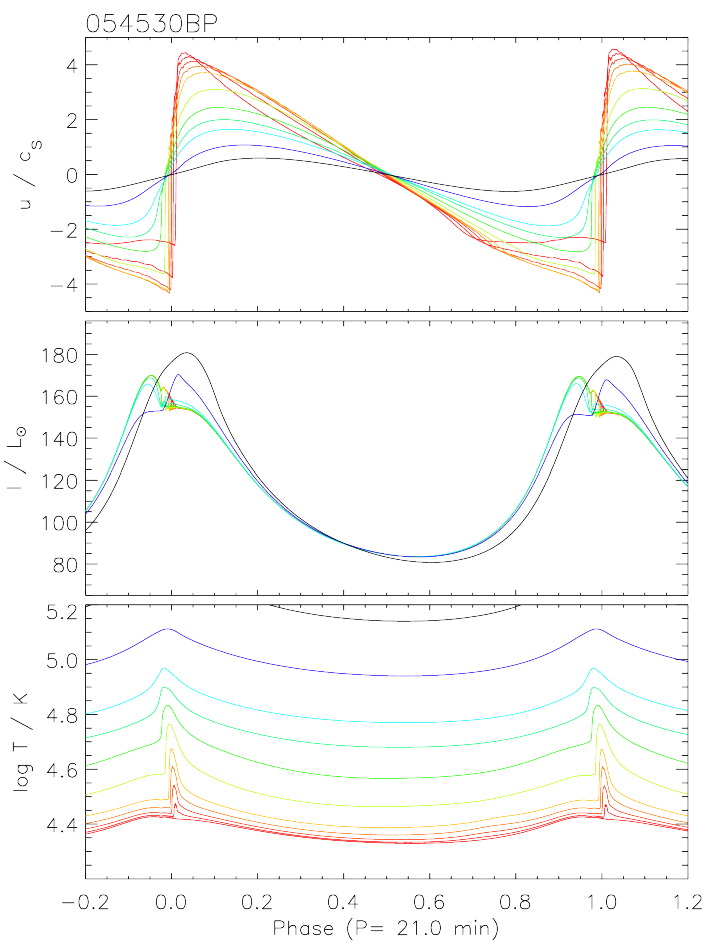}
\includegraphics[width=0.32\linewidth]{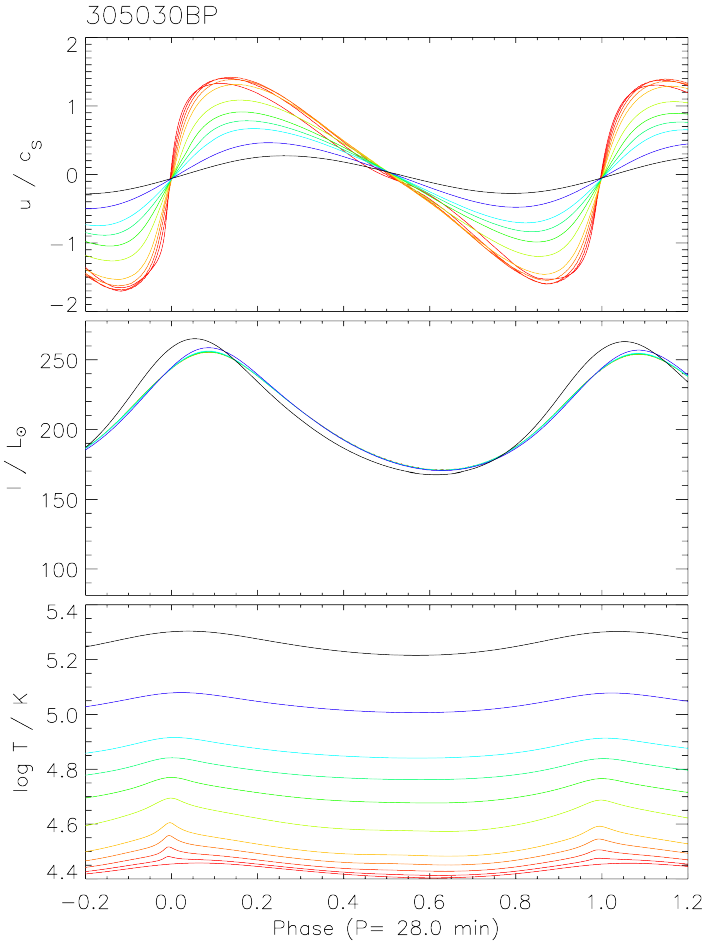}
\caption{
The runs of radial velocity $u$, luminosity $l$ and temperature $T$ with phase for selected zones in the outer part of the stellar envelope. Velocity has been scaled to the local sound speed. Each shell is colour-coded identically in each panel, red being coolest and violet (black) being warmest. The panels include the reference model (254530BP: left), a low $L$ model (054530BP: middle), and a low-amplitude model (305030BP: right).}
	\label{f:layers}
\end{figure*}

\begin{figure*}
\centering
\includegraphics[width=0.32\linewidth]{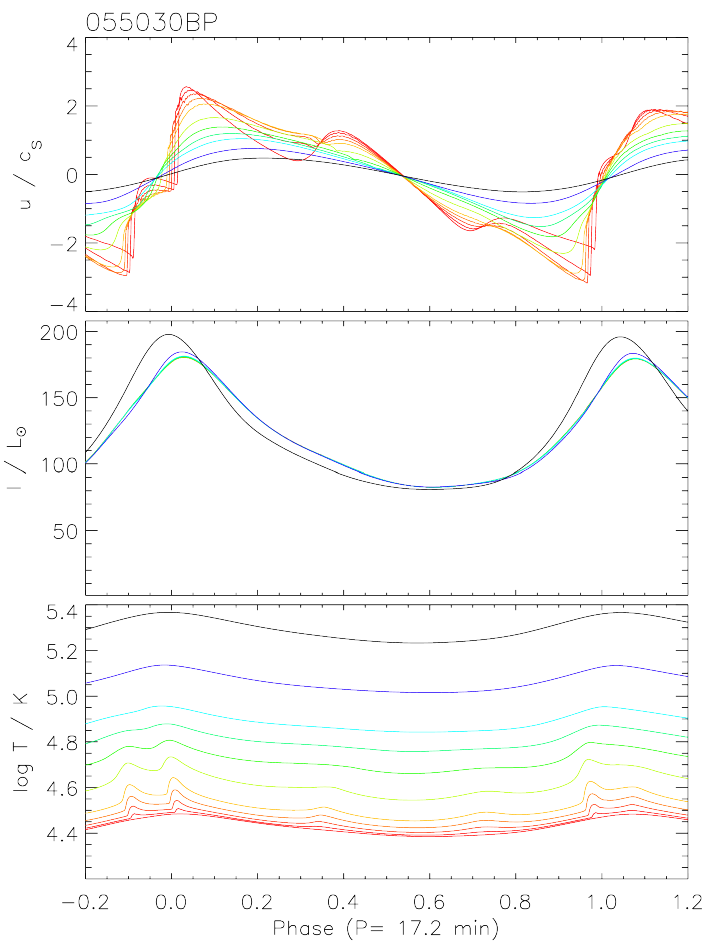}
\includegraphics[width=0.32\linewidth]{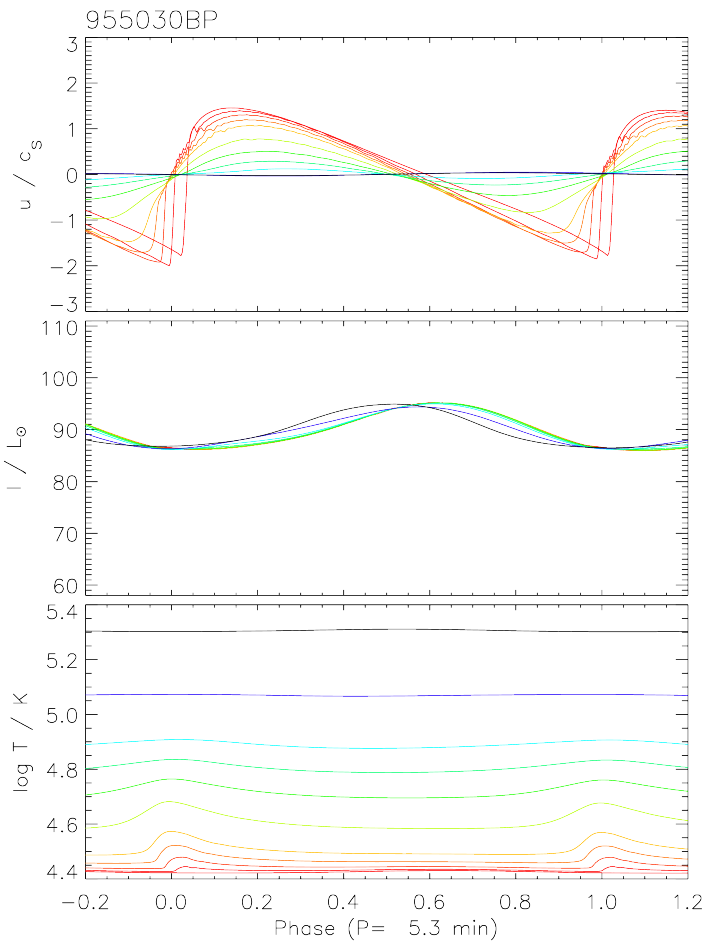}
\includegraphics[width=0.32\linewidth]{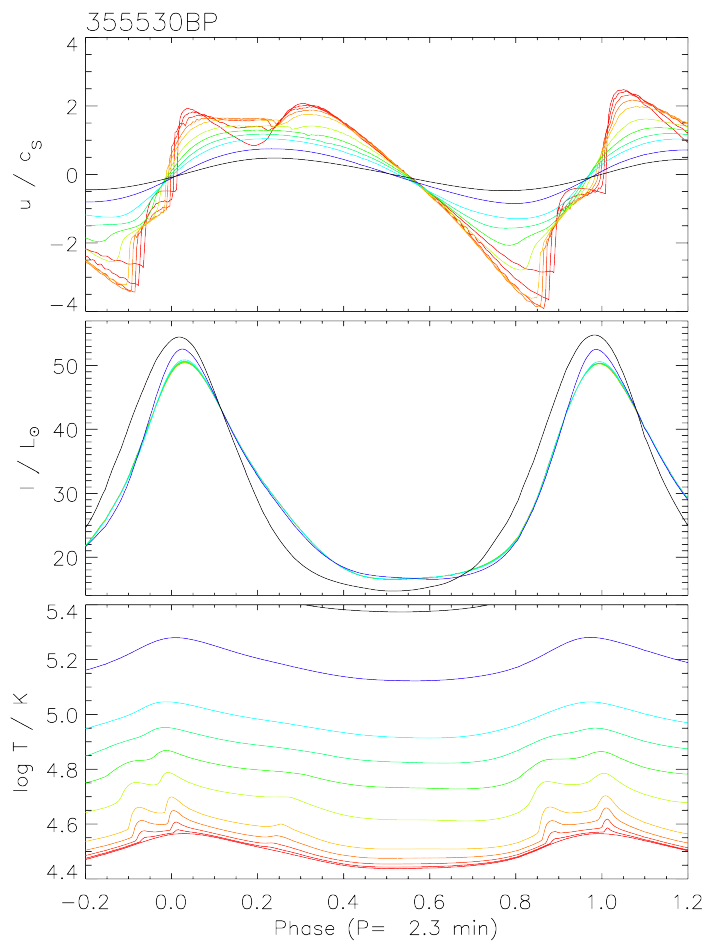}
\caption{
As Fig.\,\ref{f:layers} for a selection of unusual models including: 
055030BP (left) which shows a standstill in $u$ just before $R_{\rm min}$ in Fig:\,\ref{f:mods30}, 955030BP (middle) which is likely a high-order overtone pulsation, and 355530BP (right) which appears to be a high-order overtone pulsation with shocks.}
	\label{f:weird}
\end{figure*}

\subsection{Shocks}
\label{s:shocks}

\citet{jeffery15b}, \citet{jeffery22a} and \citet{jeffery22b} have already discussed the observational and theoretical evidence for shocks in the pulsations of V652\,Her.
\citet{bradshaw24} provide substantial evidence that shocks occur in the pulsation of OGLE-BLAP-009.
In the case of V652\,Her, the primary criterion for a shock at minimum radius is pulsation amplitude, and this is contingent upon large enough opacity gradients in the iron-nickel bump. 
The lower amplitude pulsations in BX\,Cir do not trigger a shock. 

Owing to the high iron and nickel abundance adopted in the current models, which is supported by the OGLE-BLAP-009 observations \citep{bradshaw24} and  predicted by diffusion theory \citep{byrne20}, shocks occur at minimum radius in nearly all models pulsating in the fundamental radial mode.  
Evidence is provided by supersonic jumps in velocity, with $\delta u/c_s \gg 1$ and temperature spikes indicative of sudden heating ({\it e.g.} 254530BP, Fig.\,\ref{f:layers}). 
Exceptions include low $L/M$ models ({\it e.g.} 054540BP, Fig.\,\ref{f:mods40} and 305030BP, Fig.\,\ref{f:layers}) and some models pulsating in the 1H mode ({\it e.g.} 355035BP, Fig.\,\ref{f:mods35}). 

As noted earlier, the present models differ from the  \citet{jeffery22a} models by the presence of a reverse shock at maximum radius ({\it e.g.} 254530BP, Fig.\,\ref{f:layers}) and multiple shocks at minimum radius ({\it e.g.} 055030BP, Fig.\,\ref{f:layers}).
The reverse shock is a consequence of differential expansion in the early part of the pulsation cycle; when the outer layers of the star are accelerated ballistically but are unsupported by more slowly expanding layers beneath, they start to fall inwards whilst the lower layers are still expanding, leading to a weak shock when they collide (Fig.\,\ref{f:surf}). 
It is not immediately clear what conditions are necessary for the reverse shock to occur, since they are found in both extended (453530BP) and compact (204530BP) models, but are not always present (054530BP: Fig.\,\ref{f:layers}). 
The shock in the more compact models could be driven by a resonance with the 1H overtone in the outer layers of the star. 
It only occurs in models pulsating in the F-mode which are slightly more luminous than those in which the 1H mode is excited, and in which the reverse shock is not seen. 

\subsection{Curiosities}
\label{s:weird}

There are a number of other interesting models in the survey (cf. Fig:\,\ref{f:mods30}). 
These include 055030BP, which appears to have a step in the radial velocity curve, indicative of a double shock at minimum radius. 
Closer inspection shows this model did not converge to a limit cycle of 19\,200\,000 time steps, even though showing well-developed oscillations.  
The suspicion is that additional modes are excited and interfere in the surface layers  (Fig.\,\ref{f:weird}). 
However, the FT shows only one dominant period and its harmonic series. 
Model 955030BP shows a low-amplitude oscillation with period $5.4$min substantially shorter than the anticipated F-mode period of $15.2$m and is almost certainly a high-order radial overtone. The FT shows both modes are present, with the short-period being 5 times larger in amplitude. The period ratio of $0.36 \approx 0.81^5$ implies that the $5.4$min period is the 5H mode.   
Such stars could account for some of the short-period BLAPs identified by \citet{kupfer19}.
Curiously, the model in between (005030BP) shows a mismatch between the radius and luminosity curves. Again, the FT shows three principal modes and their harmonics.
Assuming period ratios $P_{k\rm H}/P_{\rm F} \approx 0.81^k$, these are likely 5H (5.8\,min), F (16.7\,min) and 3H (9.1\,min) in order of amplitude.

For the $M=0.30\Msolar$ BP grid, and for $\log \Teff / {\rm K} = 4.55$ and 4.60 and $\log L/\Lsolar \geq 2.25$, half of the models are stable against pulsations, and half show very short-period radial oscillations; model 355530BP is illustrated in Fig.\,\ref{f:weird}.
The short-period oscillators have large light and velocity amplitudes, but small radius amplitude (Fig:\,\ref{f:mods30}); several shocks or compression episodes occur within each cycle. 
Their periods lie between 1.5 and 3\,min and $\approx 1/10$ of the predicted F-mode period; all are likely associated with excitation of one or more high-order radial overtone. 
Although the periods are appropriate, the \Teff\ and $L$ of these models are too high to associate with the short-period BLAPs \citep{kupfer19}. 

In the $M=0.30\Msolar$ N1 grid and for $\log \Teff / {\rm K} = 4.50$, and for other mixtures, there are several models with periods from the FT which are less than half those from successive minima   (Fig.\,\ref{f:mods30N1}). 
The reason is that the search algorithm for the latter uses an estimate of the F-mode period to define the search window; it can be deceived by local minima if shorter-period oscillations are present.
Scrutiny of the internal pulsation characteristics and the FTs indicate once again that high-order radial modes are excited. 
 
The excitation of these large-amplitude high-overtone modes is of considerable interest. 
Further work should establish whether they are purely artefacts of the models, whether non-radial modes would be present as well, and under what circumstances they could be detected in real stars.
Some of these models may indeed be identified with the short-period BLAPs. 
{ There is also a corollary with RR\,Lyrae variables, where evidence for multi-mode pulsations including high-order radial and non-radial modes is emerging \citep{netzel22,benko23,netzel24}. }

\subsection{Linear versus non-linear models}

Figures\,\ref{f:nmZ02}--\ref{f:nmM} compare the number of unstable modes in the linear non-adiabatic models and the dominant pulsation mode in the non-linear models for masses and compositions where both have been calculated. 
The first general conclusion is that the red edge of the instability region coincides in both treatments; in several cases the agreement is remarkable (Fig.\,\ref{f:nmM}: $X=0.47, Z=0.047$), in others not 
(Fig.\,\ref{f:nmZ}: $X=0.47, Z=0.02$).
The second is that 1H overtone pulsations are found on the blue side of the instability region in both treatments and generally at higher masses (Fig.\,\ref{f:nmM}). 
2H overtones are predicted at high mass ($0.70\Msolar$) in the linear analysis. 2H and higher overtones are found quite frequently in the non-linear models 
where they extend the blue edge of the instability region beyond that which the linear models predict. { The lowest-order excited mode for each model is identified in Figs.\,\ref{f:254530BP}, \ref{f:mods30}, \ref{f:mcomp} (lower left of each panel), Table \ref{t:models} (last column: $k$) and their counterparts in App.\,\ref{s:app2}. }

\section{Future Work}

Whilst we have presented a broad survey of linear and non-linear radial pulsation models for faint blue stars, much work remains to be done. 

To compare the model light curves quantitatively with observation, they should be combined with a radiative transfer code to give the emergent spectral energy distribution a a function of phase, and then convolved with  appropriate filter functions to compare with the instrument or survey of interest.  
Similarly, the pulsation models should be combined with realistic model atmospheres to generate phase-dependent spectra.  
Both can be done with {\sc spectrum} \citep{jeffery22b} in due course.
This is required to compare models directly with stars like OGLE-BLAP-009 \citep{bradshaw24}, as was done for V652\,Her by \citet{jeffery22a}. 

The present models assume homogeneous envelopes; the derived periods appear to be robust to composition, but chemically-stratified envelopes will enhance opacity gradients and hence affect the stability and amplitude of given modes. 
This will be particularly important in defining the F, 1H and higher overtone boundaries. 
This uncertainty needs to be addressed by taking abundance profiles from evolution models that include the chemical diffusion of iron and nickel by radiative levitation and gravitational settling.
Since the values predicted by \citep{byrne20} are too high for our pulsation models, consideration needs to be given to how chemical diffusion might be moderated by  large-amplitude pulsations. 
This will be more important for models around $\log \Teff\approx 4.45$ than around $\log \Teff\approx 4.30$ owing to the larger displacement and shorter period of the iron+nickel driving zone when it lies closer to the stellar surface. 

The discovery of multi-mode oscillations in our models was unexpected. 
It warrants systematic exploration to establish what circumstances trigger this behaviour and how likely it is to be detected in real stars. 
Models such as 555060BP (Fig:\,\ref{f:multi} will help to interpret observations of multi-mode oscillations like those in OGLE-BLAP-030 \citep{pietrukowicz24}. 

The current models are limited by the upper $L/M$ limit that can be successfully modelled; this appears to be due to high opacity producing pulsations of such large amplitude that the models (and maybe real stars) { are numerically} unstable. Some calculations should be repeated for a lower iron and nickel content to test this.  

The presence of large-amplitude short-period pulsations in a number of hot low $L/M$ models is tantalizing. How these relate to the large-amplitude pulsations in very short-period BLAPs \citep{kupfer19} and, indeed, in some subdwarf B stars should be investigated by also extending the non-linear models into that domain. 

\section{Conclusion}

The discovery of  large-amplitude pulsations in faint blue stars (BLAPs) has opened a new window through which to explore the late stages of stellar evolution. 
However, the scarcity and distance these stars makes detailed observation challenging, even with access to the largest telescopes. 
This paper predicts a rich kaleidoscope of pulsation properties which will help to place the observations into a systematic framework. 

The linear and non-linear radial pulsation models for low-mass stars with evolved surface chemistries cover a range of masses, temperatures and luminosities appropriate for comparison with observations of BLAPs. 
The calculations are agnostic of evolution. 
The choice of parameter ranges has been guided by observations. 
The linear models are the most comprehensive available in terms of surface composition, if not of grid resolution  \citep[cf.][]{jadlovsky24}. 
The non-linear models are the first computed explicitly for comparison with BLAPs as currently understood, although not for hot evolved low-mass pulsators \citep{fadeyev96,montanes02,jeffery22a}.  

The main objectives were to: \\
(i) {\it Map the BLAP instability strip as a function of envelope composition.} 
This is done with both linear and non-linear models; a high iron+nickel abundance is key to placing the instability strip boundary, although a low hydrogen abundance will also promote pulsations.  \\
(ii) {\it Compare results for linear and non-linear pulsation calculations (Figs.\,\ref{f:nmZ02}--\ref{f:nmM}).} In general, there is good agreement at the red edge of the instability strip, but the non-linear models appear to be unstable to slightly higher \Teff\ and in higher-order modes.  \\
(iii) {\it Explore the relationship between BLAP amplitude, mass (or luminosity) and envelope composition (or opacity).}
For the majority of chemical mixtures studied, amplitudes for fundamental-mode (F) pulsations given by the non-linear models are large, typically $\sim 200\kmsec$ in stellar radial velocity, $> 10\%$ in radius and 40 -- 80\% in luminosity. The amplitudes are frequently at the limit of the models; increasing metallicity increases amplitude, but the models fail. Reducing metallicity (mix N1) reduces the amplitudes. Amplitudes drop towards instability boundaries. First overtone (1H)  pulsations occur at the low-$L$ high-\Teff\ side of the instability strip, and are more strongly excited at higher masses.  
A few models exhibit multi-mode oscillations, with overtones as high as 5H being excited. \\
(iv) {\it Interpret the diversity of BLAP light curve shapes.} 
The primary influence on light curve shape is the effective temperature of the pulsator; this determines the depth of the driving zone beneath the photosphere and consequently the phase of maximum luminosity relative to minimum radius. 
$L_{\rm max}$ precedes $R_{\rm min}$ in cool models, and follows in hot models. 
In intermediate cases where $L_{\rm max}$ and $R_{\rm min}$ overlap, shock compression in the photosphere leads to a short dip in brightness, { its precise phase relative to the overall maximum  will be a good indicator of \Teff.
A flat-bottomed light curve with a narrow light maximum generally indicates a cool BLAP.
A sinusoidal low-amplitude light curve corresponds to BLAPS with a low $L/M$ ratio or a low metallicity.
Such model properties could provide indicators for position in $L/M-\Teff$ space, especially for known $M$ and composition.
This said, the r\^oles of composition and the physical approximations used in the hydro calculations are probably not well enough understood to provide precise calibrators at this stage, and } the pulsation models need to be coupled to a radiative transfer model to explore filtered light curves and spectra more systematically.  

Although not evident in the { model light curves}, a reverse shock occurs close to radius maximum in the most compact F-mode pulsators; more compact models pulsate in the 1H-mode. The transition from F- to 1H-mode pulsations is marked by a period which is a function of mass. The reverse shock should be visible in radial-velocity measurements { with a magnitude of up to 20\kmsec \citep[allowing for a projection factor of 1/1.35:][]{montanes01}. It may also be visible in filtered photometry since phases around maximum radius should yield a secondary maximum \citep[cf. V652 Her:][]{lynasgray86b} which should respond to a sudden change in radius.   } 

In addition, various period relations have been derived from the non-linear models, as well as predictions for the 1H/F period ratio.
The classical period-mean density relation works well over the entire model range with the addition of a single term in mass and density.  
The same term can be used in providing a period-mass-radius-temperature relation which can be adapted for other parameters as required.  
The F/1H and 2H/1H period ratios are virtually the same at 0.81, the former showing a small density dependence.  
Period has a very weak dependence on the  helium/hydrogen ratio and is independent of metallicity.
An investigation of the period-luminosity relation shows that it has a strong temperature dependence, as well as a moderate mass dependence. 
These relations are likely to be more reliable than those derived from linear models, especially for large-amplitude oscillations.
Further exploration of the models will help address additional key questions raised in the \S\,\ref{s:objectives}

The current calculations are limited to the mass range 0.2 -- 0.7 \Msolar. Recent work suggests diverse origins and hence masses for BLAPs \citep[{\it e.g.\rm}][]{kolaczek24}. Hence higher-mass models should be investigated. Curiously, \citet{kolaczek24} find an allowable mass range 0.32 -- 0.70 \Msolar\ for BLAPs formed from a double white dwarf merger, a range which brackets the observations  and double white dwarf origin theory for the pulsating extreme helium star V652\,Her \citep{jeffery15b,saio00,zhang12a,jeffery22a} and BX\,Cir  \citep{kilkenny99b,woolf00,woolf02c}. Perhaps V652\,Her and BX\,Cir are the genuine BLAP prototypes. 

\section*{Data Availability} 
The outputs from the linear non-adiabatic calculations reported in \S\,\ref{s:linear} are stored as ASCII files, 1 file per mass and mixture (75 MB in total). 
A summary of the surface properties of each non-linear model as a function of phase over 1.4 pulsation cycles are stored as ASCII files, 1 file per model grid, in the format described by \citet{jeffery22a}. 
The surface properties and internal structures for the last 72\,000 timesteps of each non-linear pulsation model are stored as binary files (613 MB per model, 440 GB in total). 
These data can be made available upon reasonable request.

\section*{Acknowledgments}
The author is deeply indebted to Alan Bridger and  Pilar Monta\~n\'es-Rodr\'iguez for permission to modify and use the non-linear pulsation code {\sc puls\_nl}, and likewise to Hideyuki Saio for the linear non-adiabatic pulsation code {\sc radnad}.
Angela Jeffery suggested the inverse exponential term in the period-density relation.
{ The referee made perceptive suggestions, requested useful clarifications and caught a few lacunae, for which my thanks.}  
The Armagh Observatory and Planetarium (AOP) is funded by direct grant from the Northern Ireland Department for Communities. 
All non-linear models were run on the 480-node Sirius computing cluster at AOP. 
This research was part funded with support from the UK Science and Technology Facilities Council (STFC) grant no. ST/M000834/1.

\bibliographystyle{mnras}
\bibliography{ehe}

\appendix
\renewcommand\thefigure{A.\arabic{figure}} 
\renewcommand\thetable{A.\arabic{table}} 

\section[]{Appendices}
\subsection[]{Linear Pulsation Models}
\label{s:app1}
\begin{figure*}
\begin{center}
\vspace{-20mm}
\includegraphics[width=0.32\textwidth]{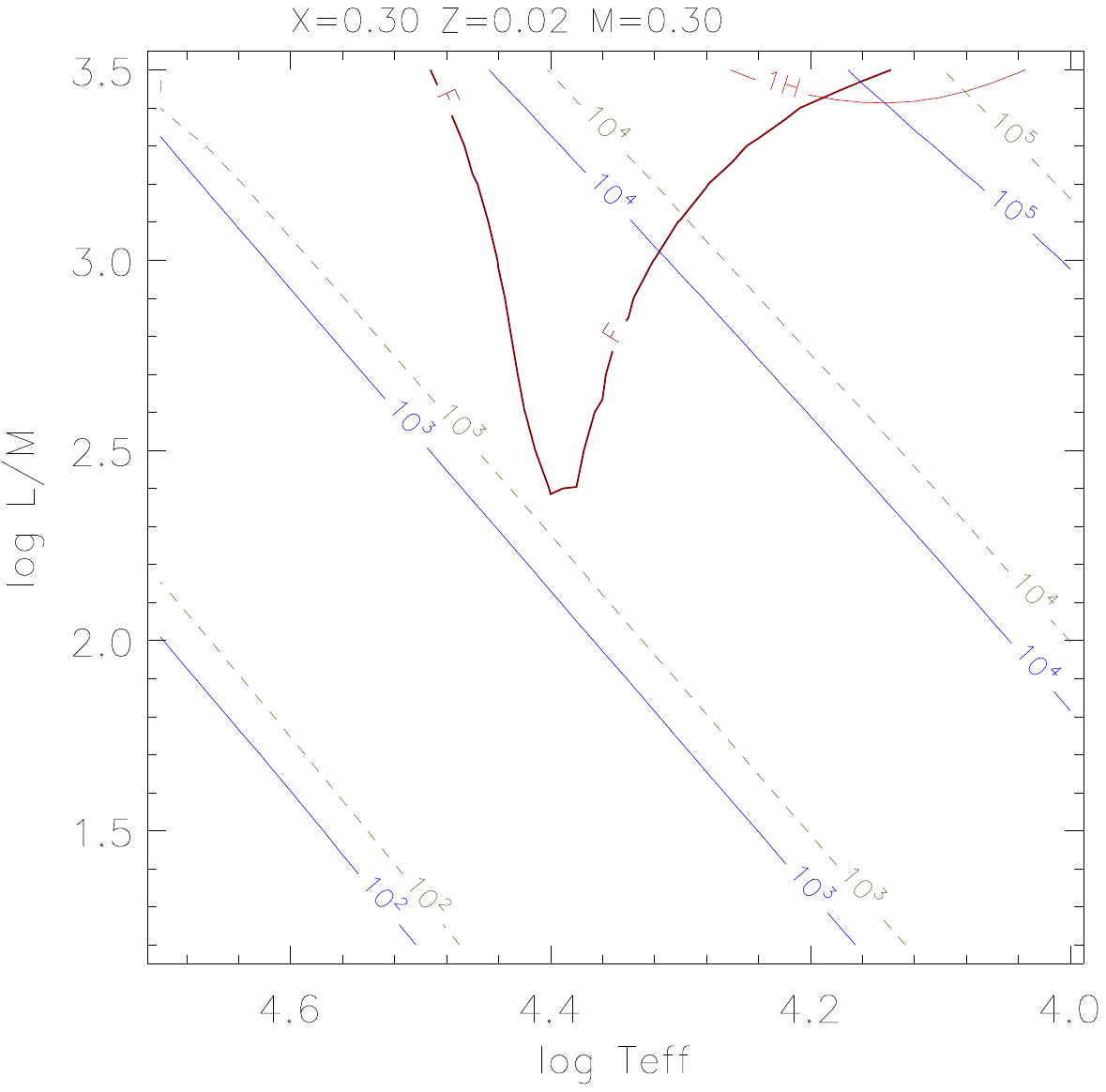}
\includegraphics[width=0.32\textwidth]{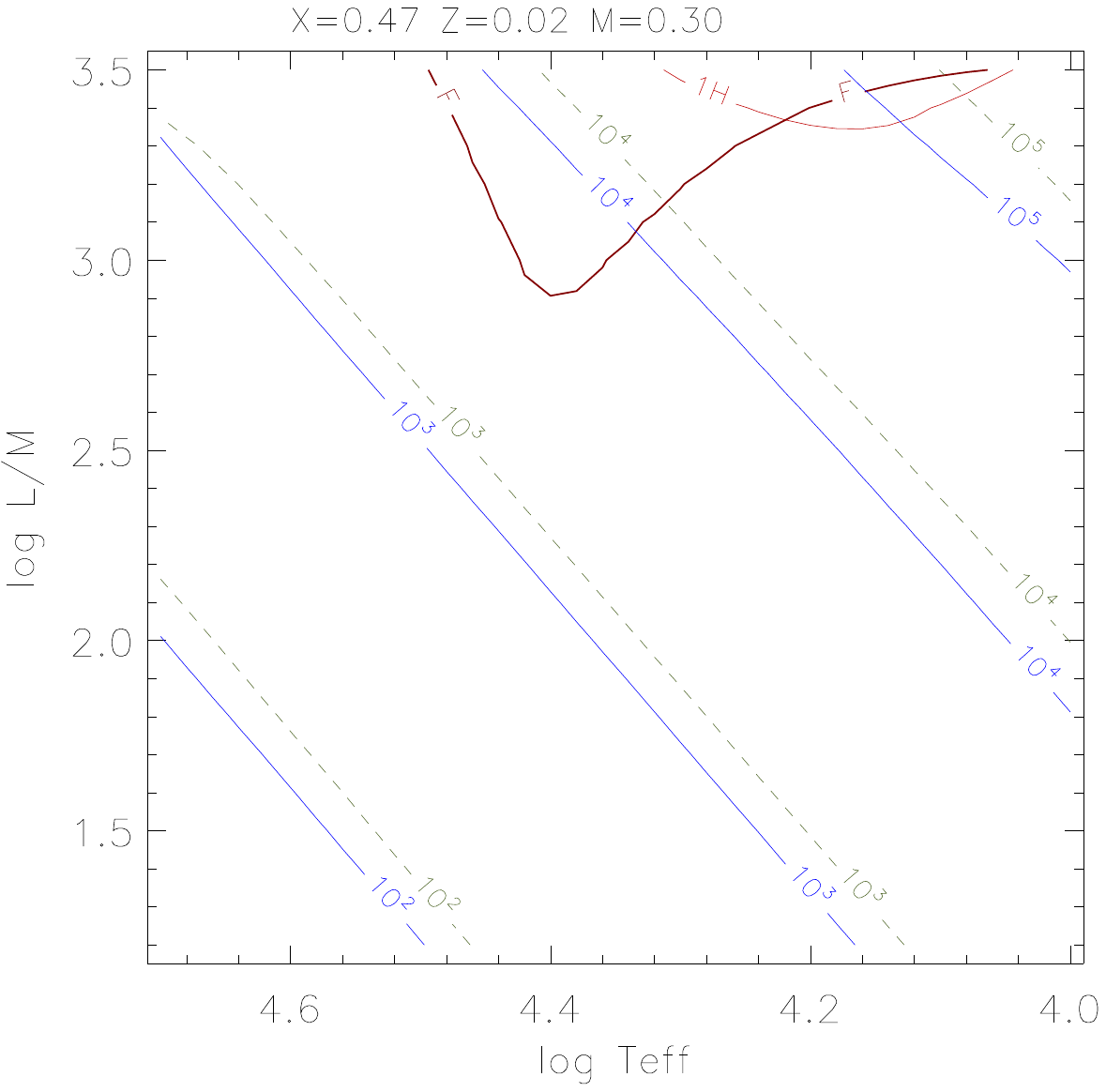}
\includegraphics[width=0.32\textwidth]{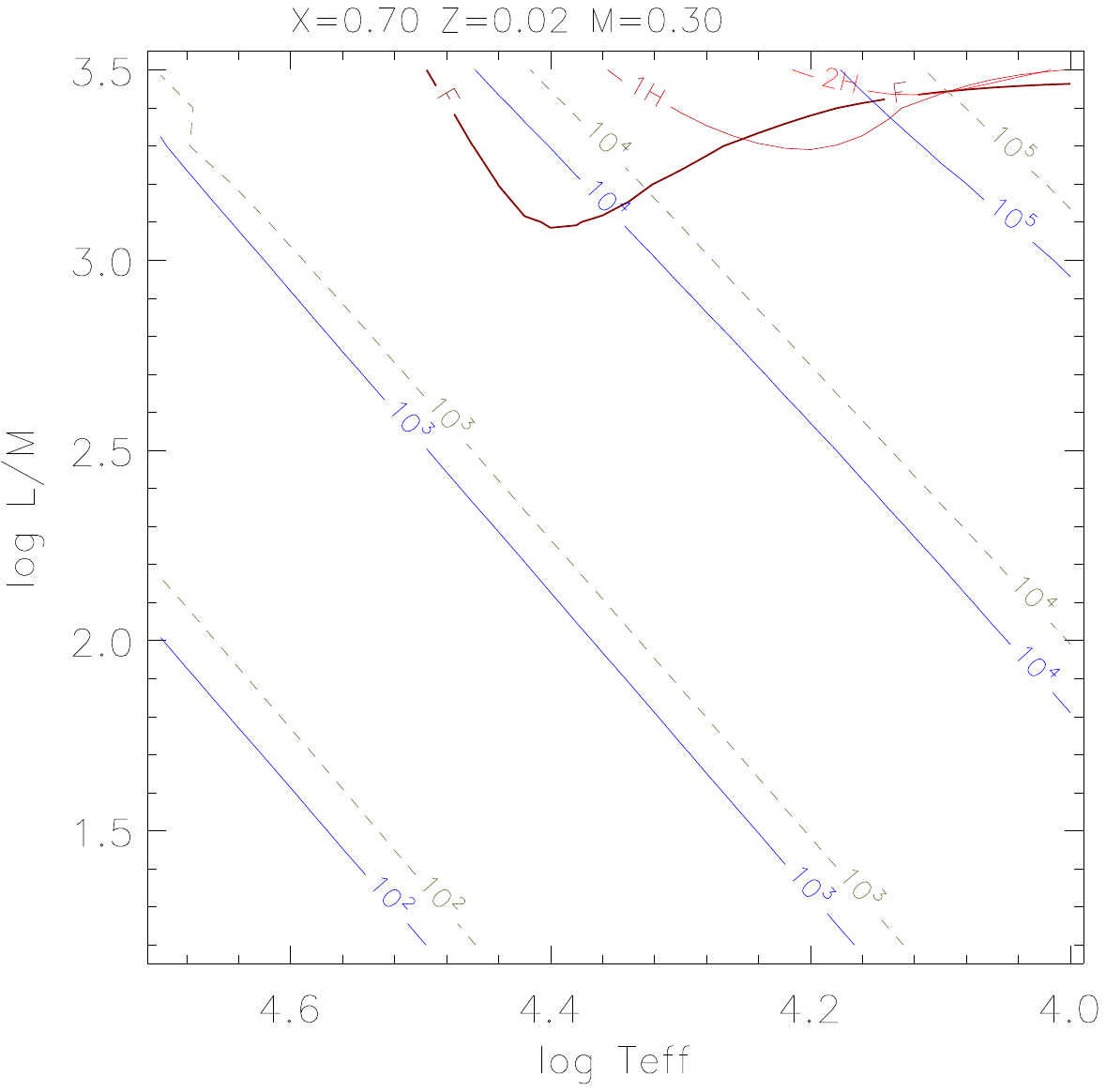}\\
\caption[Unstable mode boundaries: hydrogen]
{As Fig.\,\ref{f:nmZ02}, for models with mass $M=0.30$\Msolar, metallicity $Z=0.02$ (approximately solar) and hydrogen abundances (by mass fraction) $X=0.30, 0.475$ and $0.70$,  
but showing the 
{\it boundaries} for individual radial modes as coloured contours, with the darkest red representing
the boundary of the fundamental (F: $n=0$) mode, with increasing higher orders as labelled (1H, 2H: $n=1,2$) 
Solid blue lines represent contours of equal fundamental radial-mode (F) period in seconds spaced at decadal intervals. 
Dashed lines represent contours of equal first overtone (1H) period in seconds. 
}
\label{f:p_X}
\end{center}
\end{figure*}

\begin{figure*}
\begin{center}
\vspace{-20mm}
\includegraphics[width=0.32\textwidth]{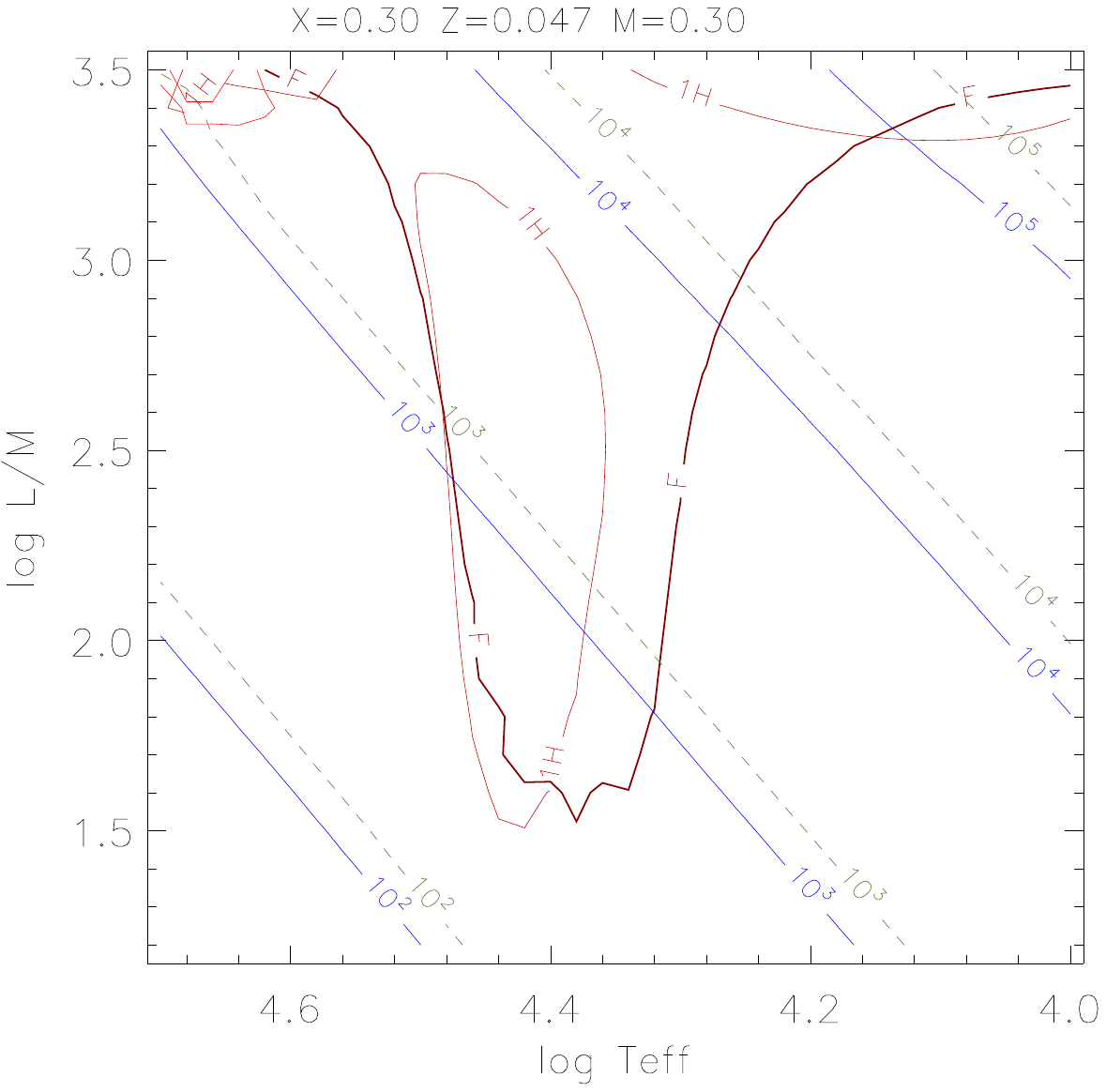}
\includegraphics[width=0.32\textwidth]{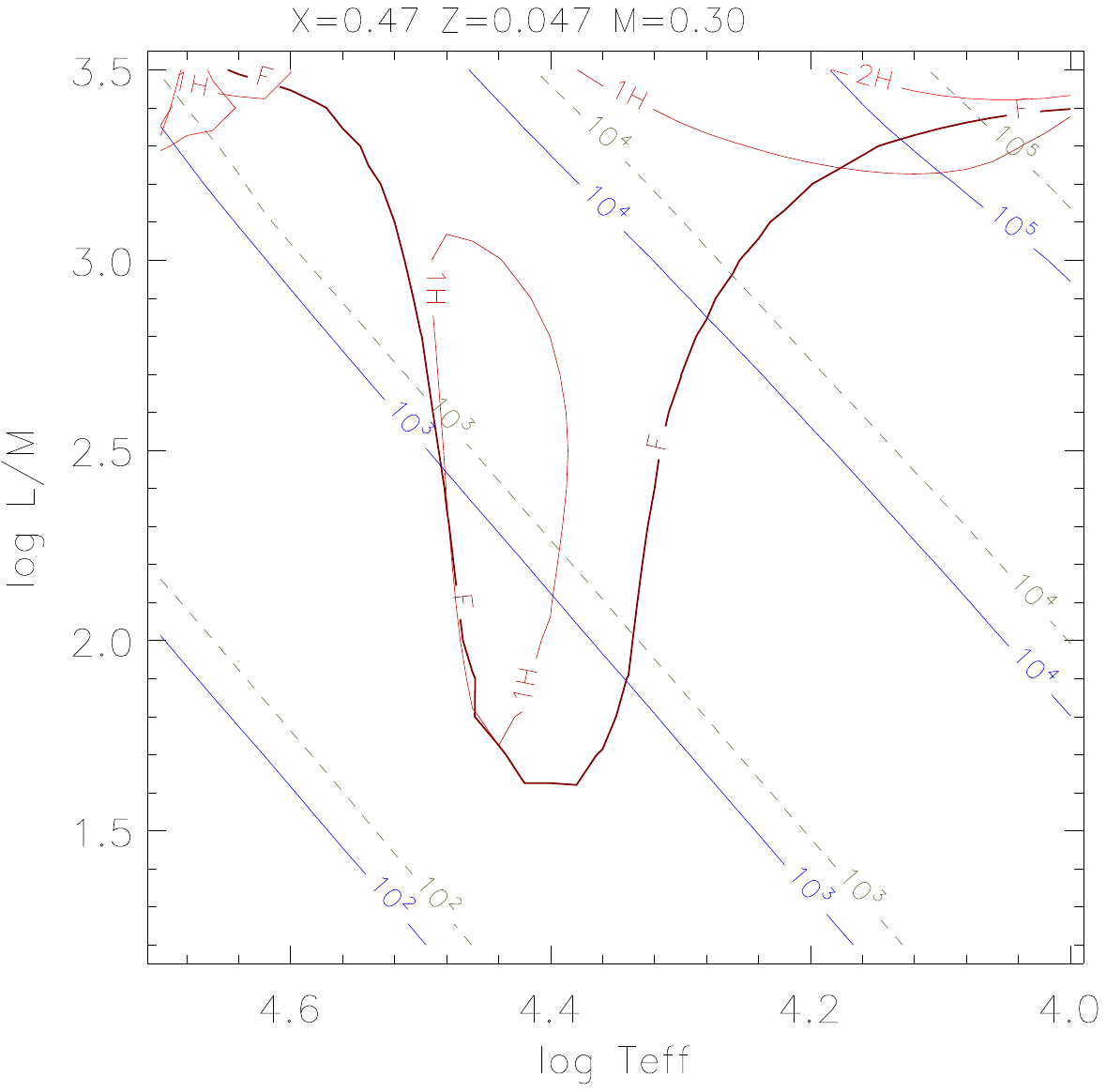}
\includegraphics[width=0.32\textwidth]{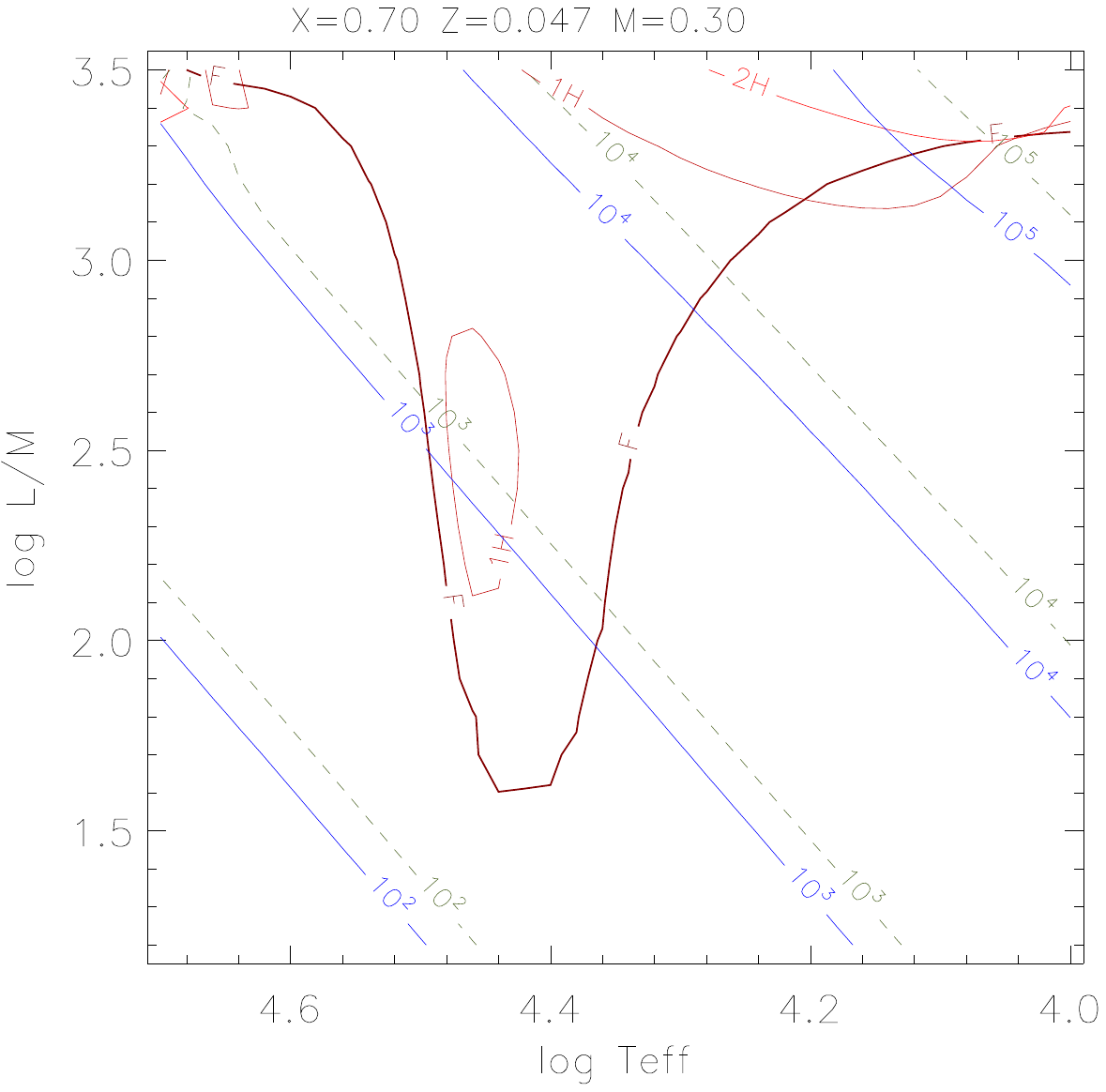}\\
\caption[Unstable mode boundaries: hydrogen with enhanced metals]
{As Fig.\,\ref{f:p_X} for models with with mass $M=0.30$\Msolar, metallicity $Z=0.047$ and hydrogen abundances (by mass fraction) $X=0.30, 0.475$ and $0.70$. 
}
\label{f:p_X_Zblap}
\end{center}
\end{figure*}

\begin{figure*}
\begin{center}
\vspace{-20mm}
\includegraphics[width=0.32\textwidth]{periods/periods_x475z02m0.30_00_opal.pdf}
\includegraphics[width=0.32\textwidth]{periods/periods_x475z047BPm0.30_00_opal.pdf}
\includegraphics[width=0.32\textwidth]{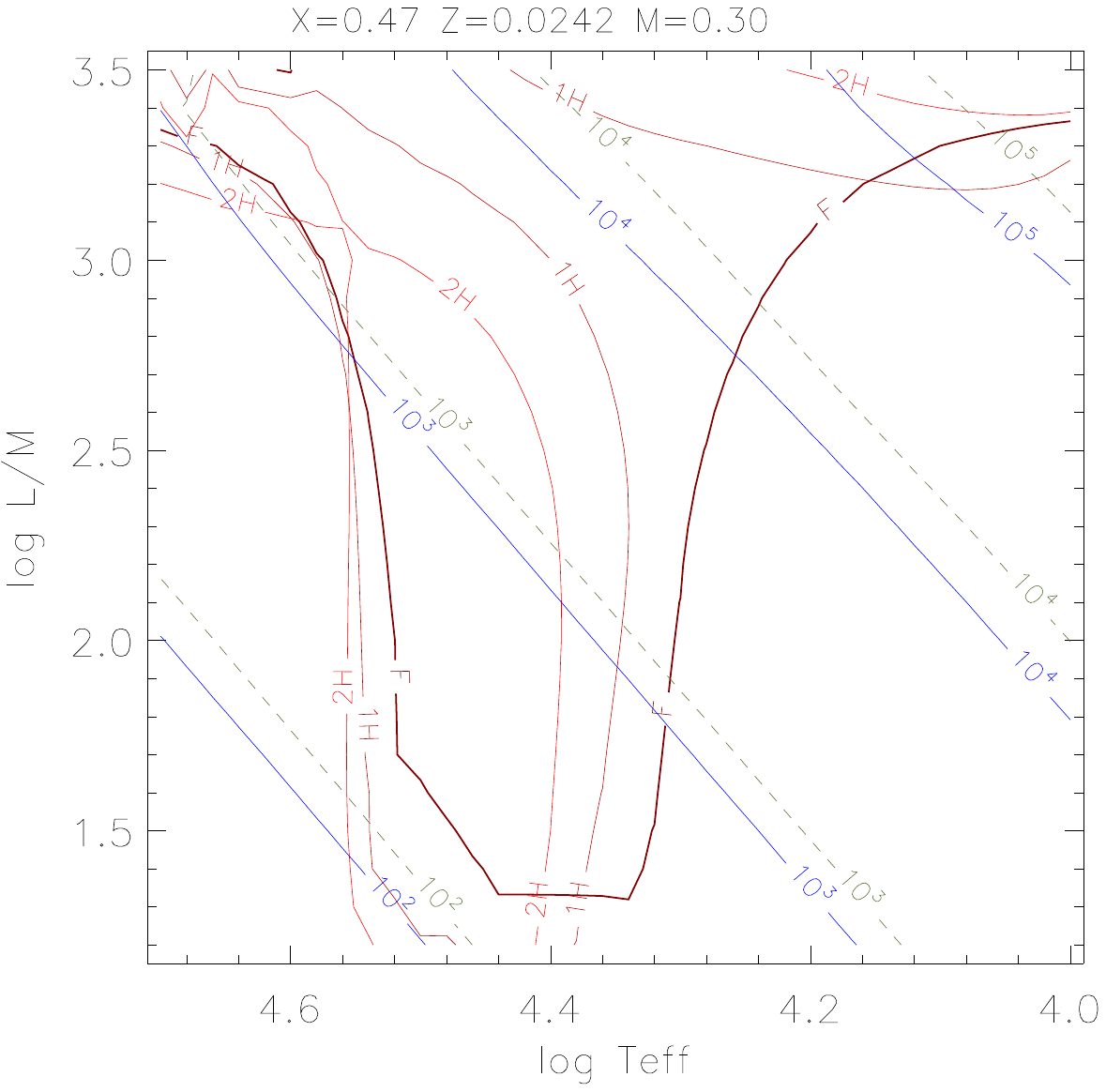}\\
\caption[Unstable modes as f(Z)]
{As Fig.\,\ref{f:p_X} for models with mass $0.30$\Msolar, hydrogen abundance (by mass fraction) $X=0.475$, and metallicity.  $Z=0.02, 0.047$ and 0.0242. 
The fraction of iron and nickel relative to other metals is solar for $Z=0.2$, $4\times$ solar for $Z=0.047$, and 10$\times$ solar for $Z=0.0242$.      
}
\label{f:p_Z}
\end{center}
\end{figure*}

Figures \ref{f:p_X} to \ref{f:p_Z} contain corollaries of Figs.\,\ref{f:nmZ02}-\ref{f:nmZ} 
showing the stability boundaries for the fundamental radial mode (F) and first two radial overtones (1H, 2H).

\begin{figure*}
\begin{center}
\vspace{-20mm}
\includegraphics[width=0.32\textwidth]{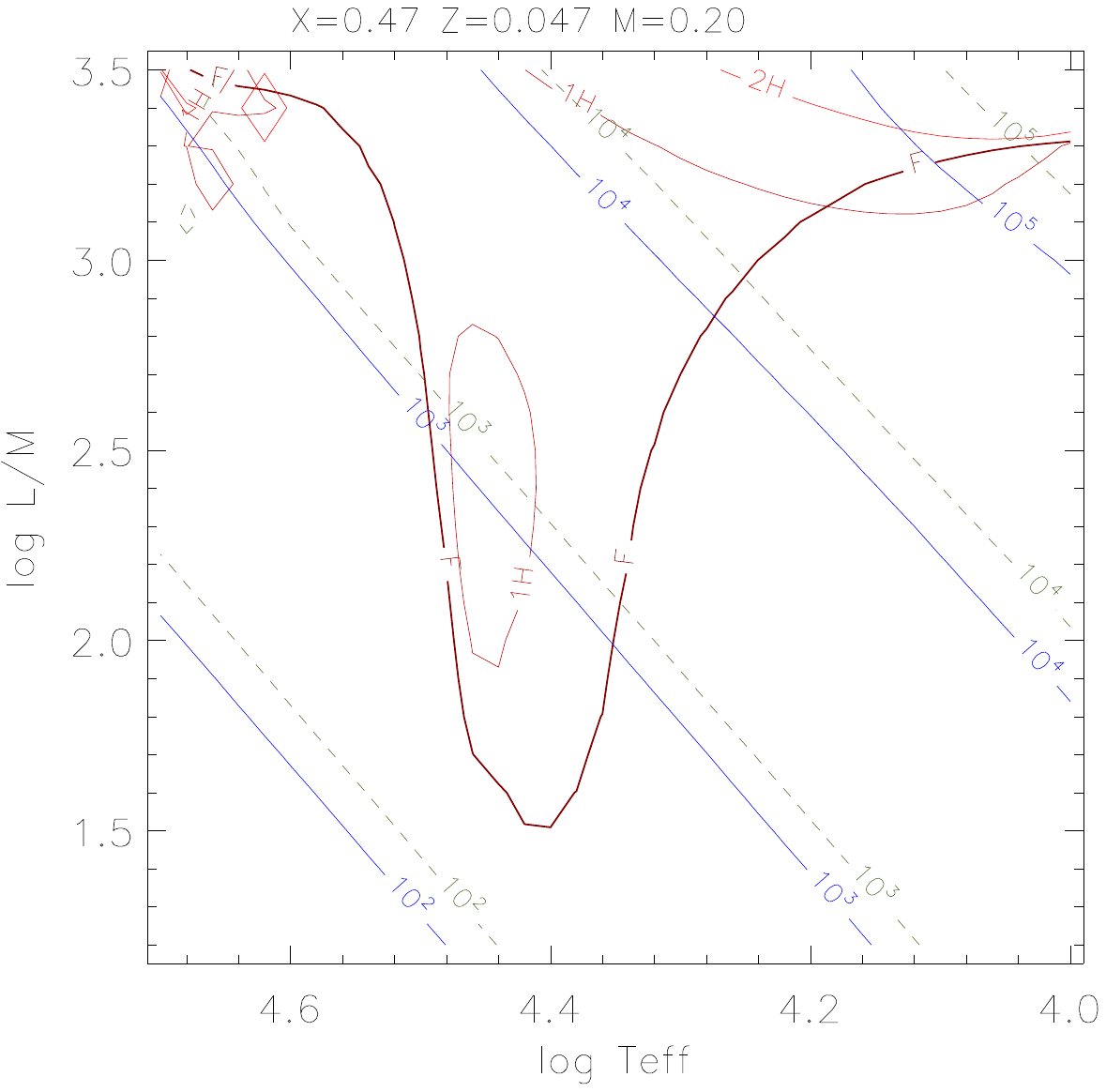}
\includegraphics[width=0.32\textwidth]{periods/periods_x475z047BPm0.30_00_opal.pdf}
\includegraphics[width=0.32\textwidth]{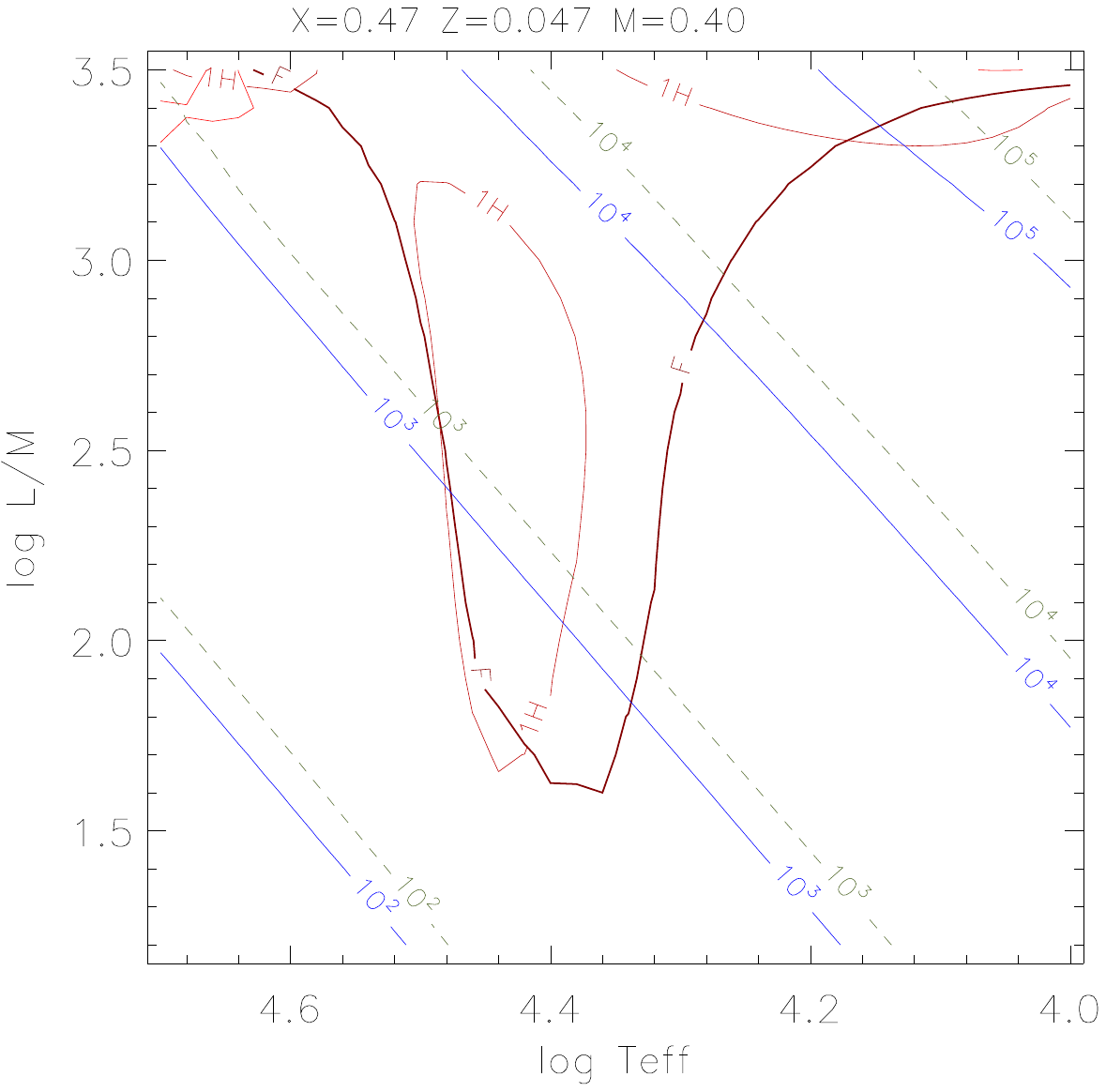}\\
\vspace{-20mm}
\includegraphics[width=0.32\textwidth]{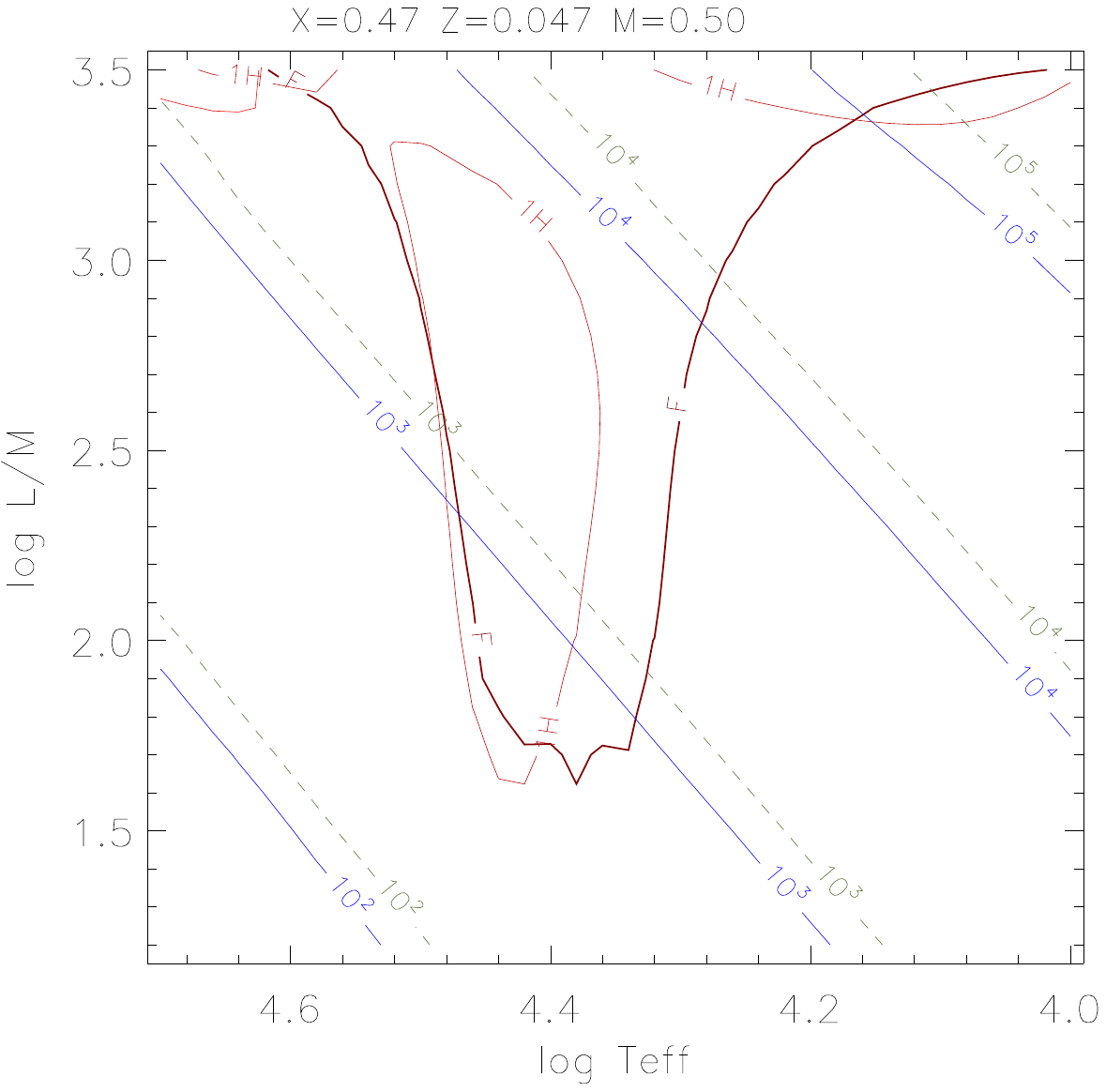}
\includegraphics[width=0.32\textwidth]{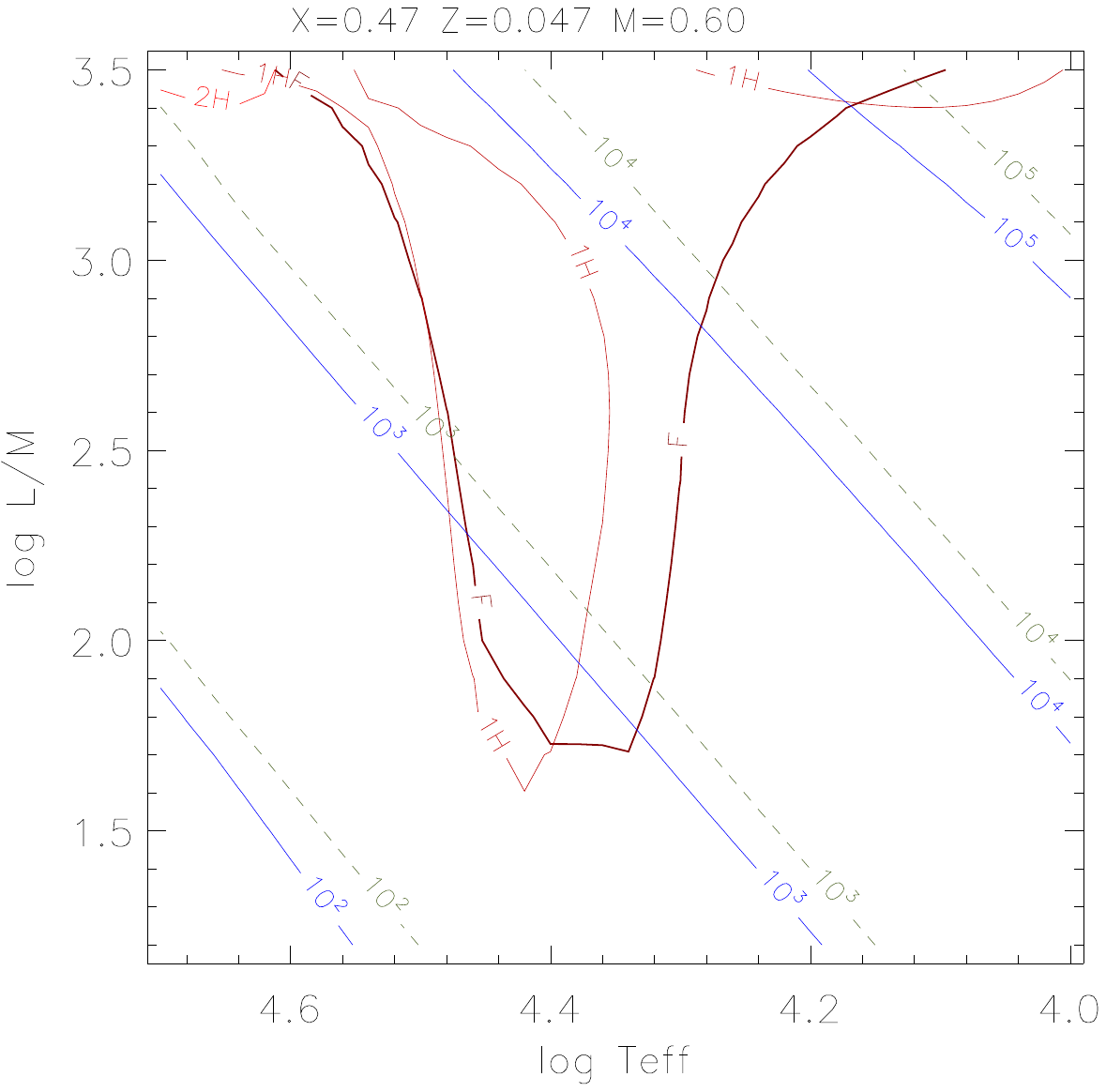}
\includegraphics[width=0.32\textwidth]{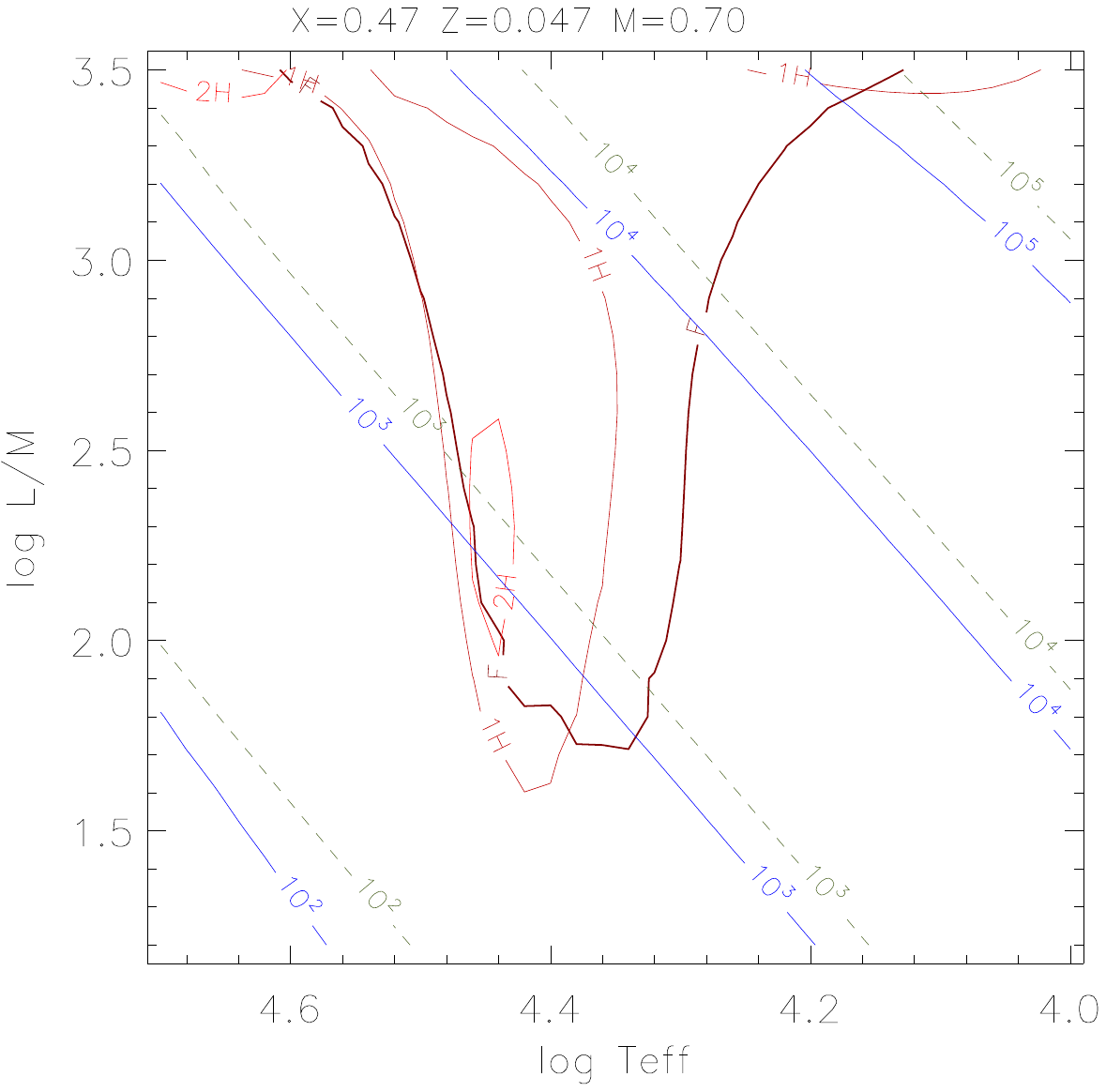}\\
\caption[Unstable mode boundaries; mass with enhanced metals]
{As Fig.\,\ref{f:p_X} for models with  mass $M=0.20, 0.30$ and 0.40\Msolar, metallicity $Z=0.047$ and hydrogen abundance (by mass fraction) $X=0.475$.}
\label{f:p_mass}
\end{center}
\end{figure*}

\subsection[]{Non-linear Pulsation Models}
\label{s:app2}
\begin{figure*}
\centering
\vspace*{-5mm}
\includegraphics[width=0.95\textheight,angle=90]{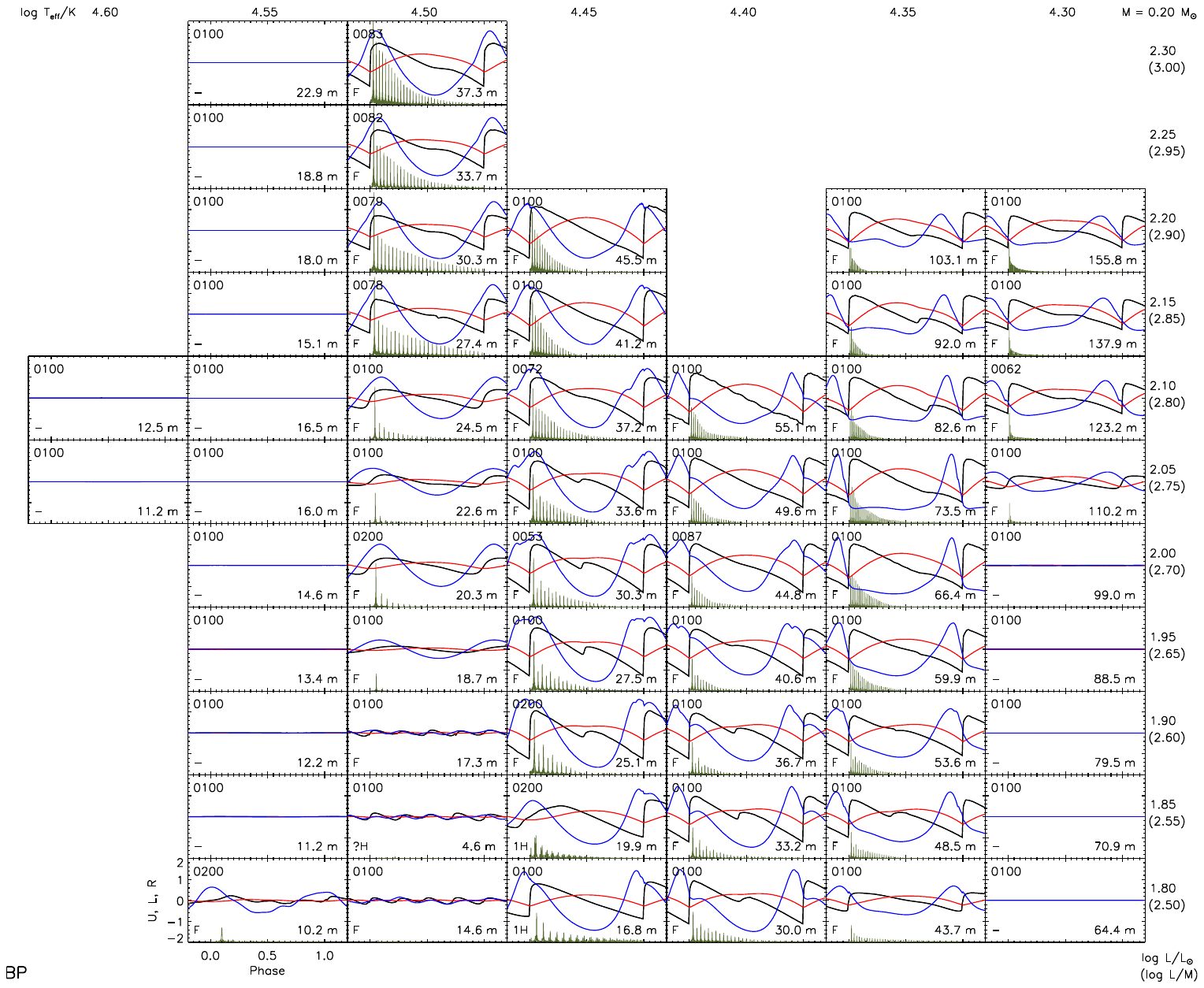}
\vspace*{-20mm}
\caption{As Fig.~\ref{f:mods30}, $M=0.20\Msolar$}
\label{f:mods20}
\end{figure*}

\begin{figure*}
\centering
\vspace*{-5mm}
\includegraphics[width=0.95\textheight,angle=90]{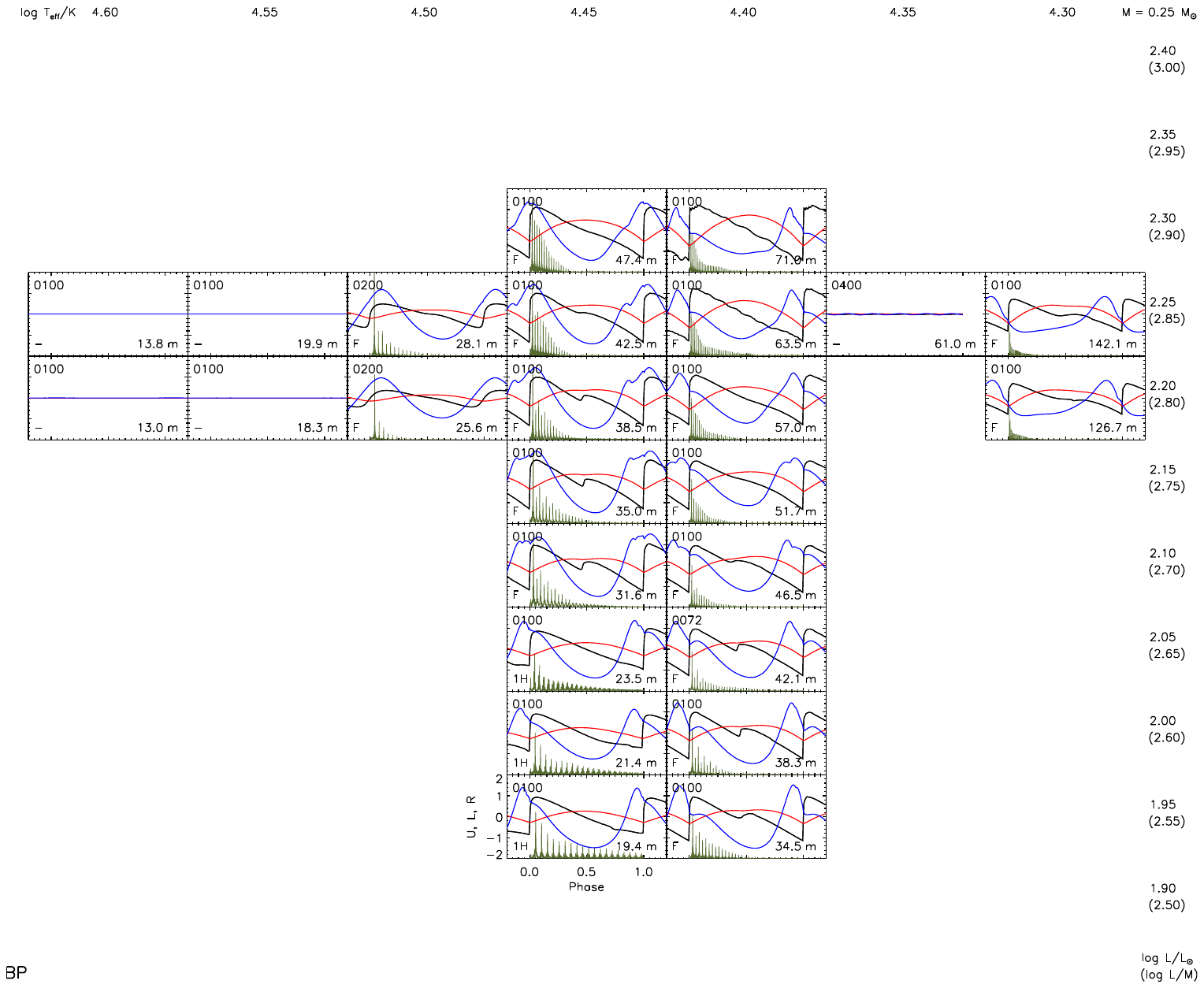}
\vspace*{-20mm}
\caption{As Fig.~\ref{f:mods30}, $M=0.25\Msolar$}
\label{f:mods25}
\end{figure*}

\begin{figure*}
\centering
\vspace*{-5mm}
\includegraphics[width=0.95\textheight,angle=90]{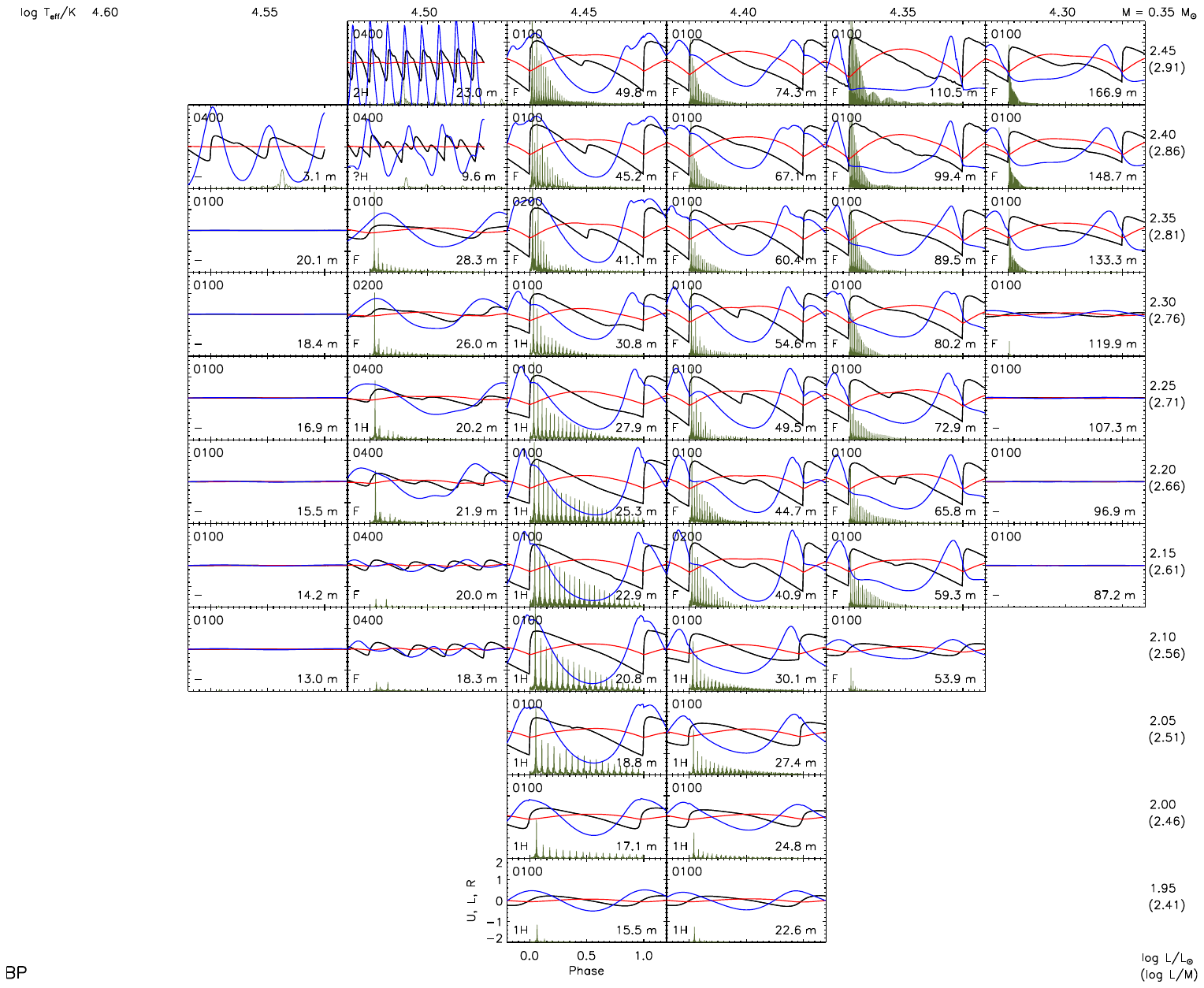}
\vspace*{-20mm}
\caption{As Fig.~\ref{f:mods30}, $M=0.35\Msolar$}
\label{f:mods35}
\end{figure*}

\begin{figure*}
\centering
\vspace*{-5mm}
\includegraphics[width=0.95\textheight,angle=90]{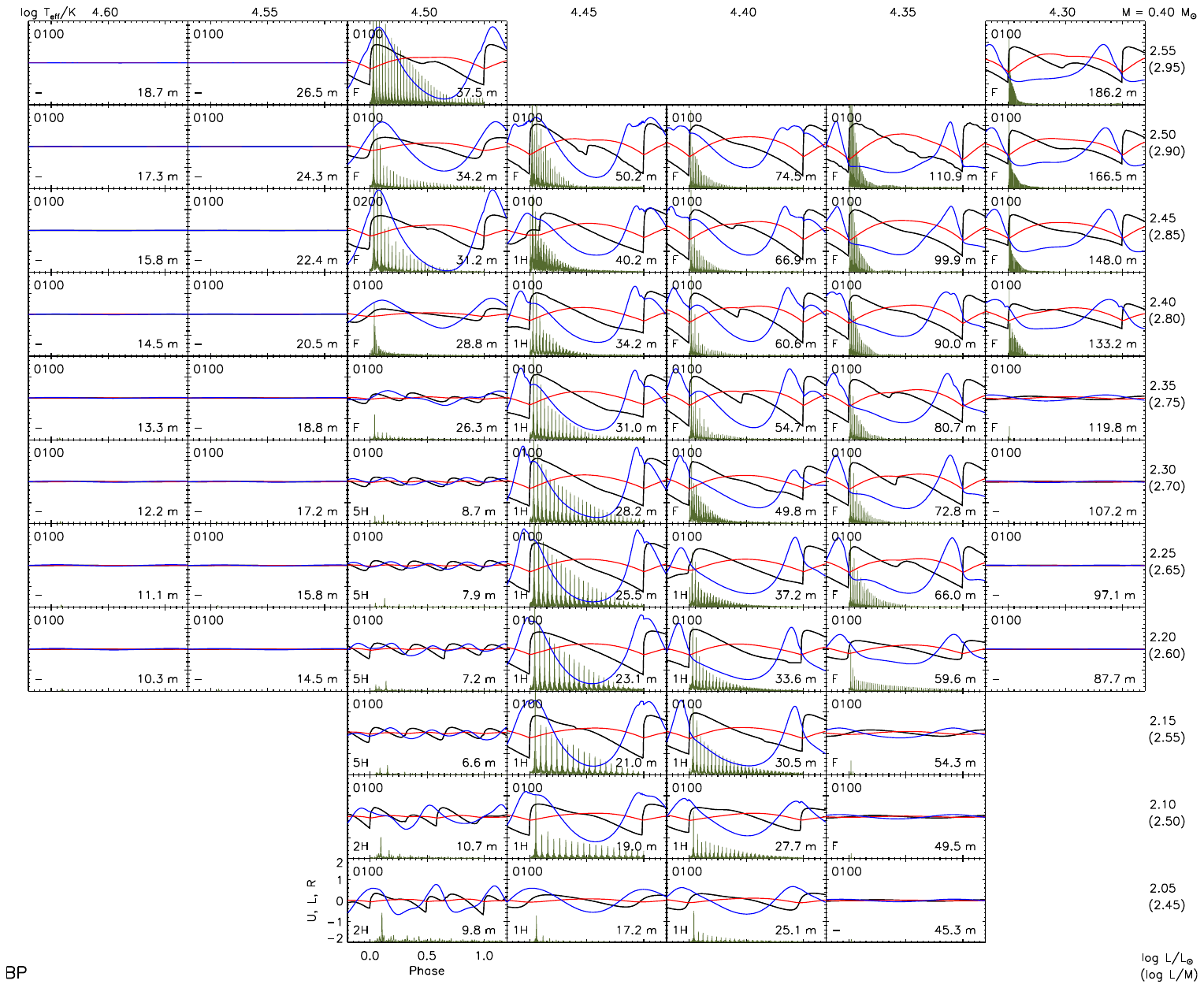}
\vspace*{-20mm}
\caption{As Fig.~\ref{f:mods30}, $M=0.40\Msolar$}
\label{f:mods40}
\end{figure*}

\begin{figure*}
\centering
\vspace*{-5mm}
\includegraphics[width=0.95\textheight,angle=90]{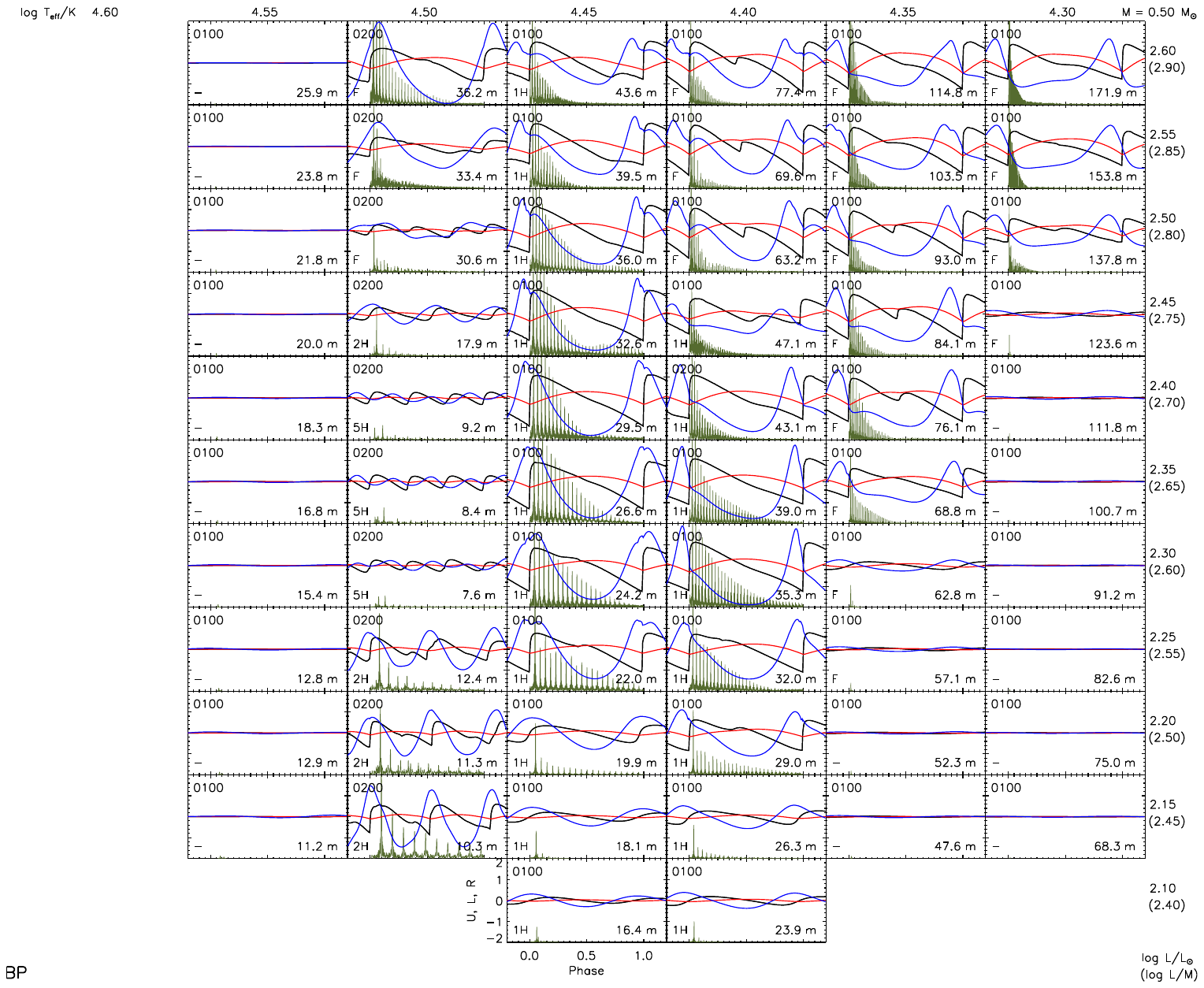}
\vspace*{-20mm}
\caption{As Fig.~\ref{f:mods30}, $M=0.50\Msolar$}
\label{f:mods50}
\end{figure*}

\begin{figure*}
\centering
\vspace*{-5mm}
\includegraphics[width=0.95\textheight,angle=90]{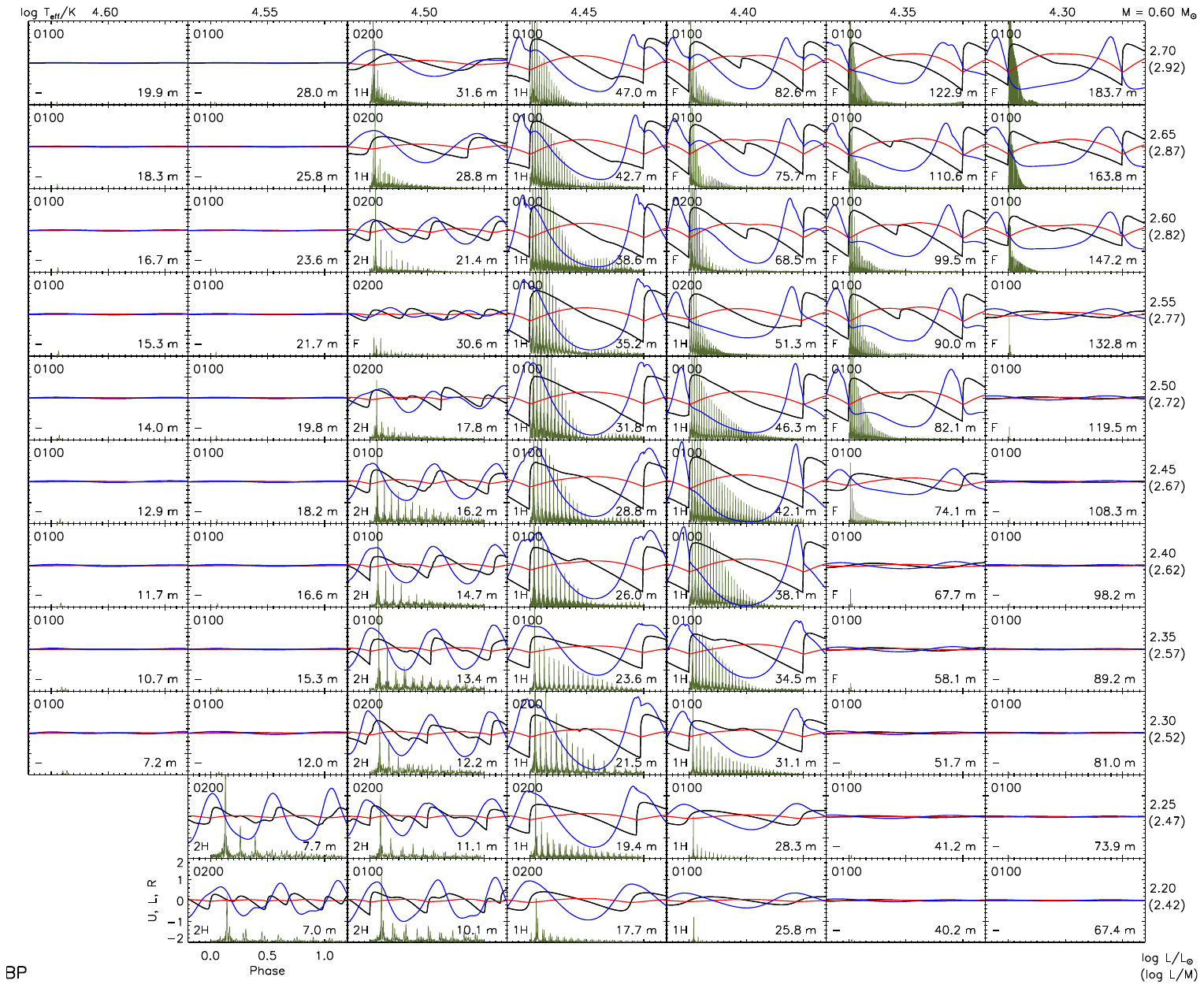}
\vspace*{-20mm}
\caption{As Fig.~\ref{f:mods30}, $M=0.60\Msolar$}
\label{f:mods60}
\end{figure*}

\begin{figure*}
\centering
\vspace*{-5mm}
\includegraphics[width=0.95\textheight,angle=90]{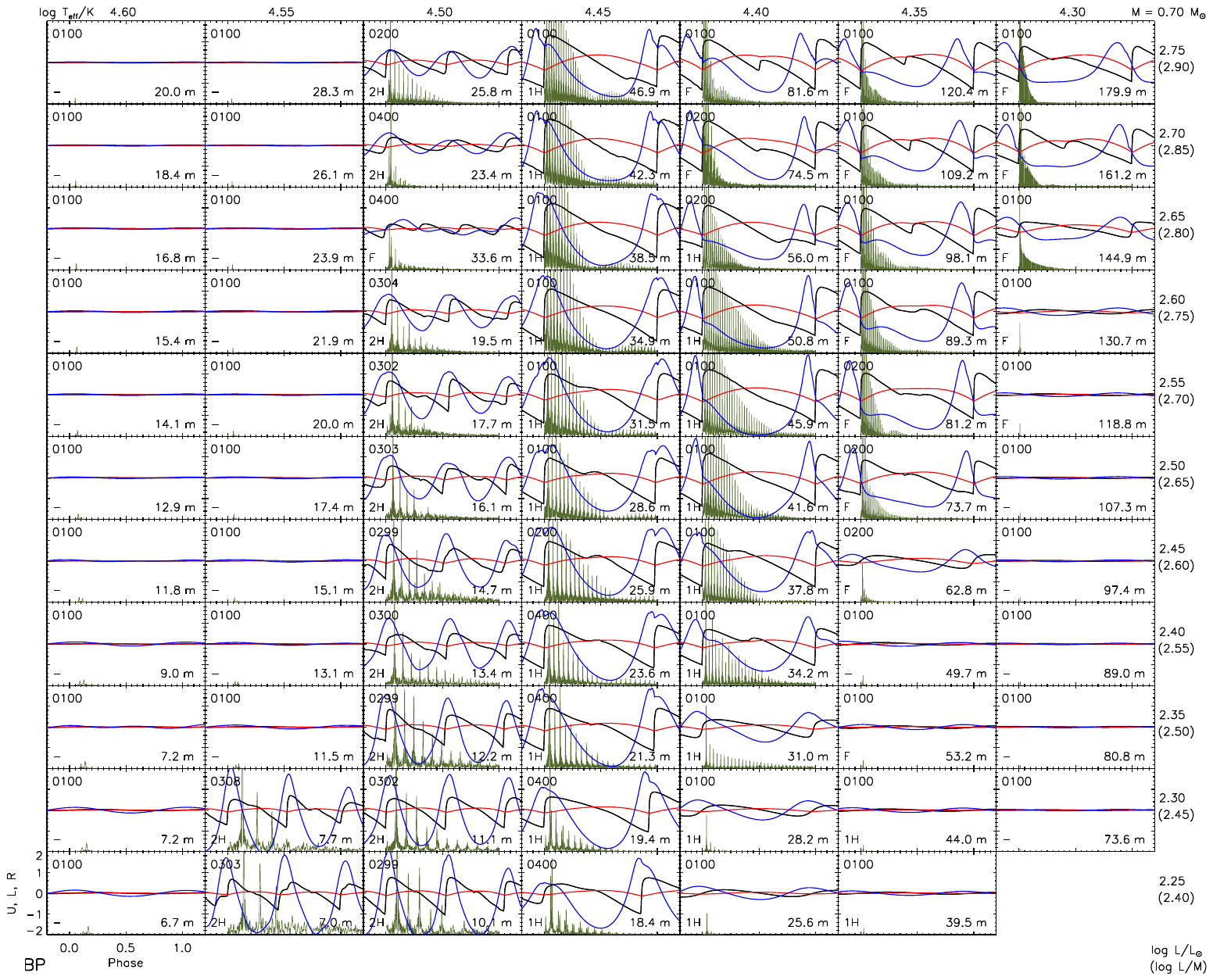}
\vspace*{-20mm}
\caption{As Fig.~\ref{f:mods30}, $M=0.70\Msolar$}
\label{f:mods70}
\end{figure*}

\begin{figure*}
\centering
\vspace*{-5mm}
\includegraphics[width=0.95\textheight,angle=90]{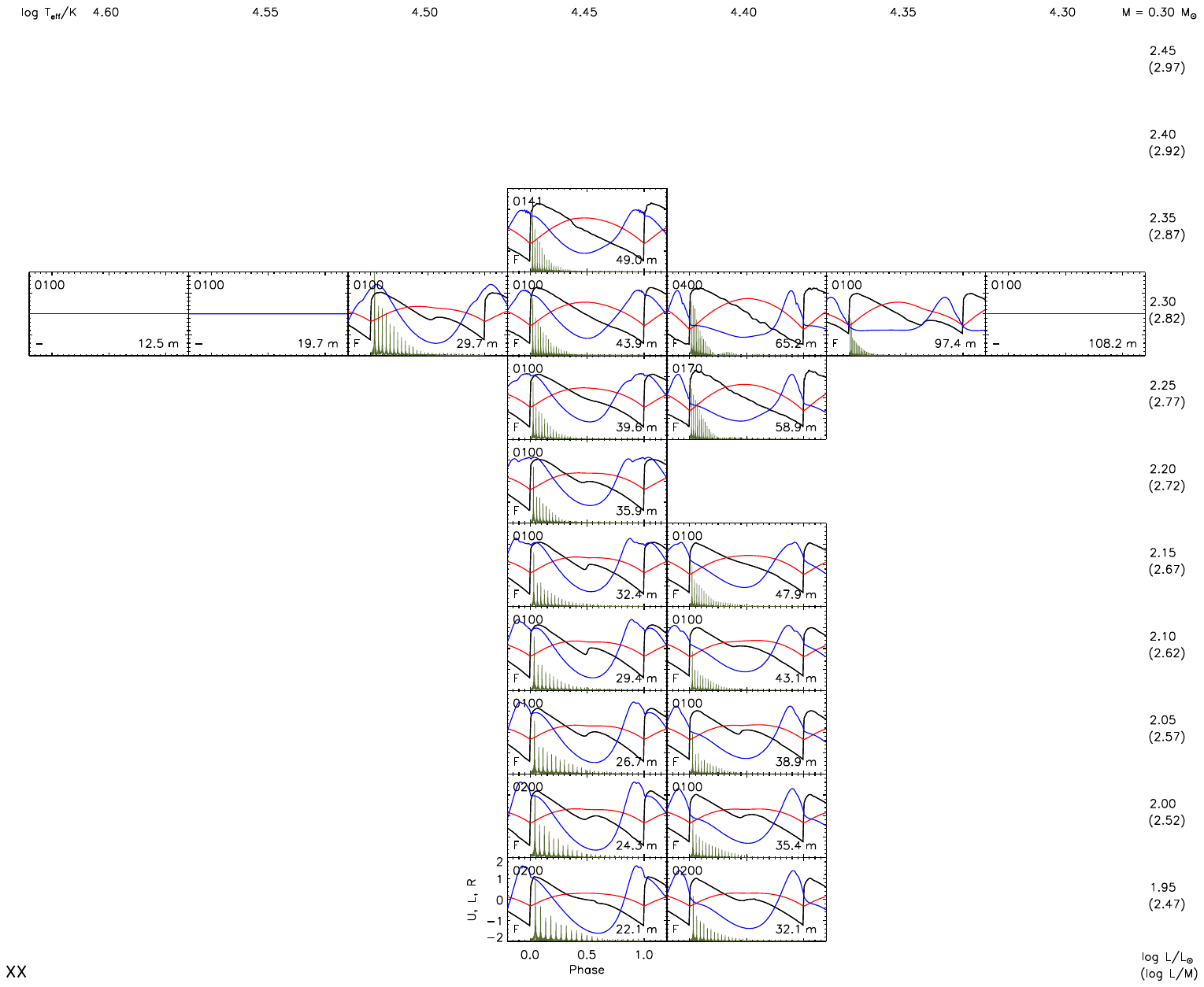}
\vspace*{-20mm}
\caption{As Fig.~\ref{f:mods30}, $X=0.9, Y=0.053$}
\label{f:mods30XX}
\end{figure*}

\begin{figure*}
\centering
\vspace*{-5mm}
\includegraphics[width=0.95\textheight,angle=90]{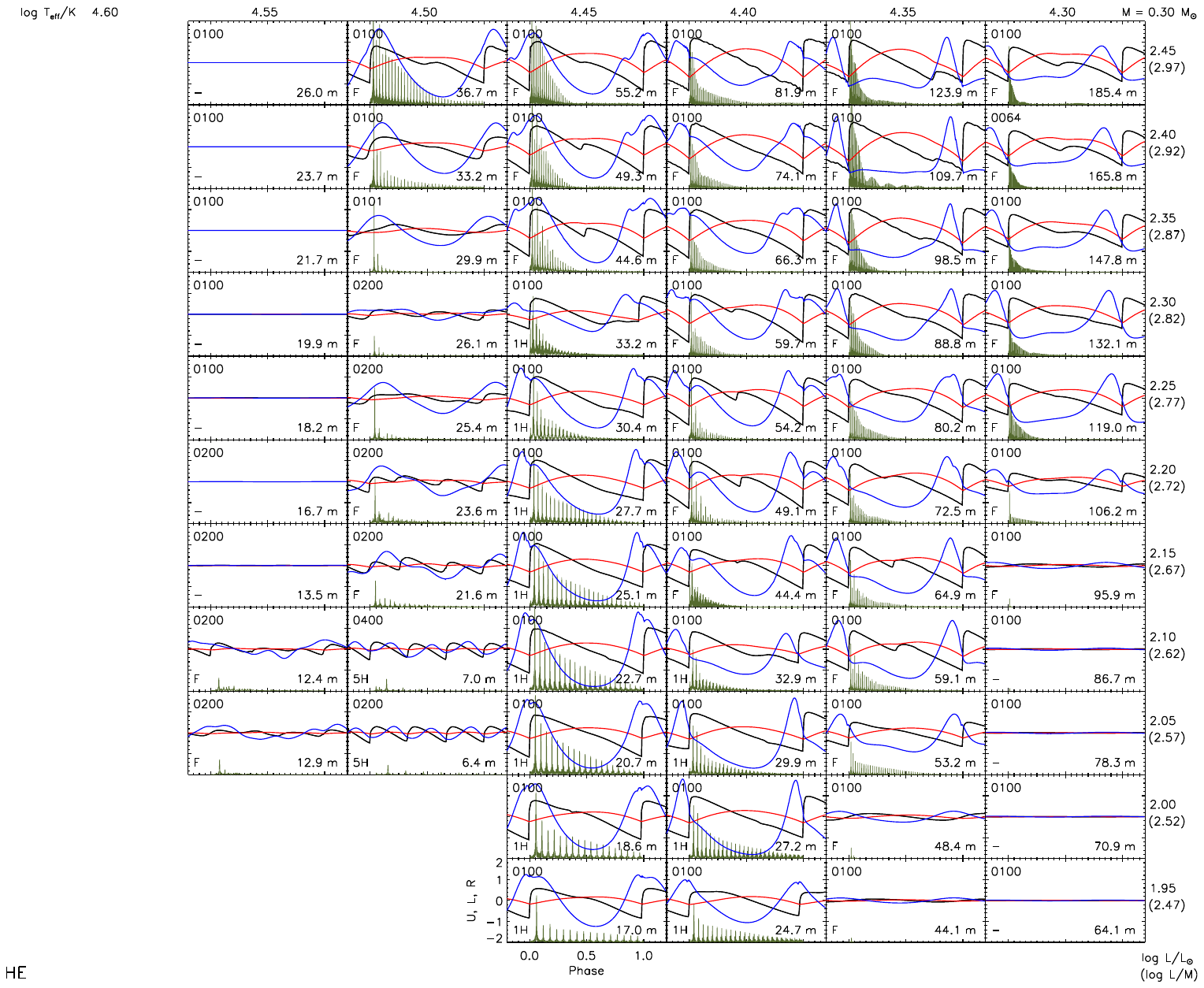}
\vspace*{-20mm}
\caption{As Fig.~\ref{f:mods30}, $X=0.353, Y=0.60$}
\label{f:mods30HE}
\end{figure*}

\begin{figure*}
\centering
\vspace*{-5mm}
\includegraphics[width=0.95\textheight,angle=90]{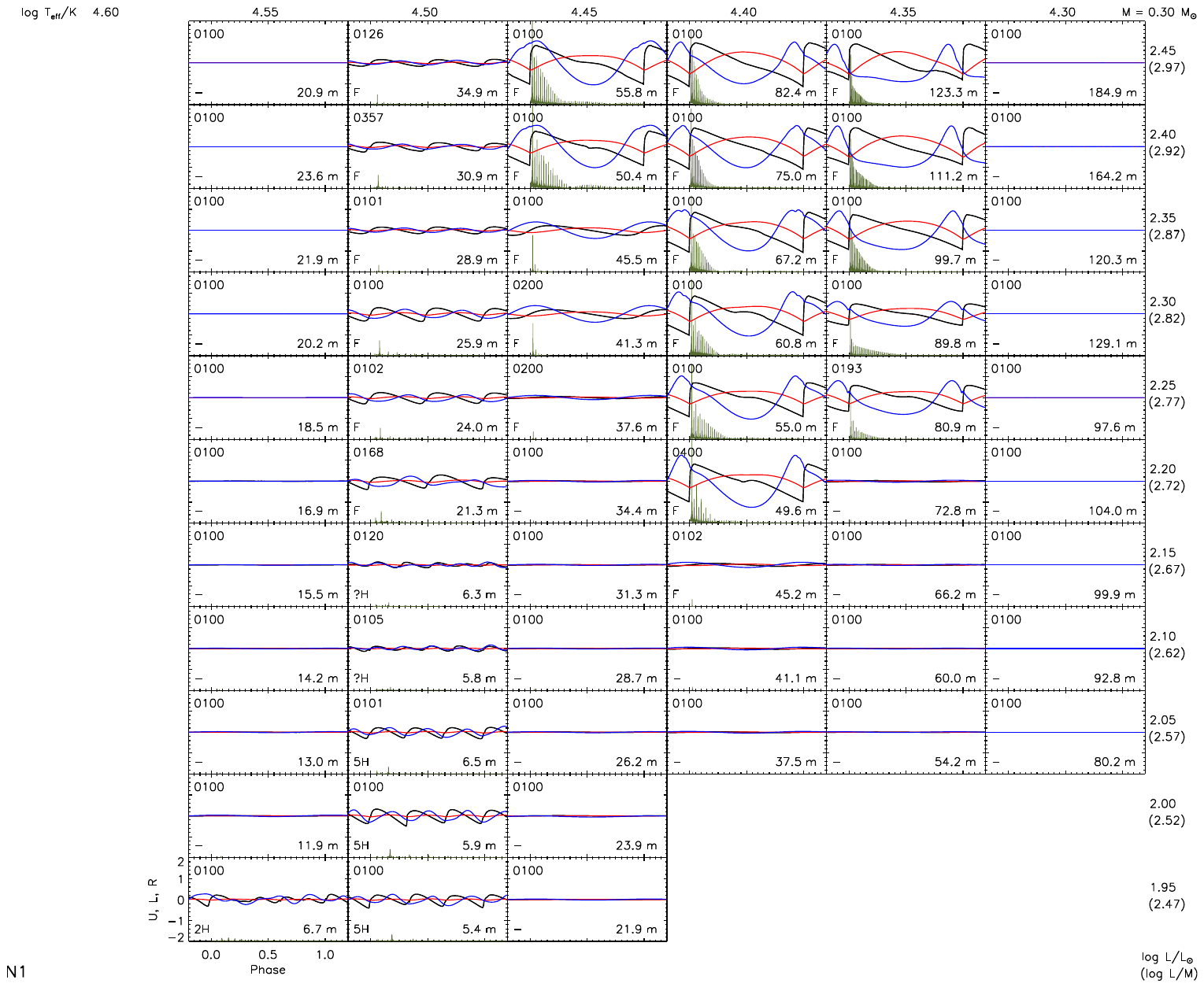}
\vspace*{-20mm}
\caption{As Fig.~\ref{f:mods30}, with metals represented by mixture N1.}
\label{f:mods30N1}
\end{figure*}

Figures \ref{f:mods20} to \ref{f:mods30N1} contain equivalent figures to Fig.\,\ref{f:mods30} for masses in the range 0.20 -- 0.70 \Msolar\ and 3 additional chemical mixtures. Tables\,\ref{a:models}ff contain the full version of Table\,\ref{t:models}.

\label{lastpage}

\begin{table*}
\begin{center}
\caption{}
\label{a:models}
\begin{tabular}{lrrrr rrrrr rrrrr}
\hline
 model  & $M$    & $\log L$ & $\log \Teff$ & $\log R$ & $\log g$  & $\log \bar{\rho}$ & $\log L/M$ & $P_{\rm nl}$ & $P_{\rm ft}$ &  $\Delta L$ &  $\Delta R$ &  $\Delta u$   & $\Delta \phi$ & $k$\\
$llttmmNN$& \Msolar & \Lsolar & K     & \Rsolar & cm\,s$^{-2}$  & $\bar{\rho_{\odot}}$ & \Lsolar/\Msolar & min &  min & $L_{\star}$ & $R_{\star}$ & \kmsec &      &     \\
\hline
803020BP&0.20&1.80&4.30&-0.18&4.10&-0.15&2.50& 64.4&  0.0&0.00&0.00&  0& 0.37&-\\
803520BP&0.20&1.80&4.35&-0.28&4.30& 0.15&2.52& 43.8& 43.7&0.09&0.28& 87& 0.08&0\\
804020BP&0.20&1.80&4.40&-0.38&4.50& 0.45&2.53& 30.0& 30.0&0.13&0.60&206& 0.08&0\\
804520BP&0.20&1.80&4.45&-0.48&4.70& 0.75&2.53& 17.2& 16.8&0.10&0.57&177& 0.05&1\\
805020BP&0.20&1.80&4.50&-0.58&4.90& 1.05&2.50& 16.9& 14.6&0.01&0.07& 31& 0.47&0\\
805520BP&0.20&1.80&4.55&-0.68&5.10& 1.35&2.52& 11.4& 10.2&0.02&0.26& 27& 0.00&0\\
853020BP&0.20&1.85&4.30&-0.16&4.05&-0.23&2.55& 70.9&  0.0&0.00&0.00&  0& 0.32&-\\
853520BP&0.20&1.85&4.35&-0.26&4.25& 0.07&2.60& 48.6& 48.5&0.14&0.48&180& 0.11&0\\
854020BP&0.20&1.85&4.40&-0.36&4.45& 0.37&2.59& 33.2& 33.2&0.15&0.58&218& 0.10&0\\
854520BP&0.20&1.85&4.45&-0.46&4.65& 0.67&2.58& 19.9& 18.4&0.09&0.53&177& 0.00&1\\
855020BP&0.20&1.85&4.50&-0.56&4.85& 0.97&2.55& 18.8&  4.6&0.01&0.06& 37& 0.46&8\\
855520BP&0.20&1.85&4.55&-0.66&5.05& 1.27&2.55& 11.2& 11.2&0.00&0.01&  2&-0.05&0\\
903020BP&0.20&1.90&4.30&-0.13&4.00&-0.30&2.60& 80.5& 79.5&0.00&0.00&  1& 0.44&0\\
903520BP&0.20&1.90&4.35&-0.23&4.20&-0.00&2.63& 53.7& 53.6&0.16&0.47&191& 0.11&0\\
904020BP&0.20&1.90&4.40&-0.33&4.40& 0.30&2.63& 36.6& 36.7&0.16&0.54&218& 0.12&0\\
904520BP&0.20&1.90&4.45&-0.43&4.60& 0.60&2.62& 25.1& 25.1&0.14&0.66&230& 0.04&0\\
905020BP&0.20&1.90&4.50&-0.53&4.80& 0.90&2.60& 19.6& 17.3&0.01&0.05& 25&-0.23&0\\
905520BP&0.20&1.90&4.55&-0.63&5.00& 1.20&2.60& 12.3& 12.2&0.00&0.01&  2& 0.04&0\\
953020BP&0.20&1.95&4.30&-0.11&3.95&-0.38&2.65& 88.9& 88.5&0.00&0.01&  0& 0.07&0\\
953520BP&0.20&1.95&4.35&-0.21&4.15&-0.08&2.70& 59.7& 59.9&0.19&0.51&204& 0.08&0\\
954020BP&0.20&1.95&4.40&-0.31&4.35& 0.22&2.67& 40.4& 40.6&0.18&0.49&224& 0.14&0\\
954520BP&0.20&1.95&4.45&-0.41&4.55& 0.52&2.66& 27.5& 27.5&0.14&0.62&230&-0.06&0\\
955020BP&0.20&1.95&4.50&-0.51&4.75& 0.82&2.67& 18.7& 18.7&0.02&0.18& 29&-0.08&0\\
955520BP&0.20&1.95&4.55&-0.61&4.95& 1.12&2.65& 13.5& 13.4&0.00&0.01&  1&-0.14&0\\
103020BP&0.20&2.10&4.30&-0.03&3.80&-0.60&2.82&123.2&123.2&0.17&0.33&145& 0.15&0\\
103520BP&0.20&2.10&4.35&-0.13&4.00&-0.30&2.86& 82.6& 82.6&0.24&0.44&187& 0.14&0\\
104020BP&0.20&2.10&4.40&-0.23&4.20&-0.00&2.84& 54.9& 55.1&0.26&0.49&251& 0.10&0\\
104520BP&0.20&2.10&4.45&-0.33&4.40& 0.30&2.83& 37.2& 37.2&0.18&0.57&228& 0.00&0\\
105020BP&0.20&2.10&4.50&-0.43&4.60& 0.60&2.83& 24.5& 24.5&0.07&0.40& 88&-0.09&0\\
105520BP&0.20&2.10&4.55&-0.53&4.80& 0.90&2.80& 16.5&  0.0&0.00&0.00&  1& 0.49&-\\
106020BP&0.20&2.10&4.60&-0.63&5.00& 1.20&2.80& 12.5&  0.0&0.00&0.00&  2&-0.31&-\\
153020BP&0.20&2.15&4.30&-0.01&3.75&-0.68&2.87&138.1&  0.0&0.18&0.31&143& 0.17&-\\
153520BP&0.20&2.15&4.35&-0.11&3.95&-0.38&2.89& 92.0& 92.0&0.22&0.38&183& 0.17&0\\
154520BP&0.20&2.15&4.45&-0.31&4.35& 0.22&2.88& 41.2& 41.2&0.22&0.58&242& 0.01&0\\
155020BP&0.20&2.15&4.50&-0.41&4.55& 0.52&2.88& 27.6& 27.4&0.11&0.57&162&-0.08&0\\
155520BP&0.20&2.15&4.55&-0.51&4.75& 0.82&2.85& 15.1&  0.0&0.00&0.04&  0& 0.38&-\\
203020BP&0.20&2.20&4.30& 0.02&3.70&-0.75&2.92&155.3&  0.0&0.19&0.28&153& 0.19&-\\
203520BP&0.20&2.20&4.35&-0.08&3.90&-0.45&2.92&103.0&103.1&0.22&0.32&173& 0.19&0\\
204520BP&0.20&2.20&4.45&-0.28&4.30& 0.15&2.93& 45.6& 45.5&0.25&0.53&258& 0.02&0\\
205020BP&0.20&2.20&4.50&-0.38&4.50& 0.45&2.94& 30.5& 30.3&0.12&0.54&164&-0.09&0\\
205520BP&0.20&2.20&4.55&-0.48&4.70& 0.75&2.90& 18.0&  0.0&0.00&0.00&  1& 0.16&-\\
255020BP&0.20&2.25&4.50&-0.36&4.45& 0.37&2.99& 33.6& 33.7&0.14&0.56&182&-0.07&0\\
255520BP&0.20&2.25&4.55&-0.46&4.65& 0.67&2.95& 18.8&  0.0&0.00&0.00&  0& 0.25&-\\
305020BP&0.20&2.30&4.50&-0.33&4.40& 0.30&3.03& 37.3& 37.3&0.17&0.62&206&-0.05&0\\
305520BP&0.20&2.30&4.55&-0.43&4.60& 0.60&3.00& 22.9&  0.0&0.00&0.00&  0&-0.31&-\\
\\[-1mm]
954025BP&0.25&1.95&4.40&-0.31&4.45& 0.32&2.59& 34.5& 34.5&0.13&0.61&211& 0.09&0\\
954525BP&0.25&1.95&4.45&-0.41&4.65& 0.62&2.60& 19.2& 19.4&0.11&0.63&168& 0.06&1\\
104025BP&0.25&2.10&4.40&-0.23&4.30& 0.09&2.73& 46.5& 46.5&0.17&0.49&224& 0.14&0\\
104525BP&0.25&2.10&4.45&-0.33&4.50& 0.39&2.71& 32.1& 31.6&0.13&0.59&227&-0.07&0\\
154025BP&0.25&2.15&4.40&-0.21&4.25& 0.02&2.77& 51.5& 51.7&0.19&0.46&232& 0.07&0\\
154525BP&0.25&2.15&4.45&-0.31&4.45& 0.32&2.76& 34.9& 35.0&0.14&0.60&234&-0.06&0\\
203025BP&0.25&2.20&4.30& 0.02&3.80&-0.66&2.83&127.2&126.7&0.15&0.34&148& 0.14&0\\
204025BP&0.25&2.20&4.40&-0.18&4.20&-0.06&2.82& 57.1& 57.0&0.22&0.49&238& 0.09&0\\
204525BP&0.25&2.20&4.45&-0.28&4.40& 0.24&2.82& 38.5& 38.5&0.16&0.58&232&-0.05&0\\
205025BP&0.25&2.20&4.50&-0.38&4.60& 0.54&2.83& 25.6& 25.6&0.06&0.39& 79&-0.09&0\\
205525BP&0.25&2.20&4.55&-0.48&4.80& 0.84&2.80& 18.3&  0.0&0.00&0.00&  0&-0.07&-\\
206025BP&0.25&2.20&4.60&-0.58&5.00& 1.14&2.80& 13.0& 13.0&0.00&0.00&  0&-0.06&0\\
253025BP&0.25&2.25&4.30& 0.04&3.75&-0.73&2.88&142.2&142.1&0.17&0.34&152& 0.15&0\\
253525BP&0.25&2.25&4.35&-0.06&3.95&-0.43&1.85& 61.0& 28.0&0.00&0.02&  2& 0.01&5\\
254025BP&0.25&2.25&4.40&-0.16&4.15&-0.13&2.89& 63.4& 63.5&0.26&0.47&255& 0.10&0\\
\hline
\end{tabular}
\end{center}
\end{table*}
\begin{table*}
\begin{center}
\caption{}
\begin{tabular}{lrrrr rrrrr rrrrr}
\hline
 model  & $M$    & $\log L$ & $\log \Teff$ & $\log R$ & $\log g$  & $\log \bar{\rho}$ & $\log L/M$ & $P_{\rm nl}$ & $P_{\rm ft}$ &  $\Delta L$ &  $\Delta R$ &  $\Delta u$   & $\Delta \phi$ & $k$\\
$llttmmNN$& \Msolar & \Lsolar & K     & \Rsolar & cm\,s$^{-2}$  & $\bar{\rho_{\odot}}$ & \Lsolar/\Msolar & min &  min & $L_{\star}$ & $R_{\star}$ & \kmsec &      &     \\
\hline
254525BP&0.25&2.25&4.45&-0.26&4.35& 0.17&2.87& 42.6& 42.5&0.17&0.57&235&-0.03&0\\
255025BP&0.25&2.25&4.50&-0.36&4.55& 0.47&2.87& 28.1& 28.1&0.08&0.48&114&-0.10&0\\
255525BP&0.25&2.25&4.55&-0.46&4.75& 0.77&2.85& 19.9&  0.0&0.00&0.00&  0&-0.38&-\\
256025BP&0.25&2.25&4.60&-0.56&4.95& 1.07&2.85& 13.8&  0.0&0.00&0.00&  0&-0.09&-\\
304025BP&0.25&2.30&4.40&-0.13&4.10&-0.21&2.95& 70.7& 71.0&0.30&0.45&274& 0.12&0\\
304525BP&0.25&2.30&4.45&-0.23&4.30& 0.09&2.93& 47.3& 47.4&0.21&0.58&245& 0.01&0\\
\\[-1mm]
953030BP&0.30&1.95&4.30&-0.11&4.13&-0.20&2.47& 65.1&  0.0&0.00&0.01&  0& 0.11&-\\
953530BP&0.30&1.95&4.35&-0.21&4.33& 0.10&2.50& 44.5& 44.5&0.10&0.44&125& 0.07&0\\
954030BP&0.30&1.95&4.40&-0.31&4.53& 0.40&2.51& 25.0& 25.0&0.07&0.48&137& 0.05&1\\
954530BP&0.30&1.95&4.45&-0.41&4.73& 0.70&2.48& 17.3& 17.2&0.08&0.50&159& 0.04&1\\
955030BP&0.30&1.95&4.50&-0.51&4.93& 1.00&2.48& 16.1&  5.4&0.02&0.12& 73& 0.13&5\\
955530BP&0.30&1.95&4.55&-0.61&5.13& 1.30&2.47& 10.9& 10.9&0.00&0.02&  1&-0.07&0\\
103030BP&0.30&2.10&4.30&-0.03&4.38&-0.43&2.62& 88.6& 88.6&0.00&0.01&  0& 0.13&0\\
103530BP&0.30&2.10&4.35&-0.13&4.58&-0.13&2.68& 60.0&  0.0&0.14&0.52&194& 0.11&-\\
104030BP&0.30&2.10&4.40&-0.23&4.78& 0.17&2.66& 40.9& 41.1&0.14&0.61&228& 0.10&0\\
104530BP&0.30&2.10&4.45&-0.33&4.98& 0.47&2.67& 23.1& 23.1&0.12&0.64&194& 0.06&1\\
105030BP&0.30&2.10&4.50&-0.43&5.18& 0.77&2.68& 20.0& 19.8&0.03&0.32& 45&-0.12&0\\
105530BP&0.30&2.10&4.55&-0.53&4.03& 1.07&2.62& 14.1& 14.1&0.00&0.01&  0&-0.09&0\\
153030BP&0.30&2.15&4.30&-0.01&4.33&-0.50&2.67& 98.2& 98.1&0.00&0.01&  0& 0.11&0\\
153530BP&0.30&2.15&4.35&-0.11&4.53&-0.20&2.72& 66.5& 66.3&0.16&0.47&198& 0.13&0\\
154030BP&0.30&2.15&4.40&-0.21&4.73& 0.10&2.70& 45.1& 45.0&0.15&0.57&232& 0.12&0\\
154530BP&0.30&2.15&4.45&-0.31&4.93& 0.40&2.71& 25.0& 25.4&0.12&0.59&171& 0.09&1\\
155030BP&0.30&2.15&4.50&-0.41&5.13& 0.70&2.73& 21.4& 21.5&0.10&0.86&149&-0.06&0\\
155530BP&0.30&2.15&4.55&-0.51&3.98& 1.00&2.67& 15.5& 15.4&0.00&0.01&  0&-0.11&0\\
156030BP&0.30&2.15&4.60&-0.61&4.18& 1.30&2.67& 10.9& 10.9&0.00&0.01&  0&-0.06&0\\
203030BP&0.30&2.20&4.30& 0.02&4.28&-0.58&2.73&109.0&109.0&0.11&0.26& 85& 0.10&0\\
203530BP&0.30&2.20&4.35&-0.08&4.48&-0.28&2.75& 73.4& 73.5&0.17&0.45&205& 0.08&0\\
204030BP&0.30&2.20&4.40&-0.18&4.68& 0.02&2.75& 49.8& 49.7&0.16&0.53&231& 0.13&0\\
204530BP&0.30&2.20&4.45&-0.28&4.88& 0.32&2.73& 34.2& 34.2&0.14&0.64&235& 0.01&0\\
205030BP&0.30&2.20&4.50&-0.38&5.08& 0.62&2.75& 23.3& 23.2&0.10&0.75&136&-0.08&0\\
205530BP&0.30&2.20&4.55&-0.48&3.93& 0.92&2.72& 16.9& 16.8&0.00&0.01&  2&-0.10&0\\
206030BP&0.30&2.20&4.60&-0.58&4.13& 1.22&2.72& 11.9& 11.9&0.00&0.01&  0&-0.03&0\\
253030BP&0.30&2.25&4.30& 0.04&4.23&-0.65&2.80&121.4&121.2&0.14&0.34&125& 0.12&0\\
253530BP&0.30&2.25&4.35&-0.06&4.43&-0.35&2.82& 81.6& 82.1&0.20&0.50&215& 0.10&0\\
254030BP&0.30&2.25&4.40&-0.16&4.63&-0.05&2.79& 55.1& 55.2&0.17&0.47&234& 0.16&0\\
254530BP&0.30&2.25&4.45&-0.26&4.83& 0.25&2.78& 37.4& 37.7&0.14&0.61&232&-0.05&0\\
255030BP&0.30&2.25&4.50&-0.36&5.03& 0.55&2.80& 25.3& 25.4&0.09&0.68&136&-0.08&0\\
255530BP&0.30&2.25&4.55&-0.46&3.88& 0.85&1.92& 13.9&  1.9&0.01&0.78& 93& 0.00&8\\
256030BP&0.30&2.25&4.60&-0.56&4.08& 1.15&2.77& 13.0& 13.0&0.00&0.01&  0&-0.04&0\\
303030BP&0.30&2.30&4.30& 0.07&4.18&-0.73&2.85&135.9&135.6&0.15&0.35&147& 0.14&0\\
303530BP&0.30&2.30&4.35&-0.03&4.38&-0.43&2.89& 90.8& 90.8&0.24&0.52&232& 0.10&0\\
304030BP&0.30&2.30&4.40&-0.13&4.58&-0.13&2.84& 61.4& 61.7&0.20&0.45&234& 0.08&0\\
304530BP&0.30&2.30&4.45&-0.23&4.78& 0.17&2.83& 41.2& 41.2&0.15&0.58&234&-0.05&0\\
305030BP&0.30&2.30&4.50&-0.33&4.98& 0.47&2.85& 27.7& 27.9&0.06&0.40& 79&-0.10&0\\
305530BP&0.30&2.30&4.55&-0.43&3.83& 0.77&2.82& 22.1&  0.0&0.00&0.00&  0&-0.33&-\\
306030BP&0.30&2.30&4.60&-0.53&4.03& 1.07&2.82& 14.1& 14.1&0.00&0.00&  1&-0.11&0\\
353030BP&0.30&2.35&4.30& 0.09&4.33&-0.80&2.90&151.8&151.8&0.17&0.35&164& 0.15&0\\
353530BP&0.30&2.35&4.35&-0.01&4.53&-0.50&2.96&101.0&101.1&0.27&0.52&230& 0.10&0\\
354030BP&0.30&2.35&4.40&-0.11&4.73&-0.20&2.90& 67.8& 68.3&0.23&0.47&257& 0.09&0\\
354530BP&0.30&2.35&4.45&-0.21&4.93& 0.10&2.88& 45.5& 45.5&0.16&0.57&236&-0.05&0\\
355030BP&0.30&2.35&4.50&-0.31&3.78& 0.40&2.90& 30.8& 30.8&0.09&0.55&134&-0.09&0\\
355530BP&0.30&2.35&4.55&-0.41&3.98& 0.70&2.06& 10.9&  2.2&0.01&1.06&159& 0.40&8\\
403030BP&0.30&2.40&4.30& 0.12&4.48&-0.88&2.95&170.3&170.7&0.18&0.33&165& 0.17&0\\
403530BP&0.30&2.40&4.35& 0.02&4.68&-0.58&2.98&113.5&113.6&0.25&0.46&198& 0.15&0\\
404030BP&0.30&2.40&4.40&-0.08&4.88&-0.28&2.96& 75.5& 75.7&0.27&0.44&260& 0.11&0\\
404530BP&0.30&2.40&4.45&-0.18&3.73& 0.02&2.94& 50.5& 50.6&0.18&0.56&241& 0.01&0\\
405030BP&0.30&2.40&4.50&-0.28&3.93& 0.32&2.96& 34.0& 34.1&0.11&0.63&190&-0.07&0\\
405530BP&0.30&2.40&4.55&-0.38&4.13& 0.62&2.92& 25.2&  0.0&0.00&0.00&  0&-0.04&-\\
453030BP&0.30&2.45&4.30& 0.14&3.68&-0.95&3.00&192.6&  0.0&0.19&0.29&161& 0.20&-\\
453530BP&0.30&2.45&4.35& 0.04&3.88&-0.65&3.01&127.4&127.5&0.23&0.36&194& 0.18&0\\
\hline
\end{tabular}
\end{center}
\end{table*}
\begin{table*}
\begin{center}
\caption{}
\begin{tabular}{lrrrr rrrrr rrrrr}
\hline
 model  & $M$    & $\log L$ & $\log \Teff$ & $\log R$ & $\log g$  & $\log \bar{\rho}$ & $\log L/M$ & $P_{\rm nl}$ & $P_{\rm ft}$ &  $\Delta L$ &  $\Delta R$ &  $\Delta u$   & $\Delta \phi$ & $k$\\
$llttmmNN$& \Msolar & \Lsolar & K     & \Rsolar & cm\,s$^{-2}$  & $\bar{\rho_{\odot}}$ & \Lsolar/\Msolar & min &  min & $L_{\star}$ & $R_{\star}$ & \kmsec &      &     \\
\hline
455030BP&0.30&2.45&4.50&-0.26&4.08& 0.25&3.01& 37.9& 37.8&0.14&0.64&206&-0.07&0\\
455530BP&0.30&2.45&4.55&-0.36&4.28& 0.55&2.09& 15.2&  2.7&0.01&0.99&135& 0.00&8\\
\\[-1mm]
954035BP&0.35&1.95&4.40&-0.31&4.63& 0.47&2.41& 22.8& 22.6&0.03&0.19& 51& 0.04&1\\
954535BP&0.35&1.95&4.45&-0.41&4.83& 0.77&2.40& 15.8& 15.5&0.03&0.21& 49& 0.00&1\\
103535BP&0.35&2.10&4.35&-0.13&4.50&-0.06&2.56& 53.8&  0.0&0.05&0.18& 57& 0.09&-\\
104035BP&0.35&2.10&4.40&-0.23&4.70& 0.24&2.60& 30.1& 30.1&0.09&0.52&122& 0.09&1\\
104535BP&0.35&2.10&4.45&-0.33&4.55& 0.54&2.60& 20.9& 20.8&0.10&0.66&210& 0.02&1\\
105035BP&0.35&2.10&4.50&-0.43&4.75& 0.84&2.57& 19.6& 18.3&0.02&0.16& 76& 0.10&0\\
105535BP&0.35&2.10&4.55&-0.53&4.60& 1.14&2.56& 13.0& 13.0&0.00&0.02&  3&-0.05&0\\
153035BP&0.35&2.15&4.30&-0.01&5.00&-0.43&2.61& 86.9& 87.2&0.00&0.01&  2& 0.19&0\\
153535BP&0.35&2.15&4.35&-0.11&4.25&-0.13&2.64& 59.2&  0.0&0.12&0.48&178& 0.08&-\\
154035BP&0.35&2.15&4.40&-0.21&4.45& 0.17&2.66& 40.9& 40.9&0.13&0.69&231& 0.10&0\\
154535BP&0.35&2.15&4.45&-0.31&4.65& 0.47&2.66& 22.8& 22.9&0.12&0.70&208& 0.05&1\\
155035BP&0.35&2.15&4.50&-0.41&4.85& 0.77&2.61& 21.3& 20.0&0.01&0.12& 48& 0.14&0\\
155535BP&0.35&2.15&4.55&-0.51&5.05& 1.07&2.61& 14.2& 14.2&0.00&0.01&  3&-0.07&0\\
203035BP&0.35&2.20&4.30& 0.02&4.95&-0.51&2.66& 97.4& 96.9&0.00&0.01&  2& 0.10&0\\
203535BP&0.35&2.20&4.35&-0.08&4.00&-0.21&2.71& 65.7&  0.0&0.14&0.52&200& 0.11&-\\
204035BP&0.35&2.20&4.40&-0.18&4.20& 0.09&2.69& 45.1& 44.7&0.14&0.61&211& 0.09&0\\
204535BP&0.35&2.20&4.45&-0.28&4.40& 0.39&2.70& 25.1& 25.3&0.12&0.66&207& 0.06&1\\
205035BP&0.35&2.20&4.50&-0.38&4.60& 0.69&2.70& 21.6& 21.9&0.04&0.34& 63& 0.03&0\\
205535BP&0.35&2.20&4.55&-0.48&4.80& 0.99&2.66& 15.5& 15.5&0.00&0.01&  3&-0.01&0\\
253035BP&0.35&2.25&4.30& 0.04&4.90&-0.58&2.71&108.0&107.3&0.00&0.01&  4& 0.15&0\\
253535BP&0.35&2.25&4.35&-0.06&3.95&-0.28&2.75& 72.8& 72.9&0.16&0.48&201& 0.13&0\\
254035BP&0.35&2.25&4.40&-0.16&4.15& 0.02&2.74& 49.4& 49.5&0.14&0.57&234& 0.12&0\\
254535BP&0.35&2.25&4.45&-0.26&4.35& 0.32&2.74& 28.0& 27.9&0.13&0.60&197& 0.08&1\\
255035BP&0.35&2.25&4.50&-0.36&4.55& 0.62&2.74& 20.2& 23.8&0.04&0.32& 47& 0.09&0\\
255535BP&0.35&2.25&4.55&-0.46&4.75& 0.92&2.71& 17.0& 16.9&0.00&0.01&  2&-0.11&0\\
303035BP&0.35&2.30&4.30& 0.07&4.85&-0.66&2.76&119.2&119.9&0.03&0.07& 20& 0.14&0\\
303535BP&0.35&2.30&4.35&-0.03&3.90&-0.36&2.78& 80.3& 80.2&0.17&0.44&204& 0.14&0\\
304035BP&0.35&2.30&4.40&-0.13&4.10&-0.06&2.78& 54.5&  0.0&0.15&0.53&234& 0.13&-\\
304535BP&0.35&2.30&4.45&-0.23&4.30& 0.24&2.78& 30.2& 30.8&0.13&0.55&174& 0.10&1\\
305035BP&0.35&2.30&4.50&-0.33&4.50& 0.54&2.77& 26.7& 26.0&0.04&0.32& 57&-0.07&0\\
305535BP&0.35&2.30&4.55&-0.43&4.70& 0.84&2.76& 18.5& 18.4&0.00&0.01&  2&-0.05&0\\
353035BP&0.35&2.35&4.30& 0.09&4.80&-0.73&2.83&133.1&133.3&0.14&0.37&137& 0.12&0\\
353535BP&0.35&2.35&4.35&-0.01&3.85&-0.43&2.85& 89.5& 89.5&0.20&0.48&217& 0.10&0\\
354035BP&0.35&2.35&4.40&-0.11&4.05&-0.13&2.82& 60.3&  0.0&0.17&0.48&237& 0.16&-\\
354535BP&0.35&2.35&4.45&-0.21&4.25& 0.17&2.81& 41.1& 41.1&0.14&0.61&236& 0.01&0\\
355035BP&0.35&2.35&4.50&-0.31&4.45& 0.47&2.82& 28.2& 28.3&0.05&0.34& 58&-0.13&0\\
355535BP&0.35&2.35&4.55&-0.41&4.65& 0.77&2.81& 20.0& 20.1&0.00&0.01&  1&-0.08&0\\
403035BP&0.35&2.40&4.30& 0.12&4.00&-0.81&2.89&149.0&148.7&0.15&0.37&148& 0.14&0\\
403535BP&0.35&2.40&4.35& 0.02&4.20&-0.51&2.92& 99.5& 99.4&0.23&0.50&238& 0.11&0\\
404035BP&0.35&2.40&4.40&-0.08&4.40&-0.21&2.87& 67.2& 67.1&0.19&0.42&237& 0.08&0\\
404535BP&0.35&2.40&4.45&-0.18&4.60& 0.09&2.86& 45.1& 45.2&0.15&0.58&237&-0.05&0\\
453035BP&0.35&2.45&4.30& 0.14&4.15&-0.88&2.93&166.0&166.9&0.17&0.35&168& 0.16&0\\
453535BP&0.35&2.45&4.35& 0.04&4.35&-0.58&2.99&110.4&110.5&0.27&0.53&241& 0.10&0\\
454035BP&0.35&2.45&4.40&-0.06&4.55&-0.28&2.92& 74.1& 74.3&0.22&0.45&243& 0.09&0\\
454535BP&0.35&2.45&4.45&-0.16&4.75& 0.02&2.92& 49.9& 49.8&0.16&0.57&239&-0.05&0\\
455035BP&0.35&2.45&4.50&-0.26&3.80& 0.32&1.94& 23.0&  3.5&0.01&0.98&157& 0.00&8\\
\\[-1mm]
103540BP&0.40&2.10&4.35&-0.13&4.95&-0.00&2.50& 49.4&  0.0&0.01&0.05& 12& 0.09&-\\
104040BP&0.40&2.10&4.40&-0.23&3.70& 0.30&2.51& 27.8& 27.7&0.07&0.40&107&-0.00&1\\
104540BP&0.40&2.10&4.45&-0.33&3.90& 0.60&2.50& 19.3& 19.0&0.07&0.48&146& 0.03&1\\
105040BP&0.40&2.10&4.50&-0.43&4.10& 0.90&2.50& 19.1& 10.7&0.02&0.24& 62& 0.08&2\\
153540BP&0.40&2.15&4.35&-0.11&4.90&-0.08&2.55& 54.3&  0.0&0.03&0.10& 30& 0.09&-\\
154040BP&0.40&2.15&4.40&-0.21&4.35& 0.22&2.62& 30.5& 30.5&0.11&0.65&175& 0.07&1\\
154540BP&0.40&2.15&4.45&-0.31&4.55& 0.52&2.58& 21.1& 21.0&0.09&0.63&204& 0.01&1\\
155040BP&0.40&2.15&4.50&-0.41&4.75& 0.82&2.55& 19.6&  6.6&0.02&0.13& 60& 0.48&5\\
203040BP&0.40&2.20&4.30& 0.02&4.25&-0.45&2.60& 87.6& 87.7&0.00&0.01&  1& 0.18&0\\
203540BP&0.40&2.20&4.35&-0.08&4.45&-0.15&2.61& 59.6&  0.0&0.08&0.28& 90& 0.08&-\\
204040BP&0.40&2.20&4.40&-0.18&4.65& 0.15&2.66& 33.5& 33.6&0.11&0.64&155& 0.09&1\\
204540BP&0.40&2.20&4.45&-0.28&4.85& 0.45&2.65& 23.2& 23.1&0.10&0.66&209& 0.01&1\\
\hline
\end{tabular}
\end{center}
\end{table*}
\begin{table*}
\begin{center}
\caption{}
\begin{tabular}{lrrrr rrrrr rrrrr}
\hline
 model  & $M$    & $\log L$ & $\log \Teff$ & $\log R$ & $\log g$  & $\log \bar{\rho}$ & $\log L/M$ & $P_{\rm nl}$ & $P_{\rm ft}$ &  $\Delta L$ &  $\Delta R$ &  $\Delta u$   & $\Delta \phi$ & $k$\\
$llttmmNN$& \Msolar & \Lsolar & K     & \Rsolar & cm\,s$^{-2}$  & $\bar{\rho_{\odot}}$ & \Lsolar/\Msolar & min &  min & $L_{\star}$ & $R_{\star}$ & \kmsec &      &     \\
\hline
205040BP&0.40&2.20&4.50&-0.38&4.30& 0.75&2.60& 22.0&  7.2&0.02&0.13& 55& 0.11&5\\
205540BP&0.40&2.20&4.55&-0.48&4.50& 1.05&2.60& 14.5& 14.5&0.00&0.02&  3&-0.04&0\\
206040BP&0.40&2.20&4.60&-0.58&4.70& 1.35&2.60& 10.3& 10.3&0.00&0.02&  3&-0.07&0\\
253040BP&0.40&2.25&4.30& 0.04&4.00&-0.53&2.65& 96.9& 97.1&0.00&0.01&  2& 0.15&0\\
253540BP&0.40&2.25&4.35&-0.06&4.20&-0.23&2.69& 65.8&  0.0&0.13&0.53&196& 0.09&-\\
254040BP&0.40&2.25&4.40&-0.16&4.40& 0.07&2.70& 37.6& 37.2&0.10&0.54&180& 0.07&1\\
254540BP&0.40&2.25&4.45&-0.26&4.60& 0.37&2.70& 25.6& 25.5&0.12&0.70&233& 0.05&1\\
255040BP&0.40&2.25&4.50&-0.36&4.80& 0.67&2.65& 23.8&  7.9&0.01&0.08& 49&-0.19&5\\
255540BP&0.40&2.25&4.55&-0.46&5.00& 0.97&2.65& 15.8& 15.8&0.00&0.02&  3&-0.05&0\\
256040BP&0.40&2.25&4.60&-0.56&5.20& 1.27&2.65& 11.2& 11.1&0.00&0.02&  2&-0.05&0\\
303040BP&0.40&2.30&4.30& 0.07&3.95&-0.60&2.70&107.4&107.2&0.00&0.02&  4& 0.14&0\\
303540BP&0.40&2.30&4.35&-0.03&4.15&-0.30&2.75& 73.0&  0.0&0.14&0.51&205& 0.11&-\\
304040BP&0.40&2.30&4.40&-0.13&4.35&-0.00&2.74& 48.8&  0.0&0.14&0.59&159& 0.08&-\\
304540BP&0.40&2.30&4.45&-0.23&4.55& 0.30&2.75& 28.2& 28.2&0.13&0.69&227& 0.06&1\\
305040BP&0.40&2.30&4.50&-0.33&4.75& 0.60&2.70& 25.9&  8.7&0.01&0.09& 41&-0.15&5\\
305540BP&0.40&2.30&4.55&-0.43&4.95& 0.90&2.70& 17.3& 17.2&0.00&0.01&  2&-0.11&0\\
306040BP&0.40&2.30&4.60&-0.53&5.15& 1.20&2.70& 12.2& 12.2&0.00&0.02&  1&-0.06&0\\
353040BP&0.40&2.35&4.30& 0.09&3.90&-0.68&2.75&119.2&119.8&0.02&0.06& 17& 0.14&0\\
353540BP&0.40&2.35&4.35&-0.01&4.10&-0.38&2.80& 80.9& 80.7&0.16&0.48&205& 0.14&0\\
354040BP&0.40&2.35&4.40&-0.11&4.30&-0.08&2.78& 54.8&  0.0&0.14&0.57&238& 0.12&-\\
354540BP&0.40&2.35&4.45&-0.21&4.50& 0.22&2.79& 30.9& 31.0&0.13&0.63&202& 0.08&1\\
355040BP&0.40&2.35&4.50&-0.31&4.70& 0.52&2.76& 28.0& 26.3&0.02&0.14& 43&-0.15&0\\
355540BP&0.40&2.35&4.55&-0.41&4.90& 0.82&2.75& 18.7& 18.8&0.00&0.01&  2&-0.09&0\\
356040BP&0.40&2.35&4.60&-0.51&5.10& 1.12&2.75& 13.4& 13.3&0.00&0.01&  2&-0.08&0\\
403040BP&0.40&2.40&4.30& 0.12&3.85&-0.75&2.81&133.0&133.2&0.12&0.28&118& 0.12&0\\
403540BP&0.40&2.40&4.35& 0.02&4.05&-0.45&2.83& 89.5& 90.0&0.17&0.43&210& 0.15&0\\
404040BP&0.40&2.40&4.40&-0.08&4.25&-0.15&2.82& 60.5&  0.0&0.16&0.52&238& 0.14&-\\
404540BP&0.40&2.40&4.45&-0.18&4.45& 0.15&2.83& 34.3& 34.2&0.12&0.54&199& 0.08&1\\
405040BP&0.40&2.40&4.50&-0.28&4.65& 0.45&2.81& 29.1& 28.8&0.04&0.27& 57&-0.24&0\\
405540BP&0.40&2.40&4.55&-0.38&4.85& 0.75&2.80& 20.6& 20.5&0.00&0.01&  1&-0.07&0\\
406040BP&0.40&2.40&4.60&-0.48&5.05& 1.05&2.80& 14.5& 14.5&0.00&0.01&  1&-0.06&0\\
453040BP&0.40&2.45&4.30& 0.14&3.80&-0.83&2.88&148.1&148.0&0.15&0.41&147& 0.12&0\\
453540BP&0.40&2.45&4.35& 0.04&4.00&-0.53&2.89& 99.7& 99.9&0.19&0.47&222& 0.10&0\\
454040BP&0.40&2.45&4.40&-0.06&4.20&-0.23&2.87& 67.0&  0.0&0.17&0.47&239& 0.16&-\\
454540BP&0.40&2.45&4.45&-0.16&4.40& 0.07&2.88& 40.2& 37.6&0.11&0.47&213& 0.02&1\\
455040BP&0.40&2.45&4.50&-0.26&4.60& 0.37&2.90& 31.2& 31.2&0.10&0.78&165&-0.07&0\\
455540BP&0.40&2.45&4.55&-0.36&4.80& 0.67&2.85& 22.6& 22.4&0.00&0.01&  2&-0.11&0\\
456040BP&0.40&2.45&4.60&-0.46&5.00& 0.97&2.85& 15.8& 15.8&0.00&0.01&  2&-0.08&0\\
503040BP&0.40&2.50&4.30& 0.17&3.75&-0.90&2.93&166.5&166.5&0.16&0.37&162& 0.15&0\\
503540BP&0.40&2.50&4.35& 0.07&3.95&-0.60&2.96&110.7&110.9&0.24&0.50&240& 0.11&0\\
504040BP&0.40&2.50&4.40&-0.03&4.15&-0.30&2.91& 74.6& 74.5&0.19&0.41&242& 0.08&0\\
504540BP&0.40&2.50&4.45&-0.13&4.35&-0.00&2.90& 50.0&  0.0&0.15&0.59&244& 0.00&-\\
505040BP&0.40&2.50&4.50&-0.23&4.55& 0.30&2.92& 34.4& 34.2&0.07&0.47& 99&-0.11&0\\
505540BP&0.40&2.50&4.55&-0.33&4.75& 0.60&2.90& 24.3& 24.3&0.00&0.01&  2&-0.12&0\\
506040BP&0.40&2.50&4.60&-0.43&4.95& 0.90&2.90& 17.4& 17.3&0.00&0.01&  1&-0.03&0\\
553040BP&0.40&2.55&4.30& 0.19&4.30&-0.98&2.98&185.7&186.2&0.17&0.36&170& 0.17&0\\
555040BP&0.40&2.55&4.50&-0.21&4.50& 0.22&2.99& 37.5& 37.5&0.11&0.69&193&-0.08&0\\
555540BP&0.40&2.55&4.55&-0.31&4.70& 0.52&2.95& 26.5&  0.0&0.00&0.00&  2&-0.25&-\\
556040BP&0.40&2.55&4.60&-0.41&4.90& 0.82&2.95& 18.6& 18.7&0.00&0.01&  2&-0.08&0\\
\\[-1mm]
104050BP&0.50&2.10&4.40&-0.23&3.90& 0.39&2.40& 24.4& 23.9&0.02&0.17& 39& 0.01&1\\
104550BP&0.50&2.10&4.45&-0.33&4.10& 0.69&2.40& 17.0& 16.4&0.01&0.16& 22&-0.12&1\\
153050BP&0.50&2.15&4.30&-0.01&4.80&-0.28&2.45& 68.1&  0.0&0.00&0.01&  2& 0.12&-\\
153550BP&0.50&2.15&4.35&-0.11&3.65& 0.02&2.45& 46.7& 47.6&0.00&0.03&  5&-0.00&0\\
154050BP&0.50&2.15&4.40&-0.21&4.45& 0.32&2.45& 26.7& 26.3&0.04&0.24& 64&-0.00&1\\
154550BP&0.50&2.15&4.45&-0.31&4.65& 0.62&2.44& 18.6& 18.1&0.02&0.22& 39&-0.06&1\\
155050BP&0.50&2.15&4.50&-0.41&4.85& 0.92&2.49& 19.5& 10.3&0.04&0.60&119& 0.00&2\\
155550BP&0.50&2.15&4.55&-0.51&3.70& 1.22&2.45& 11.2& 11.8&0.00&0.03&  3& 0.00&0\\
203050BP&0.50&2.20&4.30& 0.02&4.35&-0.36&2.50& 75.0&  0.0&0.00&0.01&  3& 0.12&-\\
203550BP&0.50&2.20&4.35&-0.08&4.55&-0.06&2.50& 54.5&  0.0&0.01&0.03&  8& 0.01&-\\
204050BP&0.50&2.20&4.40&-0.18&4.75& 0.24&2.52& 29.3& 29.0&0.07&0.44&143& 0.07&1\\
\hline
\end{tabular}
\end{center}
\end{table*}
\begin{table*}
\begin{center}
\caption{}
\begin{tabular}{lrrrr rrrrr rrrrr}
\hline
 model  & $M$    & $\log L$ & $\log \Teff$ & $\log R$ & $\log g$  & $\log \bar{\rho}$ & $\log L/M$ & $P_{\rm nl}$ & $P_{\rm ft}$ &  $\Delta L$ &  $\Delta R$ &  $\Delta u$   & $\Delta \phi$ & $k$\\
$llttmmNN$& \Msolar & \Lsolar & K     & \Rsolar & cm\,s$^{-2}$  & $\bar{\rho_{\odot}}$ & \Lsolar/\Msolar & min &  min & $L_{\star}$ & $R_{\star}$ & \kmsec &      &     \\
\hline
204550BP&0.50&2.20&4.45&-0.28&4.95& 0.54&2.49& 20.4& 19.9&0.04&0.32& 82&-0.05&1\\
205050BP&0.50&2.20&4.50&-0.38&5.15& 0.84&2.52& 21.3& 11.3&0.04&0.44&106& 0.49&2\\
205550BP&0.50&2.20&4.55&-0.48&4.60& 1.14&2.50& 12.7& 12.9&0.00&0.03&  2& 0.00&0\\
253050BP&0.50&2.25&4.30& 0.04&4.30&-0.43&2.55& 83.0&  0.0&0.00&0.01&  3& 0.15&-\\
253550BP&0.50&2.25&4.35&-0.06&4.50&-0.13&2.55& 57.0&  0.0&0.01&0.05& 15& 0.09&-\\
254050BP&0.50&2.25&4.40&-0.16&4.70& 0.17&2.59& 32.3& 32.0&0.10&0.57&188& 0.05&1\\
254550BP&0.50&2.25&4.45&-0.26&4.90& 0.47&2.56& 22.3& 22.0&0.08&0.57&185& 0.05&1\\
255050BP&0.50&2.25&4.50&-0.36&5.10& 0.77&2.56& 23.3& 12.4&0.04&0.44& 91&-0.45&2\\
255550BP&0.50&2.25&4.55&-0.46&4.15& 1.07&2.55& 12.8& 14.1&0.00&0.02&  3& 0.01&0\\
303050BP&0.50&2.30&4.30& 0.07&4.25&-0.51&2.60& 91.2&  0.0&0.00&0.01&  3& 0.14&-\\
303550BP&0.50&2.30&4.35&-0.03&4.45&-0.21&2.61& 62.8&  0.0&0.03&0.11& 36& 0.07&-\\
304050BP&0.50&2.30&4.40&-0.13&4.65& 0.09&2.70& 35.4& 35.3&0.12&0.76&232& 0.06&1\\
304550BP&0.50&2.30&4.45&-0.23&4.85& 0.39&2.64& 24.4& 24.2&0.09&0.65&212&-0.07&1\\
305050BP&0.50&2.30&4.50&-0.33&5.05& 0.69&2.60& 23.0&  7.6&0.01&0.11& 52& 0.49&5\\
305550BP&0.50&2.30&4.55&-0.43&4.10& 0.99&2.60& 15.2& 15.4&0.00&0.02&  2&-0.06&0\\
353050BP&0.50&2.35&4.30& 0.09&4.20&-0.58&2.65&101.0&  0.0&0.00&0.02&  5& 0.12&-\\
353550BP&0.50&2.35&4.35&-0.01&4.40&-0.28&2.68& 68.7&  0.0&0.10&0.39&140& 0.08&-\\
354050BP&0.50&2.35&4.40&-0.11&4.60& 0.02&2.74& 39.4& 39.0&0.13&0.70&216& 0.06&1\\
354550BP&0.50&2.35&4.45&-0.21&4.80& 0.32&2.70& 26.9& 26.6&0.11&0.69&228& 0.01&1\\
355050BP&0.50&2.35&4.50&-0.31&5.00& 0.62&2.66& 25.5&  8.4&0.02&0.12& 71& 0.12&5\\
355550BP&0.50&2.35&4.55&-0.41&4.05& 0.92&2.65& 16.8& 16.8&0.00&0.02&  3&-0.06&0\\
403050BP&0.50&2.40&4.30& 0.12&4.15&-0.66&2.70&111.8&111.8&0.01&0.03&  7& 0.12&0\\
403550BP&0.50&2.40&4.35& 0.02&4.35&-0.36&2.75& 75.8&  0.0&0.14&0.54&206& 0.10&-\\
404050BP&0.50&2.40&4.40&-0.08&4.55&-0.06&2.77& 44.9&  0.0&0.13&0.64&209& 0.08&-\\
404550BP&0.50&2.40&4.45&-0.18&4.75& 0.24&2.76& 29.7& 29.5&0.12&0.72&247& 0.04&1\\
405050BP&0.50&2.40&4.50&-0.28&4.95& 0.54&2.70& 27.5&  9.2&0.02&0.12& 59&-0.20&5\\
405550BP&0.50&2.40&4.55&-0.38&4.00& 0.84&2.70& 18.4& 18.3&0.00&0.02&  2&-0.09&0\\
453050BP&0.50&2.45&4.30& 0.14&4.10&-0.73&2.75&124.2&123.6&0.02&0.07& 20& 0.14&0\\
453550BP&0.50&2.45&4.35& 0.04&4.30&-0.43&2.80& 84.0&  0.0&0.14&0.52&214& 0.11&-\\
454050BP&0.50&2.45&4.40&-0.06&4.50&-0.13&2.80& 46.4&  0.0&0.10&0.53&121& 0.15&-\\
454550BP&0.50&2.45&4.45&-0.16&4.70& 0.17&2.81& 32.5& 32.6&0.13&0.69&241& 0.05&1\\
455050BP&0.50&2.45&4.50&-0.26&4.90& 0.47&2.75& 33.4& 17.9&0.03&0.19& 54& 0.00&2\\
455550BP&0.50&2.45&4.55&-0.36&3.95& 0.77&2.75& 20.0& 20.0&0.00&0.01&  3&-0.07&0\\
503050BP&0.50&2.50&4.30& 0.17&4.05&-0.81&2.83&137.7&137.8&0.10&0.30&111& 0.12&0\\
503550BP&0.50&2.50&4.35& 0.07&4.25&-0.51&2.85& 92.9& 93.0&0.16&0.49&214& 0.13&0\\
504050BP&0.50&2.50&4.40&-0.03&4.45&-0.21&2.83& 63.1&  0.0&0.14&0.58&239& 0.12&-\\
504550BP&0.50&2.50&4.45&-0.13&4.65& 0.09&2.85& 35.9& 36.0&0.13&0.65&227& 0.06&1\\
505050BP&0.50&2.50&4.50&-0.23&4.85& 0.39&2.81& 32.6& 30.6&0.02&0.17& 44&-0.13&0\\
505550BP&0.50&2.50&4.55&-0.33&3.90& 0.69&2.80& 21.7& 21.8&0.00&0.01&  2&-0.06&0\\
553050BP&0.50&2.55&4.30& 0.19&4.00&-0.88&2.89&153.2&153.8&0.15&0.41&164& 0.10&0\\
553550BP&0.50&2.55&4.35& 0.09&4.20&-0.58&2.89&103.2&103.5&0.17&0.43&219& 0.17&0\\
554050BP&0.50&2.55&4.40&-0.01&4.40&-0.28&2.88& 69.7&  0.0&0.15&0.52&240& 0.14&-\\
554550BP&0.50&2.55&4.45&-0.11&4.60& 0.02&2.88& 39.2& 39.5&0.14&0.58&200& 0.09&1\\
555050BP&0.50&2.55&4.50&-0.21&4.80& 0.32&2.88& 34.1& 33.4&0.06&0.47& 59&-0.11&0\\
555550BP&0.50&2.55&4.55&-0.31&3.85& 0.62&2.85& 23.7& 23.8&0.00&0.03&  1&-0.08&0\\
603050BP&0.50&2.60&4.30& 0.22&3.95&-0.96&2.94&171.1&171.9&0.17&0.46&166& 0.12&0\\
603550BP&0.50&2.60&4.35& 0.12&4.15&-0.66&2.94&115.0&114.8&0.19&0.44&225& 0.11&0\\
604050BP&0.50&2.60&4.40& 0.02&4.35&-0.36&2.92& 77.2&  0.0&0.16&0.47&244& 0.16&-\\
604550BP&0.50&2.60&4.45&-0.08&4.55&-0.06&2.91& 42.3&  0.0&0.13&0.47&181& 0.12&-\\
605050BP&0.50&2.60&4.50&-0.18&4.75& 0.24&2.95& 36.2& 36.2&0.11&0.78&156&-0.08&0\\
605550BP&0.50&2.60&4.55&-0.28&3.80& 0.54&2.90& 26.1& 25.9&0.00&0.01&  1&-0.08&0\\
\\[-1mm]
203060BP&0.60&2.20&4.30& 0.02&3.90&-0.28&2.42& 67.2&  0.0&0.00&0.02&  3& 0.12&-\\
203560BP&0.60&2.20&4.35&-0.08&4.10& 0.02&2.42& 40.2& 37.3&0.00&0.04&  4& 0.11&1\\
204060BP&0.60&2.20&4.40&-0.18&4.30& 0.32&2.42& 26.3& 25.8&0.02&0.17& 38&-0.00&1\\
204560BP&0.60&2.20&4.45&-0.28&4.50& 0.62&2.42& 18.4& 17.7&0.04&0.38& 71& 0.01&1\\
205060BP&0.60&2.20&4.50&-0.38&4.70& 0.92&2.44& 19.3& 10.1&0.03&0.44& 91&-0.48&2\\
205560BP&0.60&2.20&4.55&-0.48&3.75& 1.22&2.43& 13.1&  7.0&0.02&0.34& 79&-0.00&2\\
253060BP&0.60&2.25&4.30& 0.04&4.38&-0.35&2.47& 73.7&  0.0&0.00&0.02&  3& 0.14&-\\
253560BP&0.60&2.25&4.35&-0.06&4.58&-0.05&2.47& 42.3&  0.0&0.01&0.04&  9& 0.16&-\\
254060BP&0.60&2.25&4.40&-0.16&4.78& 0.25&2.47& 28.8& 28.3&0.04&0.25& 70& 0.05&1\\
\hline
\end{tabular}
\end{center}
\end{table*}
\begin{table*}
\begin{center}
\caption{}
\begin{tabular}{lrrrr rrrrr rrrrr}
\hline
 model  & $M$    & $\log L$ & $\log \Teff$ & $\log R$ & $\log g$  & $\log \bar{\rho}$ & $\log L/M$ & $P_{\rm nl}$ & $P_{\rm ft}$ &  $\Delta L$ &  $\Delta R$ &  $\Delta u$   & $\Delta \phi$ & $k$\\
$llttmmNN$& \Msolar & \Lsolar & K     & \Rsolar & cm\,s$^{-2}$  & $\bar{\rho_{\odot}}$ & \Lsolar/\Msolar & min &  min & $L_{\star}$ & $R_{\star}$ & \kmsec &      &     \\
\hline
254560BP&0.60&2.25&4.45&-0.26&4.98& 0.55&2.48& 20.1& 19.4&0.06&0.58&156& 0.00&1\\
255060BP&0.60&2.25&4.50&-0.36&5.18& 0.85&2.48& 21.1& 11.1&0.03&0.40& 99&-0.46&2\\
255560BP&0.60&2.25&4.55&-0.46&3.70& 1.15&2.49& 14.7&  7.7&0.02&0.51& 85&-0.50&2\\
303060BP&0.60&2.30&4.30& 0.07&4.13&-0.43&2.52& 81.2&  0.0&0.00&0.02&  4& 0.12&-\\
303560BP&0.60&2.30&4.35&-0.03&4.33&-0.13&2.52& 51.7&  0.0&0.01&0.04&  8& 0.07&-\\
304060BP&0.60&2.30&4.40&-0.13&4.53& 0.17&2.54& 31.6& 31.1&0.07&0.47&152& 0.06&1\\
304560BP&0.60&2.30&4.45&-0.23&4.73& 0.47&2.54& 22.0& 21.5&0.08&0.70&207&-0.01&1\\
305060BP&0.60&2.30&4.50&-0.33&4.93& 0.77&2.54& 23.0& 12.2&0.03&0.45&114&-0.49&2\\
305560BP&0.60&2.30&4.55&-0.43&5.13& 1.07&2.52& 12.0& 14.0&0.00&0.03&  4& 0.01&0\\
306060BP&0.60&2.30&4.60&-0.53&4.18& 1.37&2.52& 11.3&  7.2&0.00&0.04&  4&-0.31&1\\
353060BP&0.60&2.35&4.30& 0.09&4.08&-0.50&2.57& 88.7&  0.0&0.00&0.02&  3& 0.13&-\\
353560BP&0.60&2.35&4.35&-0.01&4.28&-0.20&2.57& 58.1&  0.0&0.01&0.05& 16& 0.07&-\\
354060BP&0.60&2.35&4.40&-0.11&4.48& 0.10&2.60& 34.6& 34.5&0.09&0.55&191& 0.01&1\\
354560BP&0.60&2.35&4.45&-0.21&4.68& 0.40&2.57& 24.1& 23.6&0.06&0.50&153&-0.02&1\\
355060BP&0.60&2.35&4.50&-0.31&4.88& 0.70&2.58& 25.2& 13.4&0.04&0.42& 94&-0.46&2\\
355560BP&0.60&2.35&4.55&-0.41&5.08& 1.00&2.57& 17.1& 15.3&0.00&0.03&  2&-0.23&0\\
356060BP&0.60&2.35&4.60&-0.51&5.28& 1.30&2.57& 12.3& 10.7&0.00&0.03&  2&-0.28&0\\
403060BP&0.60&2.40&4.30& 0.12&4.03&-0.58&2.62& 98.2&  0.0&0.00&0.02&  4& 0.12&-\\
403560BP&0.60&2.40&4.35& 0.02&4.23&-0.28&2.62& 69.2&  0.0&0.02&0.08& 25& 0.01&-\\
404060BP&0.60&2.40&4.40&-0.08&4.43& 0.02&2.71& 38.2& 38.1&0.12&0.77&242& 0.05&1\\
404560BP&0.60&2.40&4.45&-0.18&4.63& 0.32&2.64& 26.3& 26.0&0.09&0.64&216&-0.08&1\\
405060BP&0.60&2.40&4.50&-0.28&4.83& 0.62&2.63& 27.5& 14.7&0.03&0.37& 92&-0.48&2\\
405560BP&0.60&2.40&4.55&-0.38&5.03& 0.92&2.62& 16.1& 16.6&0.00&0.02&  4& 0.00&0\\
406060BP&0.60&2.40&4.60&-0.48&5.23& 1.22&2.62& 11.2& 11.7&0.00&0.03&  3& 0.00&0\\
453060BP&0.60&2.45&4.30& 0.14&3.98&-0.65&2.67&107.7&  0.0&0.00&0.03&  6& 0.16&-\\
453560BP&0.60&2.45&4.35& 0.04&4.18&-0.35&2.68& 73.5&  0.0&0.07&0.25& 84& 0.08&-\\
454060BP&0.60&2.45&4.40&-0.06&4.38&-0.05&2.77& 41.8&  0.0&0.13&0.77&240& 0.07&-\\
454560BP&0.60&2.45&4.45&-0.16&4.58& 0.25&2.71& 28.8& 28.8&0.10&0.67&222&-0.05&1\\
455060BP&0.60&2.45&4.50&-0.26&4.78& 0.55&2.68& 30.1& 16.2&0.04&0.39&108& 0.01&2\\
455560BP&0.60&2.45&4.55&-0.36&4.98& 0.85&2.67& 18.2& 18.2&0.00&0.02&  3&-0.08&0\\
456060BP&0.60&2.45&4.60&-0.46&5.18& 1.15&2.67& 12.7& 12.9&0.00&0.03&  3& 0.00&0\\
503060BP&0.60&2.50&4.30& 0.17&3.93&-0.73&2.72&119.7&  0.0&0.01&0.04& 11& 0.13&-\\
503560BP&0.60&2.50&4.35& 0.07&4.13&-0.43&2.77& 81.4&  0.0&0.13&0.55&205& 0.10&-\\
504060BP&0.60&2.50&4.40&-0.03&4.33&-0.13&2.80& 46.1&  0.0&0.12&0.67&211& 0.07&-\\
504560BP&0.60&2.50&4.45&-0.13&4.53& 0.17&2.77& 31.8& 31.8&0.11&0.71&239& 0.01&1\\
505060BP&0.60&2.50&4.50&-0.23&4.73& 0.47&2.73& 30.2& 17.8&0.04&0.28& 91&-0.47&2\\
505560BP&0.60&2.50&4.55&-0.33&4.93& 0.77&2.72& 19.8& 19.8&0.00&0.02&  3&-0.07&0\\
506060BP&0.60&2.50&4.60&-0.43&5.13& 1.07&2.72& 14.0& 14.0&0.00&0.02&  4&-0.05&0\\
553060BP&0.60&2.55&4.30& 0.19&3.88&-0.80&2.77&132.7&132.8&0.04&0.11& 32& 0.12&0\\
553560BP&0.60&2.55&4.35& 0.09&4.08&-0.50&2.82& 90.1&  0.0&0.14&0.51&216& 0.11&-\\
554060BP&0.60&2.55&4.40&-0.01&4.28&-0.20&2.82& 49.8&  0.0&0.11&0.53&161& 0.12&-\\
554560BP&0.60&2.55&4.45&-0.11&4.48& 0.10&2.83& 34.8& 35.2&0.13&0.70&249& 0.05&1\\
555060BP&0.60&2.55&4.50&-0.21&4.68& 0.40&2.78& 32.1& 30.6&0.02&0.13& 49&-0.14&0\\
555560BP&0.60&2.55&4.55&-0.31&4.88& 0.70&2.77& 21.7& 21.7&0.00&0.02&  2&-0.08&0\\
556060BP&0.60&2.55&4.60&-0.41&5.08& 1.00&2.77& 15.3& 15.3&0.00&0.02&  3&-0.06&0\\
603060BP&0.60&2.60&4.30& 0.22&3.83&-0.88&2.86&147.4&147.2&0.11&0.36&141& 0.11&0\\
603560BP&0.60&2.60&4.35& 0.12&4.03&-0.58&2.87& 99.5&  0.0&0.15&0.50&220& 0.12&-\\
604060BP&0.60&2.60&4.40& 0.02&4.23&-0.28&2.87& 68.1&  0.0&0.14&0.63&240& 0.12&-\\
604560BP&0.60&2.60&4.45&-0.08&4.43& 0.02&2.88& 38.8& 38.6&0.14&0.70&256& 0.06&1\\
605060BP&0.60&2.60&4.50&-0.18&4.63& 0.32&2.83& 39.5& 21.4&0.04&0.27& 88&-0.49&2\\
605560BP&0.60&2.60&4.55&-0.28&4.83& 0.62&2.82& 23.5& 23.6&0.00&0.01&  3&-0.10&0\\
606060BP&0.60&2.60&4.60&-0.38&5.03& 0.92&2.82& 16.6& 16.7&0.00&0.02&  3&-0.04&0\\
653060BP&0.60&2.65&4.30& 0.24&3.78&-0.95&2.90&163.9&163.8&0.14&0.38&162& 0.13&0\\
653560BP&0.60&2.65&4.35& 0.14&3.98&-0.65&2.91&110.3&110.6&0.16&0.46&219& 0.15&0\\
654060BP&0.60&2.65&4.40& 0.04&4.18&-0.35&2.90& 74.9&  0.0&0.14&0.56&239& 0.13&-\\
654560BP&0.60&2.65&4.45&-0.06&4.38&-0.05&2.92& 42.6&  0.0&0.14&0.64&227& 0.08&-\\
655060BP&0.60&2.65&4.50&-0.16&4.58& 0.25&2.88& 28.8& 23.4&0.05&0.31& 93& 0.00&2\\
655560BP&0.60&2.65&4.55&-0.26&4.78& 0.55&2.87& 25.5& 25.8&0.00&0.01&  1&-0.07&0\\
656060BP&0.60&2.65&4.60&-0.36&4.98& 0.85&2.87& 18.3& 18.3&0.00&0.01&  2&-0.04&0\\
703060BP&0.60&2.70&4.30& 0.27&3.73&-1.03&2.98&183.9&183.7&0.18&0.50&198& 0.10&0\\
\hline
\end{tabular}
\end{center}
\end{table*}
\begin{table*}
\begin{center}
\caption{}
\begin{tabular}{lrrrr rrrrr rrrrr}
\hline
 model  & $M$    & $\log L$ & $\log \Teff$ & $\log R$ & $\log g$  & $\log \bar{\rho}$ & $\log L/M$ & $P_{\rm nl}$ & $P_{\rm ft}$ &  $\Delta L$ &  $\Delta R$ &  $\Delta u$   & $\Delta \phi$ & $k$\\
$llttmmNN$& \Msolar & \Lsolar & K     & \Rsolar & cm\,s$^{-2}$  & $\bar{\rho_{\odot}}$ & \Lsolar/\Msolar & min &  min & $L_{\star}$ & $R_{\star}$ & \kmsec &      &     \\
\hline
703560BP&0.60&2.70&4.35& 0.17&3.93&-0.73&2.95&122.9&122.9&0.17&0.40&229& 0.18&0\\
704060BP&0.60&2.70&4.40& 0.07&4.13&-0.43&2.94& 82.6&  0.0&0.15&0.50&241& 0.15&-\\
704560BP&0.60&2.70&4.45&-0.03&4.33&-0.13&2.95& 47.1&  0.0&0.14&0.55&214& 0.09&-\\
705060BP&0.60&2.70&4.50&-0.13&4.53& 0.17&2.94& 31.6& 39.3&0.04&0.26& 52& 0.00&0\\
705560BP&0.60&2.70&4.55&-0.23&4.73& 0.47&2.92& 27.9& 28.0&0.00&0.01&  2&-0.11&0\\
706060BP&0.60&2.70&4.60&-0.33&4.93& 0.77&2.92& 19.9& 19.9&0.00&0.01&  0&-0.04&0\\
\\[-1mm]
253570BP&0.70&2.25&4.35&-0.06&4.08& 0.02&2.41& 39.5& 37.3&0.01&0.05& 10& 0.08&1\\
254070BP&0.70&2.25&4.40&-0.16&4.28& 0.32&2.40& 26.4& 25.6&0.02&0.14& 30& 0.03&1\\
254570BP&0.70&2.25&4.45&-0.26&4.48& 0.62&2.45& 18.4& 14.6&0.04&0.68&173& 0.00&2\\
255070BP&0.70&2.25&4.50&-0.36&4.68& 0.92&2.47& 19.4& 10.1&0.05&0.83&143& 0.50&2\\
255570BP&0.70&2.25&4.55&-0.46&4.88& 1.22&2.46& 13.3&  7.0&0.03&0.80&125& 0.01&2\\
303070BP&0.70&2.30&4.30& 0.07&4.80&-0.36&2.45& 74.1&  0.0&0.00&0.02&  3& 0.09&-\\
303570BP&0.70&2.30&4.35&-0.03&5.00&-0.06&2.45& 44.0&  0.0&0.01&0.05& 11& 0.08&-\\
304070BP&0.70&2.30&4.40&-0.13&5.20& 0.24&2.45& 28.8& 28.2&0.03&0.20& 51& 0.02&1\\
304570BP&0.70&2.30&4.45&-0.23&5.40& 0.54&2.49& 20.1& 19.4&0.06&0.74&193& 0.00&1\\
305070BP&0.70&2.30&4.50&-0.33&3.68& 0.84&2.50& 21.1& 11.1&0.04&0.69&145& 0.49&2\\
305570BP&0.70&2.30&4.55&-0.43&3.88& 1.14&2.50& 14.8&  7.7&0.03&0.84& 99& 0.00&2\\
353070BP&0.70&2.35&4.30& 0.09&4.55&-0.43&2.50& 81.0&  0.0&0.00&0.02&  4& 0.08&-\\
353570BP&0.70&2.35&4.35&-0.01&4.75&-0.13&2.50& 53.2&  0.0&0.01&0.05& 13& 0.00&-\\
354070BP&0.70&2.35&4.40&-0.11&4.95& 0.17&2.50& 31.5& 31.0&0.04&0.30& 84& 0.05&1\\
354570BP&0.70&2.35&4.45&-0.21&5.15& 0.47&2.53& 22.0& 21.3&0.07&0.76&207& 0.02&1\\
355070BP&0.70&2.35&4.50&-0.31&5.35& 0.77&2.55& 23.1& 12.2&0.05&0.75&165&-0.48&2\\
355570BP&0.70&2.35&4.55&-0.41&4.40& 1.07&2.50& 11.5& 10.3&0.00&0.04&  5& 0.01&1\\
356070BP&0.70&2.35&4.60&-0.51&4.60& 1.37&2.50& 11.4&  7.2&0.00&0.06&  3&-0.34&1\\
403070BP&0.70&2.40&4.30& 0.12&4.70&-0.51&2.55& 88.7&  0.0&0.00&0.02&  5& 0.14&-\\
403570BP&0.70&2.40&4.35& 0.02&4.90&-0.21&2.56& 49.6&  0.0&0.01&0.06&  9& 0.15&-\\
404070BP&0.70&2.40&4.40&-0.08&5.10& 0.09&2.58& 34.5& 34.2&0.07&0.53&177& 0.06&1\\
404570BP&0.70&2.40&4.45&-0.18&5.30& 0.39&2.60& 24.1& 23.6&0.08&0.72&207& 0.00&1\\
405070BP&0.70&2.40&4.50&-0.28&4.15& 0.69&2.57& 25.3& 13.4&0.04&0.50&119& 0.00&2\\
405570BP&0.70&2.40&4.55&-0.38&4.35& 0.99&2.55& 13.1& 15.4&0.00&0.03&  3& 0.00&0\\
453070BP&0.70&2.45&4.30& 0.14&4.65&-0.58&2.60& 97.3&  0.0&0.00&0.02&  5& 0.13&-\\
453570BP&0.70&2.45&4.35& 0.04&4.85&-0.28&2.61& 62.8&  0.0&0.04&0.22& 66& 0.08&-\\
454070BP&0.70&2.45&4.40&-0.06&5.05& 0.02&2.64& 38.1& 37.8&0.10&0.59&213& 0.03&1\\
454570BP&0.70&2.45&4.45&-0.16&5.25& 0.32&2.63& 26.4& 25.9&0.08&0.68&215&-0.08&1\\
455070BP&0.70&2.45&4.50&-0.26&4.10& 0.62&2.63& 27.6& 14.7&0.05&0.56&109&-0.47&2\\
455570BP&0.70&2.45&4.55&-0.36&4.30& 0.92&2.60& 15.1& 16.8&0.00&0.03&  3& 0.01&0\\
456070BP&0.70&2.45&4.60&-0.46&4.50& 1.22&2.60& 13.5& 11.8&0.00&0.04&  4&-0.24&0\\
503070BP&0.70&2.50&4.30& 0.17&4.60&-0.66&2.65&107.1&  0.0&0.01&0.03&  6& 0.13&-\\
503570BP&0.70&2.50&4.35& 0.07&4.80&-0.36&2.71& 73.3&  0.0&0.11&0.58&158& 0.07&-\\
504070BP&0.70&2.50&4.40&-0.03&5.00&-0.06&2.74& 41.3&  0.0&0.12&0.79&246& 0.06&-\\
504570BP&0.70&2.50&4.45&-0.13&5.20& 0.24&2.68& 28.8& 28.6&0.09&0.64&221&-0.08&1\\
505070BP&0.70&2.50&4.50&-0.23&4.05& 0.54&2.67& 30.2& 16.1&0.04&0.45&147&-0.46&2\\
505570BP&0.70&2.50&4.55&-0.33&4.25& 0.84&2.65& 17.4& 18.4&0.00&0.02&  4& 0.00&0\\
506070BP&0.70&2.50&4.60&-0.43&4.45& 1.14&2.65& 14.4& 12.9&0.00&0.03&  3&-0.23&0\\
553070BP&0.70&2.55&4.30& 0.19&4.55&-0.73&2.70&118.4&  0.0&0.01&0.04& 10& 0.13&-\\
553570BP&0.70&2.55&4.35& 0.09&4.75&-0.43&2.79& 81.4&  0.0&0.12&0.67&220& 0.10&-\\
554070BP&0.70&2.55&4.40&-0.01&4.95&-0.13&2.81& 46.2&  0.0&0.13&0.78&256& 0.06&-\\
554570BP&0.70&2.55&4.45&-0.11&5.15& 0.17&2.75& 31.6& 31.5&0.10&0.71&239&-0.05&1\\
555070BP&0.70&2.55&4.50&-0.21&4.00& 0.47&2.71& 33.0& 17.7&0.04&0.46&102& 0.00&2\\
555570BP&0.70&2.55&4.55&-0.31&4.20& 0.77&2.70& 20.2& 20.0&0.00&0.02&  2&-0.09&0\\
556070BP&0.70&2.55&4.60&-0.41&4.40& 1.07&2.70& 14.3& 14.1&0.00&0.02&  1&-0.09&0\\
603070BP&0.70&2.60&4.30& 0.22&4.50&-0.81&2.76&130.6&  0.0&0.02&0.08& 21& 0.13&-\\
603570BP&0.70&2.60&4.35& 0.12&4.70&-0.51&2.81& 89.3&  0.0&0.12&0.57&211& 0.11&-\\
604070BP&0.70&2.60&4.40& 0.02&4.90&-0.21&2.85& 50.4&  0.0&0.14&0.72&234& 0.07&-\\
604570BP&0.70&2.60&4.45&-0.08&5.10& 0.09&2.81& 34.7& 34.9&0.11&0.72&248& 0.01&1\\
605070BP&0.70&2.60&4.50&-0.18&3.95& 0.39&2.76& 36.1& 19.5&0.04&0.37& 85& 0.00&2\\
605570BP&0.70&2.60&4.55&-0.28&4.15& 0.69&2.75& 21.9& 21.9&0.00&0.03&  1&-0.10&0\\
606070BP&0.70&2.60&4.60&-0.38&4.35& 0.99&2.75& 15.3& 15.4&0.00&0.02&  2&-0.03&0\\
653070BP&0.70&2.65&4.30& 0.24&4.45&-0.88&2.82&145.0&144.9&0.07&0.22& 73& 0.10&0\\
653570BP&0.70&2.65&4.35& 0.14&4.65&-0.58&2.85& 98.6&  0.0&0.14&0.52&220& 0.12&-\\
\hline
\end{tabular}
\end{center}
\end{table*}
\begin{table*}
\begin{center}
\caption{}
\begin{tabular}{lrrrr rrrrr rrrrr}
\hline
 model  & $M$    & $\log L$ & $\log \Teff$ & $\log R$ & $\log g$  & $\log \bar{\rho}$ & $\log L/M$ & $P_{\rm nl}$ & $P_{\rm ft}$ &  $\Delta L$ &  $\Delta R$ &  $\Delta u$   & $\Delta \phi$ & $k$\\
$llttmmNN$& \Msolar & \Lsolar & K     & \Rsolar & cm\,s$^{-2}$  & $\bar{\rho_{\odot}}$ & \Lsolar/\Msolar & min &  min & $L_{\star}$ & $R_{\star}$ & \kmsec &      &     \\
\hline
654070BP&0.70&2.65&4.40& 0.04&4.85&-0.28&2.88& 55.2&  0.0&0.13&0.64&188& 0.11&-\\
654570BP&0.70&2.65&4.45&-0.06&5.05& 0.02&2.86& 38.2& 38.5&0.13&0.72&262& 0.06&1\\
655070BP&0.70&2.65&4.50&-0.16&3.90& 0.32&2.81& 36.3& 33.6&0.02&0.18& 48&-0.48&0\\
655570BP&0.70&2.65&4.55&-0.26&4.10& 0.62&2.80& 23.8& 23.9&0.00&0.02&  3&-0.08&0\\
656070BP&0.70&2.65&4.60&-0.36&4.30& 0.92&2.80& 16.8& 16.8&0.00&0.02&  2&-0.03&0\\
703070BP&0.70&2.70&4.30& 0.27&4.40&-0.96&2.90&161.1&161.2&0.13&0.41&167& 0.11&0\\
703570BP&0.70&2.70&4.35& 0.17&4.60&-0.66&2.90&108.8&  0.0&0.14&0.49&224& 0.12&-\\
704070BP&0.70&2.70&4.40& 0.07&4.80&-0.36&2.91& 73.4&  0.0&0.15&0.68&228& 0.11&-\\
704570BP&0.70&2.70&4.45&-0.03&5.00&-0.06&2.91& 42.1&  0.0&0.14&0.69&260& 0.07&-\\
705070BP&0.70&2.70&4.50&-0.13&3.85& 0.24&2.86& 43.2& 23.4&0.03&0.25& 65& 0.00&2\\
705570BP&0.70&2.70&4.55&-0.23&4.05& 0.54&2.85& 26.1& 26.1&0.00&0.02&  2&-0.11&0\\
706070BP&0.70&2.70&4.60&-0.33&4.25& 0.84&2.85& 18.4& 18.4&0.00&0.02&  1&-0.04&0\\
753070BP&0.70&2.75&4.30& 0.29&4.35&-1.03&2.94&179.8&179.9&0.15&0.38&169& 0.13&0\\
753570BP&0.70&2.75&4.35& 0.19&4.55&-0.73&2.95&120.8&  0.0&0.16&0.46&224& 0.15&-\\
754070BP&0.70&2.75&4.40& 0.09&4.75&-0.43&2.94& 81.7&  0.0&0.14&0.58&244& 0.13&-\\
754570BP&0.70&2.75&4.45&-0.01&4.95&-0.13&2.95& 46.7&  0.0&0.15&0.67&247& 0.08&-\\
755070BP&0.70&2.75&4.50&-0.11&3.80& 0.17&2.92& 47.3& 25.8&0.05&0.32&118&-0.48&2\\
755570BP&0.70&2.75&4.55&-0.21&4.00& 0.47&2.90& 28.4& 28.3&0.00&0.01&  3&-0.05&0\\
756070BP&0.70&2.75&4.60&-0.31&4.20& 0.77&2.90& 20.0& 20.0&0.00&0.01&  2&-0.06&0\\
\\[-1mm]
\hline
\end{tabular}
\end{center}
\end{table*}


\end{document}